\documentclass[aps,pra,eqsecnum,superscriptaddress,groupedaddress,nofootinbib]{revtex4-2}

\usepackage{amssymb,amsmath}
\usepackage{subdepth}
\usepackage{graphicx}
\usepackage{sansmath}
\usepackage{bm}
\usepackage{soul}
\usepackage{framed}
\usepackage{marvosym}
\usepackage{upgreek}

\usepackage[breaklinks=true]{hyperref}
\hypersetup{
	colorlinks   = true, 
	urlcolor     = blue, 
	linkcolor    = red, 
	citecolor   = green 
}

\DeclareMathOperator{\tr}{tr}
\DeclareMathOperator{\Tr}{Tr}

\newcommand{\ket}[1]{\vert#1\rangle}
\newcommand{\bra}[1]{\langle#1\vert}
\newcommand{\braket}[2]{\left\langle #1 | #2 \right\rangle}
\newcommand{\proj}[1]{| #1\rangle\!\langle #1 |}
\newcommand{\ketbra}[2]{| #1\rangle\!\langle #2 |}

\newcommand{\inv}{{\,\text{-}\hspace{-1pt}1}}
\newcommand{\ad}{\mathrm{ad}}
\newcommand{\Ad}{\mathrm{Ad}}
\newcommand{\Z}{\mathcal{Z}}

\newcommand{\I}{\mathcal{I}}

\newcommand{\D}{\mathcal{D}}
\newcommand{\X}{\mathcal{X}}
\newcommand{\Y}{\mathcal{Y}}

\newcommand{\UU}{\mathcal{U}}

\newcommand{\E}{\mathcal{E}}
\newcommand{\Linv}[1]{\underrightarrow{#1}}
\newcommand{\Rinv}[1]{\underleftarrow{#1}}

\renewcommand{\O}{\mathcal{O}}

\newcommand{\R}{\mathbb R}
\newcommand{\C}{\mathbb C}

\newcommand{\OO}{\mathrm{O}}

\newcommand{\SL}{\mathrm{SL}}

\newcommand{\Hb}{\mathcal H}

\newcommand{\ankh}{\text{\Ankh}}
\newcommand{\fL}{\mathfrak{L}}
\newcommand{\fR}{\mathfrak{R}}
\newcommand{\fZ}{\mathfrak{Z}}
\newcommand{\fD}{\mathfrak{D}}
\newcommand{\rL}{\mathrm{L}}
\newcommand{\rR}{\mathrm{R}}

\newcommand{\Odot}{{\mathcal{O}\hspace{-0.575em}\raisebox{0.5pt}{$\boldsymbol{\cdot}$}\hspace{.275em}}}
\newcommand{\qddag}{{\rotatebox[origin=c]{90}{$\ddagger$}}}

\usepackage{color}

\begin{document}
	
	\title{Sequential Quantum Measurements and the Instrumental Group Algebra}
	
	\author{Christopher S. Jackson}
	\email{omgphysics@gmail.com}
	\affiliation{Perimeter Institute, Waterloo, Ontario N2L 6B9 Canada}

	\date{\today}
\begin{abstract}
	Many of the most fundamental observables | position, momentum, phase-point, and spin-direction | cannot be measured by an instrument that obeys the orthogonal projection postulate.
	Continuous-in-time measurements provide the missing theoretical framework to make physical sense of such observables.
	The elements of the time-dependent instrument define a group called the \emph{instrumental group} (IG).
	Relative to the IG, all of the time-dependence is contained in a certain function called the \emph{Kraus-operator density} (KOD), which evolves according to a classical Kolmogorov equation.
	Unlike the Lindblad master equation, the KOD Kolmogorov equation is a direct expression of how the elements of the instrument (not just the total quantum channel) evolve.
	
	Shifting from continuous measurements to sequential measurements more generally, the structure of combining instruments in sequence is shown to correspond to the convolution of their KODs.
	This convolution promotes the IG to an \emph{involutive Banach algebra} (a structure that goes all the way back to the origins of POVM and C*-algebra theory) which will be called the \emph{instrumental group algebra} (IGA).
	The IGA is the true home of the KOD, similar to how the dual of a von Neumann algebra is the true home of the density operator.
	Operators on the IGA, which play the analogous role for KODs as superoperators play for density operators, are called \emph{ultraoperators} and various important examples are discussed.
	Certain ultraoperator-superoperator intertwining relations are also considered throughout, including the relation between the KOD Kolmogorov equation and the Lindblad master equation.
	The IGA is also shown to have actually two distinct involutions: one respected by the convolution ultraoperators and the other by the quantum channel superoperators.
	Finally, the KOD Kolmogorov generators are derived for jump processes and more general diffusive processes.
\end{abstract}
	
	\maketitle
	
\pagebreak	
\tableofcontents
\pagebreak

\pagebreak

\section{Introduction: The Instrument Manifold Program}\label{Intro}

This paper is a development of the Instrument Manifold Program.~\cite{jackson2021spin, jackson2023weylheisenberg, jackson2023positive}\\

The Instrument Manifold Program (IMP) is a method for analyzing the measuring instruments that are generated by continuous measurement, primarily of the diffusive type.
This method originally emerged from the discovery that the measurement of spin-direction corresponding to the spin-coherent positive-operator-valued measure (POVM) \cite{DAriano2002, Massar1995} can be realized universally (that is for any, arbitrarily large, total angular momentum quantum number `$j$') by a diffusive simultaneous measurement of the three orthogonal components of total angular momentum~\cite{jackson2021spin,shojaee2018optimal,deutsch2010quantum}, now called the \emph{isotropic spin measurement} (ISM)~\cite{jackson2021spin,jackson2023positive}.
Such a measurement of the spin-direction is a form of informationally-complete quantum tomography analogous to the measurement of phase-point for continuous-variable systems such as a single mode of light, where the phase space for spin is a sphere instead of a plane.
On the measurement of phase-point as a form of informationally-complete quantum tomography, there have been several approaches~\cite{leonhardt1997measuring,klauder1960action,husimi1940formal}: the Arthurs-Kelly measurement~\cite{arthurs1965simultaneous,she1966simultaneous}, optical heterodyne~\cite{personick1971image}, optical homodyne tomography~\cite{lvovsky2009continuous}, and diffusive optical heterodyne~\cite{wiseman1996measurement,Goetsch1994a}.
While \cite{DAriano2002} showed that the spin-analog of \cite{arthurs1965simultaneous} does not work for $j>1/2$, the closest phase-point analog to ISM of the approaches just mentioned would be diffusive optical heterodyne~\cite{wiseman1996measurement,Goetsch1994a}.

The exact phase-point analog of ISM seems to have first been examined in \cite{barchielli1982model}, although the authors did not understand that this measurement resulted in the standard coherent POVM.
Therefore \cite{jackson2023weylheisenberg} was written to demonstrate how the IMP is able to analyze extensively the behavior and result of this diffusive measurement, now called \emph{SPQM} (the \emph{simultaneous momentum and position or ``P and Q'' measurement}) in honor of Antoine Barchielli at the University of Milan, whom the author considers the pioneer of the theory of diffusive measuring instruments~\cite{barchielli1982model,barchielli2006continual,barchielli2009quantum}.
In the application to SPQM~\cite{jackson2023weylheisenberg}, many of the key elements of the IMP were much easier to give explicit expressions than in ISM~\cite{jackson2021spin}, because the canonical commutation relations are simpler than the angular-momentum commutation relations.

Having explored ISM and SPQM extensively, \cite{jackson2023positive} was written to explain how the IMP is able to analyze the general problem of simultaneously measuring non-commuting observables, what in this paper will be refered to as \emph{diffusively measuring non-commuting observables simultaneously} (DMNCOS) to contrast it from the other types of simultaneous measurements of noncommuting observables mentioned~\cite{arthurs1965simultaneous,personick1971image,she1966simultaneous}.
Fundamentally, the general problem of DMNCOS boils down to whether the time-ordered exponential that maps any particular continuous measurement record to its instrument element can be computed.
Geometrically, this is related to whether the manifold of instrument elements is finite-dimensional or not.
If it is, the instrument was said to be \emph{principal}, else it was deemed \emph{chaotic}~\cite{jackson2023positive}.

Mathematically, the IMP consists of one main concept, two basic tools, and then a large handful of technical details.
The main concept is what the author calls the \emph{instrumental group} ($IG$) and the basic tools are the so-called \emph{Kraus-operator density} (KOD) and the \emph{modified Maurer-Cartan stochastic differential} (MMCSD).
The IG is a Lie group that can be identified to contain the elements of the measuring instrument at all times.
The KOD is a distribution function of the IG with respect to its (left-invariant) Haar measure and describes at every moment in time how the Kraus operators (or more general elements of the instrument) are distributed.
The MMCSD is a tool that is particularly well-suited for handling the stochastic differential equations that describe Brownian motion within a group, allowing one to compute/locate within the IG the instrument element that results from any given continuous measurement record.\\

An honorable mention must be given to a work by Prahlad Warszawski, et al. \cite{Warszawski2020a}.
Although the authors were not able to uncover the calculus and analysis native to the IG, they certainly recognized its existence for continuous variable systems.
With this recognition, they had some key insights into the nature of the KOD (or the closely related \emph{ostensible distribution} as they call it) and how to calculate the POVM of a continuous measurement.
The author of this paper is incredibly relieved to find that there is at least one set of authors out there who has begun to recognize the significance of the IG.\\

Returning to the IMP, the tools and perspectives that come with it still require almost anyone who works in quantum measurement to step out of their comfort zone because these basic tools are both non-standard and rather sophisticated.
The purpose of this paper is to focus on the KOD and explain it from a different perspective, centering the discussion on the more discrete concept of sequential measurement rather than on continuous measurement as considered previously.
By doing this, much of the basic structure of combining instruments in sequence (either discretely or continuously) can be addressed more directly.
By stepping away from continuous processes and into sequential processes, the tools of stochastic calculus are replaced by the tools of functional analysis.
Specifically, stochastic differential equations and Kolmogorov equations~\cite{jackson2023positive} are replaced by the multiplication of IG elements and the group-convolution of functions on the IG.

Wonderfully, every Lie group defines a form of analysis that generalizes standard real and complex analysis.
If real analysis could be boiled down to only five essential tools, arguably they are the derivative, the integral, the Taylor series, the Fourier expansion, and the convolution.
All five of these tools obtain their structure from one essential property of the real line, $\R$, that it forms a group under addition, thus giving rise to the notion of translation, $x \mapsto x +a$.
The first four are perhaps the most fundamental:
The usual derivative $d/dx$ is the unique translation-invariant vector field,
the integral $\int dx$ is the unique translation-invariant (Haar) measure,
the exponential operator $e^{a \frac{d}{dx}}$ of the Taylor series is the translation generated by $d/dx$ for variable $a$,
and the exponential kernel $e^{ikx}$ of the Fourier transform as a function of $x\in \R$ is an irreducible representation of $\R$ for every $k$.
The convolution is then the crown within which these fundamental jewels sit, combining and intertwining in glorious ways.
All of these tools still exist if the real line is replaced by a general finite-dimensional Lie group and this paper demonstrates how these tools of general group analysis are directly relevant to the problem of sequential measurements, both discrete and continuous.\\

The flow of this paper is therefore as follows:

For those in the earlier stages of research in theoretical quantum science, a quick summary of all the more standard tools in quantum science that will be needed | instruments, channels, and Lindbladians | can be found in appendix \ref{StandardTheory}.
Having this standard theory under one's belt, the following sections move into the newer concepts of the IMP~\cite{jackson2023positive}, where in this paper these newer concepts will be framed in terms of group analysis and held together through the central concept of group convolution, the analytic rule for combining instruments in sequence.
This machinery of the IMP is relatively heavy, so sections \ref{CommAnal} and \ref{Prelude} essentially act as a second and third introductions with increasing mathematical intensity.

Section \ref{CommAnal} revisits the much older, classical problem of invariant heat diffusion.  However, with a modern perspective and notation, the approach should make clear that exactly the same structures found in real analysis (though noncommutative in general) are present in the general problem of continuous and sequential measuring instruments.

Section \ref{Prelude} then begins to march into the general theory of continuous and sequential measurement, albeit in a streamlined fashion, where the focus is to introduce all of the details, while not digging into them just yet for the sake of displaying how all of the key ideas follow from one another.
Section \ref{A} starts off with a discussion of instantaneous measurements and their limitations.
Section \ref{B} then moves into the general problem of continuous measurement in terms of the basic time-ordered exponential that maps continuous measurement records to their corresponding instrument elements or Kraus operators, emphasizing the fundamental concept of the instrumental group (IG) and its universal nature.
Section \ref{C} then introduces the Kraus-operator density (KOD) and the Kolmogorov equation that governs its evolution in time when the measurements are continuous.
Section \ref{D} then begins to explain how the more standard decoherence channels and their Lindblad master equations are in fact representations of the KOD and its Kolmogorov equation.
Section \ref{E} then begins to move the discussion away from continuous measurements and their differential equations to sequential measurements and their group convolutions, in particular explaining that the composition of decoherence channels is a representation of the group convolution of KODs.
Finally, section \ref{F} explains how the group convolution promotes the space of absolutely-integrable functions of the IG (which include the KOD) to a group algebra, the so-called \emph{instrumental group algebra} (IGA).\\

Having given the reader a first exposure to all the technical elements involved in completely recasting the analysis of continuous measurement in terms of the IG and KOD, we then start over from the beginning, with the intention of explaining the details.\\

With the dimension of time, measurements can be put together in a variety of ways
| continuously, simultaneously, and sequentially | which can be considered to take place in an \emph{instrumental group} (IG), the subject of section \ref{Instruments}:

Section \ref{ContMeas} explains the fundamental problem of \emph{continuous} measurements, calculating the time-ordered product of a sequence of weak Kraus operators.
It begins by immediately considering a sequence of two instruments as the convolution of their operation sets.
The two weak instruments that generate the most common types of continuous Markov process are introduced, the jumpy Poisson process and the diffusive Wiener process.
(Appendix \ref{SMI_Models} includes a detailed derivation of these two weak instruments from basic system-meter interactions.)
The IG is then defined and the relation between the free product of groups and the convolution of operation sets is explained.
Then the fundamental problem is stated more precisely and the topics of automation and chaos are briefly discussed.

Section \ref{SimultMeas} explains the basic observation that the Kraus operators of weak measurements commute, even when the Lindblad operators that generate them do not.
With this observation in mind, it then makes perfect sense to consider multiple weak measurements with noncommuting Lindblad operators as \emph{simultaneous}.
Such simultaneous measurements can be run for finite amounts of time, thus generating new types of instruments such as those generated by the general ``DMNCOS'' procedure.
Critically, it is then explained that the finite-time instrument that is defined by such a procedure has its definition solely in the commutation relations of the Lindblad operators that generate their Kraus operators, therefore making the structure of the instrument \textbf{universal}, i.e. independent of Hilbert space.
This emphasis is very important because it reinforces the basic fact that in physical examples, these commutation relations are based in physical observables that are more fundamental than the Hilbert spaces on which any of these Kraus operators act, a point that is echoes back to a point made by Kraus, referring to `` [...] Bohr, who repeatedly stressed the importance of the classical nature of measuring instruments for the understanding of quantum mechanics.''~\cite[p.~2]{kraus1983states}
A brief mention is then made to the IG for simultaneously measuring a canonical pair of phase-space quadratures and the IG for simultaneously measuring the three orthogonal angular momentum components and the reader is referred to previous work for further details.

Section \ref{OD} then finishes with a fresh start, assuming an IG from the get-go and considering the instrument as a representation of the IG.
This introduces the notion of a Kraus-operator distribution, a precursor to the Kraus-operator density that does not yet have the properties of a function.
The convolution of two distributions over the IG is then defined and related to the operation-set convolution of two instruments in \emph{sequence}.
Finally, it is pointed out that this convolution of distributions already expands what it means for the total operations of a sequence of instruments to satisfy a group property.\\

Having reviewed and reframed the recent developments that are relevant to the current discussion, we finally dig into our main results, the analytic-algebraic structure that emerges from the IG convolution, which is given the name \emph{instrumental group algebra} (IGA).\\

Just like the real line, the IG is the domain of an entire suite of analytic tools | with invariant integrals, invariant derivatives, function spaces, and intertwined differential-algebraic equations | where functions such as the Kraus-operator density (KOD), the primary object of section \ref{KODD}, can be calculated, propogated, and varied:

Section \ref{ODF} begins by introducing the all-important Haar \emph{integral} measures.
These measures allow one in turn to define the convolution of two functions and the density function of a distribution.
The relationship between the convolution of functions and the convolution of distributions is explained.
Finally, the definition of the Kraus-operator density (KOD) is fixed relative to the left-invariant measure, a choice that stems from the convention that time is considered to move from right to left when multiplying a sequence of Kraus operators.

Section \ref{TransDiff} introduces the left- and right-translation ultraoperators.
With them, convolution by a function can also be considered as an ultraoperator.
The relations between these left- and right-convolution ultraoperators amongst themselves and each other are discussed.
Then right- and left-invariant \emph{derivatives} are defined.
It is explained how these invariant derivatives generate the translation ultraoperators in the same analytic way that the usual derivative of a real line generates the Taylor expansion.
Finally, ultraoperator commutators of the invariant derivatives are discussed in terms of the Lie algebra of the IG.

Section \ref{AdjAlg} discusses the \emph{function spaces} that are defined by the IG and its left-invariant measure.
Among the various norms one could give the IG-functional space, it is only the $\ell^1$ norm that is closed under convolution and the IG algebra (IGA) is finally defined as the Banach algebra of absolutely-integrable functions.
The IG inner product is then introduced, the Kolmogorov adjoint is defined, and the Kolmogorov adjoints of the various ultraoperators are considered.
Then the IGA is shown to have two involutions, named after Cartan and Gelfand, which respect the convolution product if the IG is unimodular.
Finally, their relations to the Kolmogorov ultraoperator and Hilbert-Schmidt superoperator adjoints are discussed.

At very last, section \ref{ODiff} returns to the problem of continuous measurement and uses the ultraoperator formalism to rederive the KOD Kolmogorov equations in a style that is much like the reverse of section \ref{CommAnal}.
The various manifestations of the Markovian group property | either by composition, convolution, or addition | are summarized.
The role of the instrument elements in intertwining the differential ultraoperators with the algebraic superoperators is then introduced.
This fundamental intertwining relation is then used to formally derive the KOD Kolmogorov forward equation, showing along the way that the instrument elements also intertwine the Markov ultraoperator with the quantum channel superoperator.
The Kolmogorov generator is then calculated for both diffusive and jump processes.
Finally, it is shown that the instrument elements also intertwine both Kolmogorov generators with the Lindbladian dissipator, as one might expect, thus transforming the differential Kolmogorov equation into the algebraic Lindblad master equations, in noncommutative analogy with the Laplace transform.\\

The paper then finishes with some concluding remarks in section \ref{Conclude}.\\

\section{A Commutative Analogy: The Real Line as an Instrument}\label{CommAnal}

The Instrumental Group (IG), Kraus-Operator Density (KOD), IG convolution, IG Algebra (IGA), and convolution ultraoperators to come, as well as their relationship to the usual density operators and channel superoperators of standard quantum theory, are a lot to swallow for those who have yet to ask for an autonomous, general theory of sequential measurement.
If, however, we consider a commutative IG, then all of these tools and concepts can easily be recognized by anyone familiar with classical real analysis.
To see the forest from the trees, namely the \textbf{universal} measuring process from the quantum channels,
it therefore seems quite instructive to go through this.
The baroque notation and strange factors of 2 will look funny, but they are chosen with the intention of making the analogy with the noncommutative case and its application to sequential measurement as obvious as possible.
It is very interesting to note how almost all of these key elements seem to stem from Laplace.\\

Consider the simplest case of diffusion, that of a function $D_t(x)$ on the real line, $\mathbb{R} \ni x$, described by the age-old heat equation,
\begin{equation}
	\frac{1}{\kappa} \partial_t D_t(x) = \frac{1}{2} \partial_x^2[D_t](x)
\end{equation}
where $t$ is the time, $\kappa$ is a diffusion constant with units of inverse time, and $\partial_x$ and $\partial_t$ are partial derivatives.
This diffusion can also be described in more modern language as a translation-invariant, differential \emph{Markov process}, in which case
\begin{equation}
	D_{t+dt}(x) = \tilde{\fZ}_{dt}^\ankh[D_t](x)
\end{equation}
with Markov kernel
\begin{align}
	d(dW) \, G_{dt}(dW) = \frac{d(dW)}{\sqrt{2\pi dt}}e^{-\frac{dW^2}{2 dt}}
\end{align}
and Markov (ultra)operator
\begin{align}\label{MarkovOp}
	\tilde{\fZ}_{dt}^\ankh &\equiv \int_\R d(dW) \, G_{dt}(dW) e^{-\sqrt{\kappa} dW \partial_x}\\
	&= e^{\kappa dt \frac{1}{2} \partial_x^2}.
\end{align}
The $\ankh$ symbol denotes the (ultra)operator adjoint, corresponding to integration-by-parts with respect to the measure $dx$ (where it assumed $D_t(x)$ goes to zero for $x\to\pm\infty$), so in particular $\partial_x^\ankh = -\partial_x$ and $x^\ankh = x$.
For any real number $\hat{\ell} \in \R^*$ (where $\R^*$ is a copy of the real line that denotes the space of positive representations, to be distinguished from the points in the original group, $\R$) the function $e^{2\hat{\ell}x}$ is an eigenfunction of the Markov operator
\begin{equation}\label{remarkable}
	\tilde{\fZ}_{dt} [e^{2\hat{\ell} x}] = \tilde{\Z}_{dt}^{(\hat{\ell})} e^{2\hat{\ell} x}
\end{equation}
with eigenvalue equal to the two-sided Laplace transform of the Markov kernel (evaluated at $\hat{\ell}$),
\begin{align}
	\tilde{\Z}_{dt}^{(\hat{\ell})}
	&\equiv \int_\R d(dW) \, G_{dt}(dW) e^{2 \hat{\ell}\sqrt\kappa dW}\\
	&=e^{2\hat{\ell}^2\kappa dt},
\end{align}
known in classical probability theory as the characteristic function of $G_{dt}$, a tool also invented by Laplace.

Equation \ref{remarkable} is rather remarkable and it exists for a very simple reason: the domain of the Markov process, the real line $\R$, with the operation of addition is a \emph{group} with a measure that is invariant under translations, $dx = d(x+\xi)$.
Such invariant measures are the so-called Haar measures and groups which have them are said to be \emph{locally compact}.
The translation (ultra)operators $e^{-\xi \partial_x}$ form the so-called \emph{regular representation} of $\R \ni \xi$ and each eigenfunction $e^{-2\hat{\ell} x}$ is said to carry a so-called \emph{irreducible representation}, $e^{-2\hat{\ell} \xi}$, of $\R \ni \xi$ .
Moreover, because it makes sense to consider linear combinations of the translation operators as seen in the Markov (ultra)operator of equation \ref{MarkovOp}, these eigenfunctions carry a representation of an even more sophisticated structure, the so-called \emph{group algebra} $L^1(\R)$.

Specifically, any real-valued function $f\in L^1(\R)$ defines a so-called convolution (ultra)operator on $L^1(\R)$,
\begin{equation}
	\fZ_f^\ankh \equiv \int_\R d\xi \, f(\xi) e^{-\xi \partial_x},
\end{equation}
and for any two such functions $g,f\in L^1(\R)$ it is easy to verify that
\begin{equation}
	\fZ_g^\ankh \circ \fZ_f^\ankh = \fZ_{g*f}^\ankh
\end{equation}
where $g*f$ denotes the convolution
\begin{equation}
	g*f(x) \equiv \int_\R d\xi \, g(\xi)f(x-\xi),
\end{equation}
also first discovered by Laplace.
Meanwhile, the two-sided Laplace transforms,
\begin{equation}
	\Z_f^{(\hat{\ell})} \equiv \int_\R d\xi f(\xi)e^{2\hat{\ell}\xi}
\end{equation}
for any fixed $\hat{\ell}$, represent convolution by scalar multiplication is easily seen by verifying
\begin{equation}
	\Z_{g*f}^{(\hat{\ell})} \equiv \Z_{g}^{(\hat{\ell})}\Z_{f}^{(\hat{\ell})},
\end{equation}
an expression that will be recognized as the convolution theorem.\\

In the case of continuous quantum measurements, the situation is \emph{exactly} the same, except that the relevant group (the IG) is generally noncommmutative, with irreducible representations (irreps) which are therefore multi-dimensional.
Imbued with the Markov process, the group $\R \ni dW$ became an \emph{instrument} with instrument elements $e^{2 \hat{\ell} dW}$, Hermitian Lindblad operators $\hat{\ell}$, and completely positive total operations $\Z_{dt}^{(\hat{\ell})}$.
Of course, because the group $\R$ is commutative, the irreps are only one-dimensional and so the only state available to occupy an irrep would be the unit operator | not nearly as interesting as when the irreps are multi-dimensional.

Uninteresting as the one-dimensional states are, the analogy between quantum instruments and diffusion on the group $\R$ should be brought to completion.
First, the total operation $\tilde{\Z}_{dt}^{(\hat{\ell})}$ can obviously be made to be trace-preserving (in this case meaning equal to one) if we divide the instrument elements by the appropriate amount,
\begin{align}
	\Z_{dt}^{(\hat{\ell})}
	&\equiv e^{-2\hat{\ell}^2\kappa dt}\tilde{\Z}_{dt}^{(\hat{\ell})}\\
	&= \int_\R d(dW) \, G_{dt}(dW) e^{-2\hat{\ell}^2\kappa dt+2 \hat{\ell} \sqrt\kappa dW}\\
	&=1.
\end{align}
In this case the instrument elements acquire a second dimension,
\begin{equation}\label{baby}
	\O_{r,x} = e^{-2\hat{\ell}^2r+2 \hat{\ell}x}
\end{equation}
and the corresponding Markov operator becomes
\begin{align}
	\fZ_{dt}^\ankh &\equiv \int_\R d(dW) \, G_{dt}(dW) e^{-\kappa dt \partial_r -\sqrt{\kappa} dW \partial_x}\\
	&= e^{\kappa dt \left(-\partial_r + \frac{1}{2} \partial_x^2\right)}.
\end{align}
Finally, the invariant derivatives $\partial_r$ and $\partial_x$ can be expressed in a more Lie-algebraic style if we denote a general element of the instrumental group by 
\begin{equation}
	g = e^{-\ell^2 r + \ell x}
\end{equation}
in which case
\begin{equation}
	\Rinv{\ell^2} = -\partial_r
	\hspace{50pt}
	\text{and}
	\hspace{50pt}
	\Rinv{\ell} = \partial_x
\end{equation}
and therefore
\begin{equation}
	\fZ_{dt}^\ankh = e^{\kappa dt \left(\Rinv{\ell^2}+ \frac{1}{2} \Rinv{\ell}\Rinv{\ell}\right)},
\end{equation}
in exact analogy with equations \ref{FancyDs} and \ref{ObviousExp}.
Meanwhile, the Kraus operators and instrument elements (equation \ref{baby}) would also be denoted
\begin{equation}
	K_g = e^{-\hat{\ell}^2 r + \hat{\ell} x}
	\hspace{50pt}
	\text{and}
	\hspace{50pt}
	\O_g = K_g K_g^\dag.
\end{equation}

\section{Prelude: Sequences of Measurement in Time and their Structure}\label{Prelude}

\subsection{There is a need to get past the idea of instantaneous measurement.}\label{A}

In quantum theory, observables and states are distinct objects, each a certain kind of operator over a Hilbert space.
In particular circumstances, an observable can correspond to directly to a choice of Hermitian operator and the act of observation can be understood to be the random appearance of one of its eigenvalues with a probability that can be calculated by the Born rule.
In this case, the appearance of the eigenvalue is also understood to signify that the state has projected onto the corresponding eigenspace.
This idea of observation, in particular where the post-measurement state is projected, is called a von Neumann measurement or the Luders rule~\cite{vonneumann1932mathematical,busch2009luders,debrota2019luders}.
Useful as it is, the von Neumann-Luders rule fails to make sense of perhaps the most fundamental of observables, the position of a dynamic particle, because the eigendistributions of the position operator are not square-integrable and therefore not sensible state-preparations.  (For a more detailed explanation, see the end of section \ref{POVMs}.)

Meanwhile, operators do not generally commute because they must also represent physical transformations such as rotations or canonical displacements.
Noncommuting observables therefore become a central feature of quantum measurement theory.
Often it is said that noncommuting observables cannot be measured simultaneously.
Of course, what this refers to is a technical point, that two Hermitian operators cannot be simultaneously diagonalized.
Yet measurements such as optical heterodyne and dual homodyne are (from an experimental point-of-view) obviously able to perform simultaneous measurements of the noncommuting quadratures of a field mode~\cite{leonhardt1997measuring,shapiro1984phase,yuen1973multiple,personick1971image}.
The von Neumann-Luders rule thus fails again to make sense of the situation. 

Despite the abject failure of the von Neumann-Luders rule,
position measurements and simultaneous measurements of noncommuting observables make perfect sense if every von Neumann measurement is not assumed to occur instantaneously, but rather considered to result from a finite-time process or the asymptotics of one.
If a pair of such continuous-in-time processes seperately result in von Neumann measurements of noncommuting observables, it nonetheless also makes sense to run these continuous processes simultaneously, resulting in a new, third measurement that is not of the von Neumann type.

Continuous-in-time processes that result in a von Neumann measurement asymptotically can be seperated into two types, both which involve the concept of an indirect measurement.
The first type is where a single meter interacts with the system observable of interest for a finite amount of time after which the meter is finally measured, registering a single random variable.
Measurements of this first type were also considered by von Neumann in the early 1930s and later revisited by others in the mid 1960s\cite{vonneumann1932mathematical,arthurs1965simultaneous,she1966simultaneous,gordon1966simultaneous}.
In the second type, at every infinitesimal time a new meter weakly interacts with the system observable and is immediately measured, so what is registered overall is a sequence of random variables, one for each infinitesimal increment of time.
Measurements of this second type were originally called forms of \emph{continual monitoring} but these days they tend to be called \emph{continuous measurements}~\cite{srinivas1981photon,barchielli1982model,Goetsch1994a,wiseman1994quantum,Silberfarb2005a,jacobs2006straightforward,barchielli2009quantum,wiseman2009quantum,carmichael2009open,cook2014single,gross2018qubit}.

\subsection{With continuous measurements, what kinds of instruments are possible?}\label{B}

The kinds of instruments that are produced by measuring noncommuting observables simultaneously are vastly underexplored, with only a handful of fundamental advances, few and far in between, mostly limited to measuring position and momentum~\cite{arthurs1965simultaneous, barchielli1982model, wiseman1996measurement, leonhardt1997measuring, lvovsky2009continuous, DAriano2002}.
For this purpose, the \emph{instrument manifold program} (IMP) was designed to begin a systematic exploration~\cite{jackson2023positive}.
So far, investigations have been focused extensively on two continuous measurements and their \textbf{universal} properties, the so-called \emph{isotropic spin measurement} (ISM)~\cite{jackson2021spin} and the \emph{simultaneous momentum and position measurement} (SPQM)~\cite{jackson2023weylheisenberg}.

ISM is a simultaneous diffusive measurement of the three orthogonal spin-components, ``$J_x$, $J_y$, and $J_z$''.
ISM stands as the first method discovered that proposes a practical realization of the so-called spin-coherent POVM for arbitrarily large spins because of its \textbf{universal} design~\cite{DAriano2002, shojaee2018optimal, jackson2021spin}.
SPQM is the simultaneous diffusive measurement of the canonical pair of phase space quadrature-components, ``$P$ and $Q$''.
SPQM appears to be the first simultaneous diffusive measurement ever considered~\cite{barchielli1982model, barchielli2006continual, barchielli2009quantum}, although it appears that an understanding of the full behavior of this measuring instrument has only found significant progress through the IMP~\cite{jackson2023weylheisenberg}.
Both of these measurements have measurement operators (a.k.a. Kraus operators) which generate 7-dimensional Lie groups, respectively called the \emph{instrumental spin group} ($\mathrm{ISpin}(3,\C)$) and the \emph{instrumental Weyl-Heisenberg group} ($\mathrm{IWH}$).

These two groups are examples of what are generally called \emph{instrumental groups} (IGs).
Every simultaneous measurement translates in general to a certain motion or process in the IG.
Put another way, the IG is to be thought of as a static space of possible measurement actions/Kraus operators.
Over time, the possible Kraus operators populate the IG in a specific manner that corresponds collectively to the stochastic motion of a certain distribution.

To analyze and navigate this motion, some sophisticated mathematical tools are required, tools that are not yet familiar to most quantum scientists, hence the recent introduction of the IMP~\cite{jackson2023positive}.
The need for this sophistication is simple, it is because the IG motions of a continuous simultaneous measurements boil down to analyzing a time-ordered product that generates the finite-time Kraus operators, a problem whose intricacy stems precisely from observable noncommutativity.
For example, if the measurements are diffusive, then these products take the form of a time-ordered exponential, the Kraus operators thus affording the expression
\begin{equation}\label{PileUp}
	\sqrt{\D\mu[d\vec{W}_{[0,T)}]} \; K^{(\vec{L})}\!\left[d\vec{W}_{[0,T)}\right]
	= \sqrt{\D\mu[d\vec{W}_{[0,T)}]}\; \mathcal{T}\!\exp \left(\int_0^{T-dt} \hat{\updelta}^{(\vec{L})}(d\vec{W}_t)\right)
\end{equation}
with
\begin{equation}
\hat{\updelta}^{(\vec{L})}(d\vec{W}) \equiv -\hat{Q}_{\vec{L}}\,\kappa dt + \sum_{\mu = 1}^n \hat{L}_\mu\sqrt\kappa dW^\mu
\hspace{25pt}
\text{and}
\hspace{25pt}
\hat{Q}_{\vec{L}} \equiv \frac12 \sum_{\mu = 1}^n \hat{L}_\mu^\dag \hat{L}_\mu + \hat{L}_\mu^2
\end{equation}
where $T$ is the overall time duration of the measurement, $\vec{L} \equiv (L_1,L_2,\ldots,L_n)$ is a finite tuple of Lindblad operators (which generalize the Hermitian observable), $d\vec{W}_{[0,T)} \equiv \{d\vec{W}_t\}_{t=0}^{T-dt}$ is a sequence of Wiener increments (often called the measurement record) representing the raw data output by the instrument, $\D\mu[d\vec{W}_{[0,T)}]$ is the Wiener measure, and $\hat{X}$ denotes a representation of the abstract Lie algebra element $X$.
(See Appendix \ref{SMI_Models} for a derivation of $\hat{\updelta}^{(L)}$.)

The IG is then the set of all Kraus operators that can result from these time-ordered exponentials and their inverses.
Because the result of the time-ordered exponential is locally determined by the commutator structure of the infinitesimal generators alone, the IG is to be considered as an abstract manifold, the so-called \textbf{universal} instrument, which encompasses all possible representations.
What this means mathematically is that the Kraus operators of equation \ref{PileUp} can be considered as representations of a \textbf{universal} time-ordered exponential $\gamma \in \text{IG}$, with the \emph{universal property}
\begin{equation}
	K^{(\vec{L})}\!\left[d\vec{W}_{[0,T)}\right] = K_{\gamma^{(\vec{L})}\!\left[d\vec{W}_{[0,T)}\right]}
\end{equation}
where
\begin{equation}
	K_{yx} = K_y K_x
	\hspace{25pt}
	\text{and}
	\hspace{25pt}
	\gamma^{(\vec{L})}\!\left[d\vec{W}_{[0,T)}\right] = \mathcal{T}\!\exp \left(\int_0^{T-dt} \updelta^{(\vec{L})}(d\vec{W}_t)\right).
\end{equation}
The abstracted Lindblad operators, $L_\mu$ without a hat, have the same commutation relations as the $\hat{L}_\mu$ but also forget whatever additional spectral information (Hilbert space, highest weights, etc.) may have been included with the $\hat{L}_\mu$.  The remaining operators without hats | $Q^{\vec{L}}$, $\updelta^{\vec{L}}$, and $\gamma^{\vec{L}}$ | are understood to exist in the (closure of the) universal enveloping algebra of the IG Lie algebra.
Technically, we would write $K = \hat{\gamma}$ if we wanted to be consistent with the hat notation denoting representation.
However, much of the purpose of this paper is to clarify the distinction between Kraus operators, $K$, and universal trajectories, $\gamma$, so it should be helpful to leave these particular notations different.

\subsection{There is an analytic structure to applying instruments in sequence.}\label{C}

The primary tools of the IMP are the so-called \emph{modified Maurer-Cartan stochastic differential} (MMCSD) and the \emph{Kraus-operator density} (KOD).
With the assistance of any of a set of standard group decompositions, the job of the MMCSD is to transform the raw measured data into a co\"ordinate system that locates the position in the IG.
In this paper, we will pay virtually no attention to the MMCSD.
This is because the instrument structure we are getting at is less about the particular outcomes that would arise in a measurement and more about where these outcomes end up collectively, which is precisely the function of the KOD.

The KOD is a positive function $D_T(x)$ defined with respect to the left-invariant Haar measure of the IG,
\begin{equation}
	d_\rL x\, D_T(x) = d_\rL x \int \D\mu[d\vec{W}_{[0,T)}] \,\delta_{\text{IG}}\!\left(\gamma[d\vec{W}_{[0,T)}]^\inv x\right)
\end{equation}
where $d_\rL x$ is the left-invariant Haar measure, $\delta_{\text{IG}}(x)$ is the IG delta-function, and the dependence on any Lindblad operators has been dropped ($\gamma = \gamma^{\hat{L}}$).
(See Appendix \ref{delta} for details on the delta-function.)
What the KOD essentially does is count the number of measurement records that give the same Kraus operator $x$ at any given time $T$, therefore collecting the actions of a continuous measuring instrument.

In the focus on ISM and SPQM, the IMP has thusfar only considered infinite sequences of weak measurements, i.e. diffusive measurements, where the KOD evolves according to a \emph{Kolmogorov forward equation}, a.k.a. a  \emph{Fokker-Planck equation}.
In general, the continuous simultaneous measurement of a set of noncommuting observables has an instrument of the form
\begin{equation}\label{introinstrument}
	\I_T = \Big\{d_Lx\,D_T(x) \O_x\Big\}_{x\in \text{IG}}
\end{equation}
where $T$ again is the time duration of the measurement, $d_\rL x \,$ is the left-invariant Haar measure of the IG, $\O_x$ is the superoperator with action $\O_x (\rho) = K_x \rho K_x^\dag$, $K_x$ is the Kraus-operator representation ($K_{yx} = K_y K_x$) of the group element $x\in \text{IG}$, and $D_T(x)$ is the KOD that satisfies the Fokker-Planck-Kolmogorov equation (FPKE)
\begin{equation}\label{forward}
	\frac1\kappa \frac{\partial }{\partial t} D_t(x) = \fD^\ankh[D_t](x)
\end{equation}
with initial condition $D_0(x) = \delta_{\text{IG}}(x)$, where $\fD^\ankh$ is the so-called the (Kolmogorov) forward generator~\cite{jackson2023positive}.
(The $\ankh$ is a type of adjoint which will be explained soon enought, but for the moment please take the symbol $\fD^\ankh$ wholesale.)
By introducing the right-invariant derivative,
\begin{equation}
	\Rinv{X}[f](x) \equiv \lim_{h\to\infty}\frac{f(e^{hX}x)-f(x)}{h},
\end{equation}
two examples can be presented which have been studied extensively:
The forward generators of ISM~\cite[p.~21]{jackson2021spin} and SPQM~\cite[p.~14]{jackson2023weylheisenberg},
\begin{equation}\label{FancyDs}
	\fD_\text{ISM}^\ankh = \Rinv{\vec{J}^{\,2}} + \frac12\left(\Rinv{J_x}\Rinv{J_x} + \Rinv{J_y}\Rinv{J_y} + \Rinv{J_z}\Rinv{J_z}\right)
	\hspace{25pt}
	\text{and}
	\hspace{25pt}
	\fD_\text{SPQM}^\ankh = 2\Rinv{{H_o}} + \frac12\left(\Rinv{Q}\Rinv{Q} + \Rinv{P}\Rinv{P}\right)
\end{equation}
where $\vec{J}^{\,2} \equiv J_x^2 + J_y^2 + J_z^2$  is the quadratic Casimir operator for spin and $H_o \equiv \frac12 (Q^2 + P^2)$ can be recognized as the same operator that would generate simple harmonic motion (if it were instead the Hamiltonian of a unitary process.)

Generators of the KOD FPKE are operators that act on functions of the IG.
To distinguish them from density operators and their superoperators, objects already familiar to quantum scientists, any operator on the space of functions of the IG will be called an \emph{ultraoperator}.
Ultraoperators are generally more analytic in purpose (like differential operators) than their quantum counterparts, so the distinct name should also help to keep this difference in mind.

Again, because the right-invariant derivatives are defined directly in terms of the IG, the KOD FPKE makes sense without any knowledge of the Hilbert space representation in which the state being measured may be and this is precisely why the evolution of the KOD must be understood as \textbf{universal}.
Of course, the instrument does not register data without a state to measure.
Nevertheless, when a state is measured, the Born rule probability of registering an element $x$ of the IG is of the form
\begin{equation}
	d_\rL x \, P(x|\rho_0;T) = d_\rL x \, D_T(x) \tr (\rho_0 \,K_x^\dag K_x).
\end{equation}
It is therefore crucial to appreciate that for diffusive measurements, the Born rule consists of three distinct factors: the KOD being the only factor that depends on time, the trace being the only factor that depends on the state, and the Haar measure itself which is neither time- nor state-dependent.

\subsection{There is a need to get past the state-centric approach to continuous measurement.}\label{D}

\begin{table}[h!]
	\begin{tabular}{rccl}
		&\multicolumn{2}{c}{\hspace{75pt}\bf Comparison of the KOD FPKE and the Lindblad master equation}&\\\hline\\
		\multicolumn{2}{c}{(\textsection \ref{KODD}) KODs are in the IGA}
		&
		\multicolumn{2}{c}{Density operators are in the dual of the von Neumann algebra}\\
		\multicolumn{2}{c}{$D_t \in L^1(\text{IG})$}
		&
		\multicolumn{2}{c}{$\rho_t \in \mathcal{B}(\Hb_\lambda)^\dag$}\\\\
		
		\multicolumn{2}{c}{\textreferencemark(\textsection \ref{ODiff}) \ul{Fokker-Planck-Kolmogorov equation}}
		&
		\multicolumn{2}{c}{\ul{Lindblad master equation}}\\
		\multicolumn{2}{c}{$\frac{1}{\kappa}\frac{\partial}{\partial t} D_t(x) = \fD_L^\ankh[D_t](x)$\phantom{$\Big|$}}
		&
		\multicolumn{2}{c}{$\frac{1}{\kappa}\frac{\partial}{\partial t} \rho_t = \D[\hat{L}](\rho_t)$}\\\\
		
		\multicolumn{2}{c}{\textreferencemark(\textsection \ref{ODiff}) \ul{Generator for a diffusive process}}
		&\\
		\multicolumn{2}{c}{$\fD_{L,\text{Diff}} = -\frac12 \Rinv{L^\dag L + L^2} + \frac12\Rinv{L}\Rinv{L}$\phantom{$\Big|$}}
		&
		\multicolumn{2}{c}{\ul{Dissipator/Lindbladian}}\\
		&&
		\multicolumn{2}{c}{$\D[\hat{L}] = - \frac12 (\hat{L}^\dag\hat{L} \!\odot\! 1 + 1 \!\odot\! \hat{L}^\dag\hat{L}) + \hat{L} \!\odot\! \hat{L}^\dag$} \\
		\multicolumn{2}{c}{(\textsection \ref{ODiff}) \ul{Generator for a jump process}}
		&\\
		\multicolumn{2}{c}{$\fD_{L,\text{Jump}} = -\frac12 \Rinv{L^\dag L} + \fL_{L}$\phantom{$\Big|$}}
		&\\\\
		\multicolumn{4}{c}{(\textsection \ref{ODiff}) \ul{Kolmogorov-Lindblad intertwining relation}}\\
		\multicolumn{4}{c}{$\fD_L[\O_x^{(\lambda)}] = \D[\hat{L}^{(\lambda)}]\circ\O_x^{(\lambda)}$\phantom{$\Big|$}}\\\\
		\multicolumn{4}{c}{(\textsection \ref{OD}) \emph{Instrument element irreps of the IG}}\\
		\multicolumn{4}{c}{$\O_{yx}^{(\lambda)} = \O_y^{(\lambda)}\circ\O_x^{(\lambda)}$}
	\end{tabular}
	\caption{The Lindblad master equation is a representation of the KOD Kolmogorov equation.
		Entries with the \textreferencemark~mark are concepts that have been introduced in previous work~\cite{jackson2023positive} but developed further in this paper.}
	\label{MarkovProcessTable}
\end{table}

A tool for considering continuous measurement that \emph{is} familiar to quantum scientists is the so-called Lindblad master equation, the equation that governs the average evolution in time of the state being measured.
In the absense of Hamiltonian system dynamics (in the interaction picture), the Lindblad master equation is
\begin{equation}\label{introLindblad_rho}
	\frac{1}{\kappa}\frac{\partial}{\partial t} \rho_t = \D(\rho_t)
\end{equation}
or equivalently
\begin{equation}\label{introLindbladZ}
	\frac{1}{\kappa}\frac{\partial}{\partial t} \Z_t = \D \circ \Z_t
\end{equation}
where $\rho_t = \Z_t(\rho_0)$ is the density operator for the state, $\Z_t$ is the channel superoperator for the  measuring process, $\D$ is the so-called dissipator or Lindbladian superoperator,
\begin{equation}
	\D(\rho) = \sum_k - \frac12\left(\hat{L}_k^\dag\hat{L}_k \rho + \rho \hat{L}_k^\dag\hat{L}_k\right) + \hat{L}_k \rho \hat{L}_k^\dag
\end{equation}
and the $L_k$ are the so-called Lindblad operators.
Of course, because the Lindblad master equation takes a state-centric view, it cannot directly address the more fundamental structure of instrument composition, while the KOD FPKE does.

The Lindblad master equation is in fact a direct consequence of the KOD FPKE (equation \ref{forward}).
Of course, the Lindblad master equation is also a direct consequence of the weak measurement that generates the KOD FPKE to begin with, so the KOD FPKE is by no means required to derive the Lindbladian.
Still, the KOD FPKE has the necessary content for describing the measuring instrument, far more content than the Lindblad master equation.

To see that the Lindblad master equation still follows from the KOD FPKE, one first needs to understand that the total operation for the time-dependent instrument defined by the KOD (equation \ref{introinstrument})
\begin{equation}\label{PositiveCase}
	\Z_t \equiv \int_{\text{IG}} d_\rL x \, D_t(x) \O_x
\end{equation}
is indeed a solution to equation \ref{introLindbladZ}.
What then remains is the use of a few group-analytic tools which are explained in section \ref{KODD}, particularly section \ref{ODiff}.
The steps involve taking a derivative of the total operation with respect to time, applying the KOD FPKE, using the left-translation invariance of the measure ($d_\rL(e^{hX} x) = d_\rL x$) to move the right-invariant derivatives from the KOD to the superoperators $\O_x$,
and finally using the Kraus-operator representation property ($K_{yx} = K_y K_x$) to observe the intertwining relations,
\begin{equation}
	\Rinv{L}[K_x] = \hat{L} K_x
\end{equation}
and therefore
\begin{equation}
	\fD\big[\O_x\big] = \D \circ \O_x.
\end{equation}
For a summary of the Kolmogorov and Lindblad representations of the measuring process and their intertwining relation, see Table \ref{MarkovProcessTable}.\\

Most of the group-analytic tools just mentioned have already been introduced, applied, and described at length in previous work~\cite{jackson2021spin,jackson2023weylheisenberg,jackson2023positive}.
However, this paper is designed to address them with a different approach that should make their relationship to the superoperator calculus (which quantum scientists expect to see) clearer.
This will be done by first directing the focus away from diffusive measurements and toward the more general structure of sequential measurement.
From the perspective of sequential measurement, the nature of the KOD should become far more apparent.

\subsection{The structure of sequential measurement is the IG convolution.}\label{E}

The solution to the superoperator Lindblad master equation (equation \ref{introLindbladZ} with initial condition $\Z_0(\rho) = \rho$) is obviously the exponential
\begin{equation}
	\Z_t = e^{\D \kappa t}
\end{equation}
and has the well-known (one-parameter) group property
\begin{equation}\label{introMarkovOne}
	\Z_{t+\Delta t} = \Z_{\Delta t} \circ \Z_t.
\end{equation}
Just as the Lindblad equation reflects the more elaborate KOD FPKE, this one-parameter group property reflects a similarly more elaborate property enjoyed by the KOD,
\begin{equation}\label{introMarkov}
	\boxed{
		\vphantom{\Bigg(}
		\hspace{10pt}
		D_{t+\Delta t} = D_{\Delta t} * D_t
		\hspace{10pt}
		\vphantom{\Bigg)}
	}
\end{equation}
where $*$ denotes the \emph{IG convolution},
\begin{equation}\label{introConvo}
	\boxed{
		\vphantom{\Bigg(}
		\hspace{10pt}
		g*f (z)
		\equiv \int_{\text{IG}} d_\rR x \, g(zx^\inv)f(x)\\
		= \int_{\text{IG}} d_\rL y \, g(y)f(y^\inv z).
		\hspace{10pt}
		\vphantom{\Bigg)}
	}
\end{equation}
where $d_\rR x$ is the right-invariant Haar measure of the IG.
In terms of the right-invariant measure, this convolution can be recognized as the Chapman-Kolmogorov equation of a right-invariant Markov process.
In terms of the left-invariant measure, this convolution allows the right-invariant Markov process to be converted into simple operator calculus, as will be explained in just a moment.
Right-invariance is the equivalent (in IG language) to the property that the measuring process is nonadaptive, i.e. invariant to the Kraus operator that just preceeds the next step in time.

In fact, this group property extends to any two instruments with a common IG.
If $D(x)$ is the KOD of an instrument and we denote it's total operation superoperator by
\begin{equation}\label{introGenTotal}
	\Z_D \equiv \int_{\text{IG}} d_\rL y \, D(y) \O_y,
\end{equation}
then for any second instrument with KOD $G(x)$
\begin{equation}\label{introDdagRep}
	\Z_G \circ \Z_D = \Z_{G*D}
\end{equation}
and the KOD of the composite instrument is $G*D(x)$.
The solution to the superoperator Lindblad master equation is thus a special case of these generalized total operation superoperators,
\begin{equation}
	\boxed{
		\vphantom{\Bigg(}
		\hspace{10pt}
		\Z_t = \Z_{D_t}.
		\hspace{10pt}
		\vphantom{\Bigg)}
	}
\end{equation}\\

In the same way that the Lindblad master equation can be lifted off the density operator by a superoperator, $\Z_t$ such that $\rho_t = \Z_t(\rho_0)$,
the FPKE (equation \ref{forward}) can be lifted off the KOD by an ultraoperator, $\fZ^\ankh_t$ such that $D_t = \fZ^\ankh_t[D_0]$, which satisfies
\begin{equation}\label{forwardUltra}
	\frac1\kappa \frac{\partial }{\partial t} \fZ^\ankh_t = \fD^\ankh \circ \fZ^\ankh_t.
\end{equation}
The solution of this equation is an ultraoperator that is also obviously exponential,
\begin{equation}\label{ObviousExp}
	\fZ^\ankh_t = e^{\fD^\ankh \kappa t}
\end{equation}
with the one-parameter group property
\begin{equation}
	\fZ^\ankh_{t+\Delta t} = \fZ^\ankh_{\Delta t} \circ \fZ^\ankh_{t}.
\end{equation}
This one-parameter group property is once again a reflection of the deeper property of the KOD (equation \ref{introMarkovOne}.)
At this level, it is even more apparent that these solutions of the KOD FPKE are just special cases of more general ultraoperators, the left-convolution ultraoperators, $\fZ^\rL_{G}$, defined such that for every KOD $D(x)$, the function $\fZ^\rL_{G}[D]$ is
\begin{equation}\label{introLeftConv}
	\fZ^\rL_{G}[D](x) \equiv G*D(x).
\end{equation}
By the associativity of the convolution, $H*(G*D) = (H*G)*D$, the left-convolution ultraoperators have the more elaborate group property
\begin{equation}\label{introAnkhRep}
	\fZ^\rL_{H} \circ \fZ^\rL_{G} \equiv \fZ^\rL_{H*G}
\end{equation}
similar to the generalized total operation superoperators (equation \ref{introDdagRep}).

The similarity between $\fZ^\rL_{D}$ and equation \ref{introGenTotal} becomes more apparent if we introduce the left-translations:
For every IG element $g$, the left-translation ultraoperator $\fL_g$ is such that for every function $D$ of the IG, the function $\fL_g[D]$ is for every point $x$ in the IG
\begin{equation}
	\fL_g[D](x) \equiv D(g^\inv x).
\end{equation}
With left-translation in hand, the left-convolution operators therefore have the expression
\begin{equation}\label{introLeftConvo}
	\fZ^\rL_{D} = \int_{\text{IG}} d_\rL x D(x) \fL_x
\end{equation}
similar to equation \ref{introGenTotal}.
Meanwhile, the ultraoperator solution to the FPKE is the special case
\begin{equation}
	\boxed{
		\vphantom{\Bigg(}
		\hspace{10pt}
		\fZ^\ankh_t = \fZ^\rL_{D_t}.
		\hspace{10pt}
		\vphantom{\Bigg)}
	}
\end{equation}\\

For a summary of the equations expressing the various group properties introduced, see Table \ref{GroupPropertyTable}.

\begin{table}[h!]
	\begin{tabular}{ccc}
		\multicolumn{3}{c}{\bf{Summary of the Various Group Properties}}\\\hline\\
		\ul{From KODs to Ultraoperators}
		&&
		\ul{From States to Superoperators}\\
		$\frac{1}{\kappa}\frac{\partial}{\partial t} D_t(x) = \fD^\ankh[D_t](x)$\phantom{$\Big|$}
		&Markovian Evolutions&
		$\frac{1}{\kappa}\frac{\partial}{\partial t} \rho_t = \D(\rho_t)$\\
		$D_t(x) = e^{\fD^\ankh \kappa t}[D_0](x)$
		&Exponential Solutions&
		$\rho_t = e^{\D \kappa t}(\rho_0)$\\
		$\fZ_t^\ankh \equiv e^{\fD^\ankh \kappa t}$\phantom{$\Big|$}
		&Evolution Operators&
		$\Z_t \equiv e^{\D \kappa t}$\\
		$\fZ_{t+\Delta t} = \fZ_{\Delta t} \circ \fZ_t$\phantom{$\Big|$}
		&Single-Parameter Group Property&
		$\Z_{t+\Delta t} = \Z_{\Delta t} \circ \Z_t$\\
		$\fZ_t^\ankh = \int_{\text{IG}} d_\rL x D_t(x) \fL_x$\phantom{$\Big|$}
		&IG Unravelings&
		$\Z_t = \int_{\text{IG}} d_\rL x D_t(x) \O_x $\\
		&&\\
		& \ul{Invariant Chapman-Kolmogorov Equation} &\\
		& $D_{\Delta t}*D_t = D_{t+\Delta t}$\phantom{$\Big|$}&\\\\
		& \ul{Instrumental Group Algebra} &\\
		$\fZ_f^\rL \equiv \int_{\text{IG}} d_\rL x\, f(x) \fL_x$\phantom{$\Big|$}
		&Expanded Representations&
		$\Z_f \equiv \int_{\text{IG}} d_\rL x\, \overline{f(x)} \O_x$\\
		$\fZ_{G*f}^\rL = \fZ_G^\rL \circ \fZ_f^\rL$\phantom{$\Big|$}
		&Expanded Group Property&
		$\Z_{G*f} = \Z_G \circ \Z_f$
	\end{tabular}
	\caption{During nonadaptive, diffusive, simultaneous measurements of any number of (possibly non-commuting) observables, both the density operator of the state being measured and the Kraus-operator density of the measuring instrument undergo a Markov process.  This Markovianity translates to evolution operators (either ultraoperators or superoperators) that satisfy a one-parameter group property.  These one-parameter group properties are actually a consequence of the Chapman-Kolmogorov equation satisfied by the time-independent, right-translation-invariant Markov kernel $d\kappa_{t, t\Delta t}(y|x) \equiv d_\rR x D_{\Delta t}(yx^\inv)$.  Because of the translation-invariance, such Chapman-Kolmogorov equations take the form of a group-covolution.  Through this, one can see that these one-parameter group properties are only specific cases of a much more expanded group satisfied by the ultraoperator and superoperator representations of the Kraus-operator density.  The formula for the superoperator ``Expanded Representation" differs from equation \ref{PositiveCase} because it can also include comp;ex-valued functions as input.}
	\label{GroupPropertyTable}
\end{table}

\subsection{The home of the KOD is the IG algebra.}\label{F}

With the IG convolution in hand (equation \ref{introConvo}), it is evident that the Kraus-operator density is a completely autonomous representation of sequential measurement.
If $D$ is the KOD of one instrument and $G$ is the KOD of a second instrument, the KOD of their composition (where the second is applied after the first) is the convolution $G*D$.
At the most basic level, what the IG convolution does is to collect all the pairs of registers from the two instruments that give the same composite instrument element.
At the most sophisticated level, what the IG convolution does is promote the space of KODs into an algebra which shall be called the \emph{instrumental group algebra} (IGA).\\

Interestingly, what we are calling IGAs are structures that go all the way back to the original considerations that gave rise to both the theory of POVMs and the theory of C*-algebras, namely what were called normed involutive rings (or also involutive Banach algebras.)
The norm (Banach property) is simply
\begin{equation}
	|| G ||_1 \equiv \int_{\text{IG}} d_\rL x \, |G(x)|,
\end{equation}
within which the KODs are certain positive functions with norm equal to unity.
Meanwhile, the IGA actually has two involutions,
\begin{equation}
	D^{\ddag}(x) \equiv \overline{D(x^\dag)}
	\hspace{50pt}
	\text{and}
	\hspace{50pt}
	D^{\ankh}(x) \equiv \overline{D(x^{-1})}
\end{equation}
where $\overline{z}$ denotes the complex conjugate of $z \in \C$, $x^\dag$ denotes Hermitian conjugation, and $x^\inv$ denotes the inverse of the group elements $x \in \text{IG}$.
The IGA involutions, $\ddag$ and $\ankh$, will respectively be called the Cartan involution\footnote{
In differential geometry, the Cartan involution would actually refer to the group homomorphism $x\mapsto x^{-\dag} = (x^\dag)^\inv = (x^\inv)^\dag$.  However, in algebra and analysis an involution usually has the property of swapping the order of multiplication and this is the convention we are following.
	
Further, the Cartan involution of a group element is here denoted by the same symbol as Hermitian conjugation because the distinction between unitary and positive transformations is also considered to be \textbf{universal}, which technically means we are making the assumption that each Kraus-operator representation intertwines the Cartan involution with the Hermitian conjugation of the Hilbert space that carries the representation, $K_x^\dag = K_{x^\dag}$.}
and the Gelfand involution.
If the left-invariant measure is also right-invariant (as is the case for the 7-dimensional $\mathrm{ISpin}$ and $\mathrm{IWH}$), the IG is called unimodular and these involutions respect the convolution in the usual way
\begin{equation}
	(G*D)^\ddag = D^\ddag * G^\ddag
	\hspace{50pt}
	\text{and}
	\hspace{50pt}
	(G*D)^\ankh = D^\ankh * G^\ankh.
\end{equation}

The superoperators $\Z_f$ defined in equation \ref{introGenTotal} and the hyperoperators $\fZ^\rL_f$ defined in equation \ref{introLeftConvo} are both representations of the IGA (by virtue of linearity and equations \ref{introDdagRep} and \ref{introAnkhRep}).
However, between the two it is only the superoperators $\Z_f$ that compose a $\ddag$-representation and only the ultraoperators $\fZ^\rL_f$ that compose an $\ankh$-representation.
The representations of these involutions are the adjoints of the Hilbert-Schmidt and Kolmogorov inner products
\begin{equation}
	(B|A)_{\Hb} \equiv \tr_\Hb B^\dag A
	\hspace{50pt}
	\text{and}
	\hspace{50pt}
	(H,G)_\text{IG} \equiv \int_\text{IG} d_\rL x \, \overline{H(x)} G(x)
\end{equation}
and we will denote the Hilbert-Schmidt and Kolmogorov adjoints by the same symbols
\begin{equation}
	\big(B\big|\Z^\ddag(A)\big)_{\Hb} \equiv \big(\Z(B)\big|A\big)_{\Hb}
	\hspace{50pt}
	\text{and}
	\hspace{50pt}
	\big(H,\fZ^\ankh[G]\big)_\text{IG} \equiv \big(\fZ[H],G\big)_\text{IG}
\end{equation}
because they represent the Cartan and Gelfand involutions, respectively
\begin{equation}
	\Z_f^\ddag = \Z_{f^\ddag}
	\hspace{50pt}
	\text{and}
	\hspace{50pt}
	(\fZ^\rL_f)^\ankh = \fZ^\rL_{f^\ankh}.
\end{equation}

With either involution, the IGA is in both senses very close to being a C*-algebra.
For unimodular IGs, the only property that is missing is the so-called C*-identity
\begin{equation}
	||f^\ddag *f||_1 \neq ||f^\ddag||_1\, ||f||_1
	\hspace{50pt}
	\text{and}
	\hspace{50pt}
	||f^\ankh *f||_1 \neq ||f^\ankh||_1\, ||f||_1.
\end{equation}
There is apparently a canonical way to derive a C*-algebra in this case, the so-called enveloping C*-algebra~\cite[p. 47]{dixmier1977c}.
The exact meaning and correspondence of the enveloping $\text{C}^\ddag$-algebra and the enveloping $\text{C}^\ankh$-algebra require further investigations that are already beyond the purpose of the current paper.\\

For a summary of the various involutive identities of the IGA, see Table \ref{Involutive}.\\

\begin{table}
	\begin{tabular}{rccl}
		\multicolumn{4}{c}{\bf{Involutive structures of the IGA and its Ultraoperator and Superoperator Representations}}\\\hline\\
		& \multicolumn{2}{c}{\ul{IG convolution} (\textsection\ref{ODF})} &\\
		& \multicolumn{2}{c}{$g*f(z) \equiv \int_{\text{IG}}d_\rL y \, \, g(y) f(y^\inv z)$} &\\\\
		(\textsection\ref{AdjAlg}) IG inner product: & $(g,f)_{\text{IG}} \equiv \int_{\text{IG}} d_\rL x \, \, \overline{g(x)} f(x)$
		&
		$(B|A)_{\Hb_\lambda} \equiv \tr_{\Hb_\lambda} B^\dag A$ & :Hilbert-Schmidt inner product\\\\
		
		(\textsection\ref{AdjAlg}) Kolmogorov adjoint: & $\big(g,\fZ^\ankh[f]\big)_{\text{IG}} \equiv \big(\fZ[g],f\big)_{\text{IG}}$
		&
		$\big(B\big|\Z^\ddag(A)\big)_{\Hb_\lambda} \equiv \big(\Z(B)\big|A\big)_{\Hb_\lambda}$ & :Hilbert-Schmidt adjoint\\
		(\textsection\ref{AdjAlg}) Gelfand involution: &$f^{\ankh}(x) \equiv \overline{f(x^\inv)}$
		&
		$f^\ddag(x) \equiv \overline{f(x^\dag)}$ & :Cartan involution (\textsection\ref{AdjAlg})\textreferencemark\\
		\emph{(if IG is unimodular)} & $(g*f)^{\ankh} = f^\ankh* g^\ankh$
		&
		$(g*f)^{\ddag} = f^\ddag * g^\ddag$ &\emph{(if IG is unimodular)}\\\\
		
		(\textsection\ref{TransDiff}) left-translation ultraoperator: & $\fL_a[f](x) \equiv f(a^\inv x)$
		&
		$\O_a \equiv K_a \odot K_a^\dag$ & :CP superoperator (\textsection\ref{OD})\textreferencemark\\
		\emph{left-regular representation of IG} & $\fL_a \circ \fL_x = \fL_{ax}$
		&
		$\O_a \circ \O_x = \O_{ax}$ & \emph{instrument-element irrep of IG}\\
		\emph{$\ankh$-representation of IG} & $\fL_x^\ankh = \fL_{x^\inv}$
		&
		$\O^\ddag_a = \O_{a^\dag}$ & \emph{$\ddag$-representation of IG}\\\\
		
		(\textsection\ref{TransDiff}) left-convolution ultraoperator: & $\fZ^\ankh_f \equiv \int_{\text{IG}} d_\rL x \, \, f(x) \, \fL_x$
		&
		$\Z_f \equiv \int_{\text{IG}} d_\rL x \, \, \overline{f(x)} \, \O_x$ & :IGA superoperator (\textsection\ref{AdjAlg})\textreferencemark\\
		\emph{IGA regular representation} & $ \fZ^\ankh_g \circ \fZ^\ankh_f = \fZ^\ankh_{g*f}$
		&
		$ \Z_g \circ \Z_f = \Z_{g*f}$ & \emph{IGA irrep}\\
		\emph{$\ankh$-representation of IGA} & $\fZ^\ankh_f = \fZ_{f^\ankh}$
		&
		$\Z^\ddag_f = \Z_{f^\ddag}$ &  \emph{$\ddag$-representation of IGA}
	\end{tabular}
	\caption{The IG convolution promotes the Banach space $L^1(\text{IG})$ to an involutive Banach algebra with two distinct involutions, one respected by the Kolmogorov evolution of the KOD and the other respected by the Lindblad evolution of the density operator.  Entries with the \textreferencemark~mark are concepts that have been introduced in previous work~\cite{jackson2023positive} but are further developed in this paper.}
	\label{Involutive}
\end{table}

\section{Instrumental Groups: Moving from Continuous to Sequential Measurement}\label{Instruments}

With the dimension of time, measurements can be put together in a variety of ways
| continuously (\ref{ContMeas}), simultaneously (\ref{SimultMeas}), and sequentially (\ref{OD}) | all of which can be considered to take place in an \emph{instrumental group} (IG).\\

\subsection{The Fundamental Problem of Continuous Measurement: Piling Kraus-Operators (Convolution 1 of 3)}\label{ContMeas}

A \emph{continuous measurement} can in principle refer to any infinite sequence of weak measurements.
Applying a sequence of two measuring instruments, $\I_1$ after $\I_0$, defines another instrument given by the \emph{convolution of two sets of operations},
\begin{equation}\label{convo}
	\boxed{
		\vphantom{\Bigg(}
		\hspace{10pt}
		\I_{1}*\I_{0} \equiv \left\{\O_1\circ\O_0: \O_1 \in \I_{1} \text{ and } \O_0 \in \I_{0}\right\}.
		\hspace{10pt}
		\vphantom{\Bigg)}
	}
\end{equation}
Applying an instrument $\I$ repeatedly $n$ times will thus be denoted as $\I^{*n}$.
The most basic continuous-time instruments are the repeated application of a single weak instrument, also known as continual monitoring.
If $\I_{dt}$ denotes the single weak instrument that occurs for a time $dt$, then the continuous instrument $\I_T$ generated over a time $T$ would have a formal expression that parallels equation \ref{DissChan2},
\begin{equation}\label{DissChan2Inst}
	\I_T = \I_{dt}^{*T/dt},
\end{equation}
which of course also has the one-parameter semigroup property
\begin{equation}\label{OneParamSemiInst}
	\I_{t+\Delta t} = \I_{\Delta t}*\I_t.
\end{equation}
Usually, the form of the weak instrument $\I_{dt}$ comes effectively in one of two effective types, generating either a jump process or a diffusive process.

The effective weak instrument that generates a jump process has only two elements,
\begin{equation}\label{Jump}
	\boxed{
		\vphantom{\Bigg(}
		\hspace{15pt}
		\I^{(\hat{L},\text{Jump})}_{dt} = \left\{ (\kappa dt)^{dN}\Odot\left[e^{-\frac12 \hat{L}^\dag \hat{L} \kappa dt}\hat{L}^{dN}\right] \right\}_{dN\in\{0,1\}}
		\hspace{10pt}
		\vphantom{\Bigg)}
	}
\end{equation}
and the ``jump'' refers to the singular binomial event $dN=1$.
For jump processes, the phase of the Lindblad operator is a symmetry of each instrument element,
\begin{equation}
	\I^{(e^{i\phi}\hat{L},\text{Jump})}_{dt} = \I^{(\hat{L},\text{Jump})}_{dt},
\end{equation}
and is therefore an irrelevant degree of freedom.

The effective weak instrument that generates a diffusive process has an infinitesimal continuum of elements,
\begin{equation}\label{Diff}
	\boxed{
		\vphantom{\Bigg(}
		\hspace{15pt}
		\I^{(\hat{L},\text{Diff})}_{dt} = \left\{\frac{d(dW)}{\sqrt{2\pi dt}}e^{-\frac{dW^2}{2dt}}\Odot\left[e^{-\frac12(\hat{L}^\dag \hat{L} + \hat{L}^2)\kappa dt + \hat{L}\sqrt{\kappa}dW}\right]\right\}_{dW\in\R}
		\hspace{10pt}
		\vphantom{\Bigg)}
	}
\end{equation}
where the Gaussian random variable $dW$ is called a standard Wiener increment and ``diffusive'' refers to the Brownian motion they generate.
For diffusive processes, the phase actually changes what is being measured.
The examples which make this most obvious are at the extreme:
If the Lindblad operator is Hermitian, then the Kraus operators are all positive (i.e. the instrument is generalized-Luders) and no POVM element is proportional to the identity.
If the Lindblad operator is antiHermitian (i.e. Hermitian times the phase $i=e^{i\pi/2}$) then the Kraus operators are all unitary and every POVM element is proportional to the identity, meaning the registers are completely insensitive to the state preparation.\\

Either of these instruments can be seen to unravel the infinitesimal dissipative channel in equation \ref{DissChan}.\\

For the effective system-meter-interaction models that give \ref{Jump} and \ref{Diff}, se Appendix \ref{SMI_Models}.\\

Every instrument $\I = \{\O_x\}$ by repetition defines a \emph{convolution semigroup}, the set of all chains of instrument elements,
\begin{equation}\label{ConvoSemi}
	\tilde{G}_\I \equiv \bigcup_{n=0}^\infty\I^{*n}
\end{equation}
where $\I^{*0}$ is the instrument with only one element, equal to the identity operation.
As mentioned in section \ref{StandardTheory}, one should note that the convolution semigroup is quite a bit more elaborate than the one-parameter semigroup of total operations mentioned after equation \ref{OneParamSemi}.
Indeed, the convolution semigroup is usually multi-dimensional and non-commutative.
In practice, the convolution semigroup can often be contained within a group, what will be called the \emph{instrumental group} (IG) and denoted $G_\I$.
In section \ref{KODD} we will use the IG in a much more systematic way that begins with a more rigorous definition.

Without specifying any commutation relations or the von Neumann algebra on which these instrument elements act,
the elements in $G_\I$ are by definition distinct and such a group is said to be \emph{free} (of constraints or relations.)
In this case, it is worth noting that the \emph{free product} of two groups (or semigroups) ${G}_1$ and ${G}_0$ is precisely the group (or semigroup) of all chains
\begin{equation}
	{G}_1 * {G}_0 \equiv \left\{\cdots \circ \O_{2} \circ \O_{1} : \forall k, \O_k \in {G}_1 \cup {G}_0\right\}.
\end{equation}
Therefore the IG of a convolution of two instruments is the free product of their individual IGs,
\begin{equation}
	G_{\I_{1}*\I_{0}} = G_{\I_{1}} *G_{\I_{0}}.
\end{equation}
In particular, $G_{\I*\I} = G_{\I} *G_{\I} = G_{\I}$ and so for any positive integer $n$
\begin{equation}
	G_{\I^{*n}} = G_{\I}.
\end{equation}
If $G_\I$ is constrained by only commutation relations, the instrument is said to be \emph{universal} \cite{jackson2023positive}.\\

\textbf{The fundamental problem of continuous measurement in general} is whether two sequences of raw registered data, $(x_1,x_2,\ldots,x_n)$ and $(y_1,y_2,\ldots,y_n)$ drawn from a repeated instrument $\I^{*n}$ for arbitrarily large $n$, can be recognized to give the same composite instrument element or not,
\begin{equation}
	\O_{x_n} \circ \cdots \circ \O_{x_2} \circ \O_{x_1} \overset{?}{=} C\; \O_{y_n} \circ \cdots \circ \O_{y_2} \circ \O_{y_1}
\end{equation}
up to some positive scalar factor $C \in \R_{>0}$ (which is added to the Kraus-operator distribution to be explained in section \ref{OD}.)
This is more commonly known as the \emph{word problem} \cite{rotman1995introduction}, a topic that was very hot in the '80s and '90s, but mostly centered around digital ideas.
Following \cite{epstein1992word}, a group is said to be \emph{automatic} if two products of group elements can be recognized as equal or distinct by a classical turing machine.
The discussion in \cite{epstein1992word, gromov1987hyperbolic, cannon1984combinatorial} was specifically interested in the much more complicated problem of finitely-generated groups,
but the concept of an automatic group can just as well be applied to diffusively-generated groups.
Here it will be left open exactly how to define what it means to ``recognize'' two chains as the same in general.
In our cases of interest, the composite instrument element of every raw chain of events can be put into a certain canonical form that coordinatizes the IG.
The ease in which these coordinatizations work thus suffices to demonstrate that such continuous IGs have the alleged form of automatic pattern recognition.

The universal IGs generated by diffusive processes were recently classified into two groups, depending on the dimension of the IG as calculated via the commutation relations of the Lindblad operators and the quadratic term~\cite{jackson2023positive}.
If the universal IG is a finite-dimensional Lie group, it is called a \emph{principal instrument}, otherwise, the IG is an infinite-dimensional Lie group and it is called a \emph{chaotic instrument}.
If the IG is finite-dimensional, it can thus be co\"ordinated by applying the MMCSD to any of a standard set of group decompositions~\cite{jackson2021spin,jackson2023weylheisenberg,jackson2023positive}.
Principal instruments are therefore clearly automatic by virtue of the ease with which this co\"ordinatization can be executed.
It may be the case that there are chaotic instruments that are also automatic, but this is an open question.\\

Of course, the topics of chaos and automation (and their relation to perception, observation, and measurement) are fascinating.
However, they are mostly beyond the current scope of this paper and the expertise of the author.
For some of the original ideas on perception and automation in general, some inspiring references are \cite{wiener1948cybernetics, bellman1957dynamic, sutton1998reinforcement, bennett1989observer,dupont2005links, bishop2006pattern}.\\

\subsection{Simultaneous Measurements of Non-Commuting Observables}\label{SimultMeas}

Continuous measurements can be done simultaneously, even when their Lindblad operators do not commute \cite{jackson2023positive, jackson2021spin, karmakar2022stochastic}.
To see this, consider the weak diffusive Kraus operators with Lindblad operator $\hat{L}$,
\begin{equation}
	K^{\hat{L}}(dW) \equiv e^{-\frac12(\hat{L}^\dag \hat{L} + \hat{L}^2)\kappa dt + \hat{L}\sqrt{\kappa}dW}.
\end{equation}
For two such Lindblad operators $\hat{L}$ and $\hat{M}$, the group commutator of one weak diffusive Kraus operator after the other is
\begin{align}
	\left(K^{\hat{L}}(dW)K^{\hat{M}}(dV)\right)^\inv K^{\hat{M}}(dV)K^{\hat{L}}(dW)
	&= e^{[\hat{M},\hat{L}]\kappa dVdW}\\
	&= 1\label{GroupComm}
\end{align}
because the Kraus-operator distribution is a product of the individual Wiener measures for $dW$ and $dV$, and so the Wiener increments are independent.

\textcolor{red}{For readers who are not comfortable with either group commutators, It\^o rules, or their combination, here is an alternative, more detailed explanation.  Recall that an operator conjugated by an exponential can be calculated as an exponential series of nested commutators,
\begin{align}
	e^{\hat{A}}\hat{B} e^{-\hat{A}} = \hat{B} + [\hat{A},\hat{B}] + \frac12 \big[\hat{A},[\hat{A},\hat{B}] \big] + \ldots
\end{align}
and that operator conjugation commutes with exponentiation,
\begin{align}
	ae^{\hat{B}}a^\inv = e^{a\hat{B}a^\inv}.
\end{align}
In particular, for $\hat{A} = -\frac12(\hat{L}^\dag \hat{L} + \hat{L}^2)\kappa dt + \hat{L}\sqrt{\kappa}dW$ and $\hat{B} = -\frac12(\hat{M}^\dag \hat{M} + \hat{M}^2)\kappa dt + \hat{M}\sqrt{\kappa}dV$, where $dW$ and $dV$ are Wiener increments, we have
\begin{align}
	[\hat{A},\hat{B}] = [\hat{L},\hat{M}]\kappa dW dV
	\hspace{50pt}
	\text{and}
	\hspace{50pt}
	\big[\hat{A},[\hat{A},\hat{B}] \big] = 0
\end{align}
where higher-order terms can be neglected by the It\^o rules $dW dt = dV dt = 0$, $dt^2 = 0$, and $dW^2=dt$.
Finally, if $dW$ and $dV$ are independent, than even the last remaining term in the algebra commutator is zero by another It\^o rule $dW dV = 0$.
Therefore even if $[\hat{L},\hat{M}] \neq 0$, we still have $[\hat{A},\hat{B}] = 0$ and so the group elements represented by $K^{\hat{L}}(dW)$ and $K^{\hat{M}}(dV)$ also commute when $dW$ and $dV$ are independent.
When $a$ and $b$ represent commuting group elements, it is not so natural to write an \emph{algebra} commutator $[b,a] = ba - ab$ because it requires defining subtraction.
More natural is the \emph{group} commutator $b^\inv a^\inv b a = 1$, which is pictured as a sequence of 4 steps which return to the identity as in equation \ref{GroupComm}.\\
}

Equation \ref{GroupComm} means that even if the two Lindblad operators are noncommuting, their weak diffusive Kraus operators still commute.
Thus two diffusive measurements can be done simultaneously, or (perhaps more descriptively) in tandem.
Therefore given any finite number of potentially noncommuting Lindblad operators $\vec{L}=(\hat{L}_1,\ldots,\hat{L}_n)$ one can define the  \emph{Diffusive Measurement of Non-Commuting Observables Simultaneously} (DMNCOS) as the diffusive process generated by the weak instrument 
\begin{align}\label{DMNCOS}
	\I^{(\vec{L},\text{Diff})}_{dt}
	&\equiv\I^{(\hat{L}_n,\text{Diff})}_{dt}*\cdots*\I^{(\hat{L}_2,\text{Diff})}_{dt}*\I^{(\hat{L}_1,\text{Diff})}_{dt}\\
	&= \left\{\frac{d^n(d\vec{W})}{(2\pi dt)^{n/2}}e^{-\frac{|d\vec{W}|^2}{2dt}}\Odot\left[e^{-\frac12(\vec{L}^\dag \!\cdot \vec{L} + \vec{L}^2)\kappa dt + \vec{L}\cdot \sqrt{\kappa}d\vec{W}}\right]\right\}_{d\vec{W}\in\R^n}
\end{align}
where some basic shorthand notation is introduced,
\begin{equation}
	|d\vec{W}|^2 \equiv \sum_\mu (dW^\mu)^2
	\hspace{25pt}
	\text{,}
	\hspace{25pt}
	\vec{L}^\dag \!\cdot \vec{L} + \vec{L}^2  \equiv \sum_\mu \hat{L}_\mu^\dag \hat{L}_\mu + \hat{L}_\mu^2
	\hspace{25pt}
	\text{, and}
	\hspace{25pt}
	\vec{L}\cdot d\vec{W} \equiv \sum_\mu \hat{L}_\mu dW^\mu.
\end{equation}

The analogous commutativity of weak jump elements is obvious because the product of two jump elements in an infinitesimal time is still negligible, regardless of whether the Lindblad operators of the two jumps commute or not.
One can also define a generating weak instrument for simultaneous jump measurements of non-commuting observables, but this will not be relevant to the current discussion.\\

If the DMNCOS procedure is carried out for a finite time, the instrument will register finitely many simultaneous sequences of Wiener increments $d\vec{W}_{[0,T)} = \{d\vec{W}_{kdt}\}_{k=0}^{T/dt - 1}$ called the measurement record, each corresponding to the Kraus operator
\begin{equation}\label{KrausRecord}
	\sqrt{\D\mu[d\vec{W}_{[0,T)}]} \; K^{(\vec{L})}\!\left[d\vec{W}_{[0,T)}\right]
	= \sqrt{\D\mu[d\vec{W}_{[0,T)}]}\; \mathcal{T}\!\exp \left(\int_0^{T-dt} \hat{\updelta}^{(\vec{L})}(d\vec{W}_t)\right)
\end{equation}
where $\D\mu[d\vec{W}_{[0,T)}] = \prod_{k=0}^{T/dt - 1} \frac{d^n(d\vec{W}_{k dt})}{(2\pi dt)^{n/2}}e^{-\frac{|d\vec{W}_{k dt}|^2}{2dt}}$ is the Wiener measure, $\mathcal{T}\!\exp$ denotes the time-ordered exponential, and the infinitesimal generator is
\begin{equation}
	\hat{\updelta}^{(\vec{L})}(d\vec{W}) \equiv -\hat{Q}_{\vec{L}}\,\kappa dt + \sum_{\mu = 1}^n \hat{L}_\mu\sqrt\kappa dW^\mu
	\hspace{25pt}
	\text{with}
	\hspace{25pt}
	\hat{Q}_{\vec{L}} \equiv \frac12 \sum_{\mu = 1}^n \hat{L}_\mu^\dag \hat{L}_\mu + \hat{L}_\mu^2.
\end{equation}
The most critical thing to understand about equation \ref{KrausRecord} is that the time-ordered exponential has a solution that only depends on the commutators of the terms within the infinitesimal generator.
Therefore, the points of the IG and the Brownian motion  defined within it by equation \ref{KrausRecord} can be considered universally, independent of the Hilbert space.
(For further explanation, see~\cite[p.~38]{jackson2023positive}.)
In effect, this simply means we can take the hats ( $\hat{}$ ) off of all the operators and write an abstract time-ordered exponential,
\begin{equation}
	\gamma^{(\vec{L})}[d\vec{W}_{[0,T)}] = \mathcal{T}\!\exp \left(\int_0^{T-dt} \updelta^{(\vec{L})}(d\vec{W}_t)\right) \in \text{IG}
\end{equation}
and consider the Kraus operators as representations of this abstract IG element,
\begin{equation}
	K^{(\vec{L})}\!\left[d\vec{W}_{[0,T)}\right] = K_{\gamma^{(\vec{L})}[d\vec{W}_{[0,T)}] }
\end{equation}
where $K$ with a subscript now denotes a representation of the IG (with the defining property $K_{yx} =K_y K_x$.)
With this insight, we can consolidate all the registers of the raw instrument (equation \ref{KrausRecord}) that actually give the same Kraus operator, defining the instrument
\begin{equation}
	\Big\{d_\rL x\, D_T(x) \Odot[K_x]\Big\}_{x\in IG}
\end{equation}
where $D_T(x)$ is the so-called Kraus-operator density, defined by
\begin{equation}
	d_\rL x\, D_T(x) = d_\rL x \int \D\mu[d\vec{W}_{[0,T)}] \,\delta_{\text{IG}}\!\left(\gamma[d\vec{W}_{[0,T)}]^\inv x\right)
\end{equation}
where $\delta_\text{IG}(x)$ is the delta function of the IG.\\

There are many types of automatic IGs that exist among the DMNCOS \cite{jackson2021spin,jackson2023weylheisenberg,jackson2023positive}.
As mentioned in section \ref{ContMeas}, there are the principal universal instruments as defined in \cite{jackson2023positive} of which two particular examples have been worked out extensively:
The \emph{Simultaneous P and Q Measurement (SPQM)} of \cite{jackson2023weylheisenberg} is the case for which $\vec{L} =(\hat{P},\hat{Q})$ (the momentum and position quadratures, satisfying the canonical commutation relations) and the resulting IG is a 7-dimensional unimodular Lie group called the \emph{instrumental Weyl-Heisenberg group}, $\mathrm{IWH}$.
The \emph{Isotropic Spin Measurement (ISM)} of \cite{jackson2021spin} is the case in which $\vec{L} = \vec{J} = (\hat{J_x},\hat{J_y},\hat{J_z})$ (the orthogonal angular momentum components) and the resulting IG is another 7-dimensional unimodular Lie group which in \cite{jackson2023positive} was called the \emph{instrumental spin group}, $\mathrm{ISpin(3)}\cong \R\times\SL(2,\C)$.

A measurement of only two angular momentum components \cite{wiseman1995su2} results in a chaotic universal instrument (\cite{jackson2023positive}, page 33).\footnote{As Howard Wiseman put it to me in 2022, ``Who knew a third measurement could make the problem infinitely simpler?!''}

Significantly, ISM is also the first measurement found to be able to practically realize the spin-coherent-state POVM \cite{jackson2021spin, shojaee2018optimal, DAriano2002} for arbitrarily large $j$, because of its universal character.
In this case, it may be worth emphasizing that universal means that the method of measurement implementation is the same no matter what spin-$j$ is being measured.
So one must be careful not to mistake the ``2'' of the ``$\SL(2,\C)$'' in the title of \cite{jackson2021spin} for a qubit. Rather, ``$\SL(2,\C)$'' refers to a universal covering group.\\

\subsection{Sequential Measurements: Kraus-Operator Distributions and the Convolution (2 of 3)}\label{OD}

A group $G$ is \emph{instrumental} if it has a \emph{Kraus-operator representation}, $\{K_x\}_{x\in G}$ (on some Hilbert space) such that
\begin{equation}
	K_y K_x = K_{yx},
\end{equation}
with a measure (a.k.a. positive distribution) $d\mu(x)$ such that the set
\begin{equation}
	\I \equiv \Big\{d\mu(x)\Odot[K_x]\Big\}_{x\in G}
\end{equation}
is an instrument, the total operation thus being
\begin{equation}\label{quantumOperation}
	\Z_\mu = \int_G d\mu(x) \Odot[K_x].
\end{equation}
Such a measure, $d\mu(x)$, will be called the \emph{instrument-element distribution} or \emph{Kraus-operator distribution}.
If the instrumental group (IG) and the Kraus operator representation elements are specified, then the Kraus-operator distribution $d\mu(x)$ is fixed and is no longer an ostensible factor of the Kraus operator.
Meanwhile, one also has the corresponding \emph{instrument-element representation} of the IG and we will use the shorthand notation
\begin{equation}
	\O_x \equiv \Odot[K_x]
\end{equation}
for which
\begin{align}
	\O_y \circ \O_x = \O_{yx}.
\end{align}
It is worth noting that often when the Kraus operator representation $\{K_x\}$ is irreducible and nonunitary, the tensor product representation $\{\O_x\}$ is also irreducible.

If two instruments representing the same IG,
\begin{equation}
	\I_\mu \equiv \{d\mu(x)\O_x\}_{x\in \text{IG}}
	\hspace{25pt}
	\text{and}
	\hspace{25pt}
	\I_\nu \equiv \{d\nu(y)\O_y\}_{y\in \text{IG}},
\end{equation}
are applied in sequence, then their composite total operation would be
\begin{align}
	\Z_\nu \circ \Z_\mu
	&= \int_\text{IG} d\nu(y) \O_y \circ \int_\text{IG} d\mu(x) \O_x\\
	&= \int_\text{IG} d\nu(y) \int_\text{IG} d\mu(x) \O_{yx}\\
	&=\int_\text{IG} \int_\text{IG} d\nu(y) d\mu(y^\inv z) \O_{z}\\
	&=\int_\text{IG} \int_\text{IG} d\nu(z x^\inv) d\mu(x) \O_{z}\\
	&=\int_\text{IG} d(\nu*\mu)(z) \O_{z},
\end{align}
where the \emph{convolution of two distributions}, $d\nu$ after $d\mu$, is
\begin{equation}\label{convoMeas}
	\boxed{
		\vphantom{\Bigg(}
		\hspace{10pt}
		d(\nu*\mu)(z) 
		\equiv \int_{y\in \text{IG}} d\nu(y) d\mu(y^\inv z)
		\hspace{10pt}
		\vphantom{\Bigg)}
	}
\end{equation}
\begin{equation}
	\hspace{50pt}
	=\int_{x\in \text{IG}} d\nu(z x^\inv) d\mu(x).
\end{equation}
Therefore, the IG provides the structure to collect all the pairs $(x,y)$ that register the same composite instrument element corresponding to the group element $z=yx$.
Sure enough, the convolution of two instruments (equation \ref{convo}) with the same IG is an instrument with Kraus-operator distribution equal to the convolution of the two seperate Kraus-operator distributions,
\begin{align}\label{SeqGrpPropInstr}
	\I_\nu * \I_\mu
	&=\I_{\nu*\mu}\\
	&\equiv \big\{d(\nu*\mu)(z)\O_z\big\}_{z\in \text{IG}}
\end{align}
with total operations that satisfy the \emph{convolution group property},
\begin{equation}\label{ConvoProp}
	\Z_\nu \circ \Z_\mu = \Z_{\nu*\mu}.
\end{equation}
Equations \ref{SeqGrpPropInstr} and \ref{ConvoProp} are deeper elaborations of equations \ref{convo} and \ref{OneParamSemi}.

\section{Kraus-Operator Densities: Ultraoperators and the IG Algebra}\label{KODD}

This paper included a tour de force in the fundamental tools of group analysis.  For further study, some recommendations are \cite{vonneumann1999invariant, barut1986theory, dieudonne1979special, dixmier1977c, nachbin1965haar, hewitt1963abstract, rudin1962fourier, montgomery1955topological, pontryagin1946topological, berezin1967remarks, kitaev2017notes, jackson2023positive}.\\

Just like the real line or the complex plane, the IG is the domain of an entire suite of analytic tools | with invariant integrals (\ref{ODF}), invariant derivatives (\ref{TransDiff}), involutive symmetries (\ref{AdjAlg}), and intertwined differential and algebraic equations (\ref{ODiff})| where functions such as the Kraus-operator density (KOD) can be calculated, propogated, and varied.\\

\subsection{The Haar Measures: Kraus-Operator Densities and the Convolution (3 of 3)}\label{ODF}

If the IG is locally compact, then it has a distinct pair of measures $d_\rL y \,$ and $d_\rR x \,$
such that for every $g \in$ IG
\begin{equation}
	\boxed{
		\vphantom{\Bigg(}
		\hspace{10pt}
		d_\rL(g y) = d_\rL y \,
		\hspace{10pt}
		\vphantom{\Bigg)}
	}
	\hspace{25pt}
	\text{and}
	\hspace{25pt}
	d_\rR(xg) = d_\rR x \,.
\end{equation}
Such a $d_\rL y \,$ is called a \emph{left-invariant Haar measure} and such a $d_\rR x \,$ is called a \emph{right-invariant Haar measure}.
The Haar measures are unique up to normalization and their normalizations may be chosen so that
\begin{equation}
	d_\rR x \, = d_\rL(x^\inv).
\end{equation}
These Haar measures in turn define the \emph{convolution of two functions}, $g$ after $f$, as
\begin{equation}\label{convoFunc}
	\boxed{
		\vphantom{\Bigg(}
		\hspace{10pt}
		g*f (z)
		\equiv \int_{y\in\text{IG}} d_\rL y \, g(y) f(y^\inv z)
		\hspace{10pt}
		\vphantom{\Bigg)}
	}
\end{equation}
\begin{equation}
	\hspace{35pt}
	=\int_{x\in\text{IG}} d_\rR x \, g(z x^\inv) f(x).
\end{equation}
The IG is said to be \emph{unimodular} if $d_\rR x= d_\rL x$.

Given a Kraus-operator distribution $d\mu(x)$, the Kraus-operator distribution functions or \emph{Kraus-operator densities} (KOD) with respect to either of these Haar measures are the Radon-Nikodym derivatives
\begin{align}\label{KODL}
	\boxed{
		\vphantom{\Bigg(}
		\hspace{10pt}
		D^{\rL}_\mu(x) \equiv \frac{d\mu(x)}{d_\rL x \,}
		\hspace{10pt}
		\vphantom{\Bigg)}
	}
	\hspace{25pt}
	\text{and}
	\hspace{25pt}
	D^{\rR}_\mu(x) \equiv \frac{d\mu(x)}{d_\rR x \,}.
\end{align}
Indeed, the KOD of a convolution of two measures is equal to the convolution of the two KODs.
To see this is a simple exercise:
on the one hand
\begin{align}
	d_\rL z \, D^{\rL}_{\nu*\mu}(z)
	&\equiv d(\nu*\mu)(z)\\
	&\equiv \int_{y\in\text{IG}} d\nu(y) d\mu(y^\inv z)\\
	&= \int_{y\in\text{IG}} d_\rL y \, D^{\rL}_\nu(y) d_\rL(y^\inv z) D^{\rL}_\mu(y^\inv z)\\
	&= \int_{y\in\text{IG}} d_\rL y \, D^{\rL}_\nu(y) d_\rL z \, D^{\rL}_\mu(y^\inv z)\\
	&= d_\rL z \, \int_{y\in\text{IG}} d_\rL y \, D^{\rL}_\nu(y)D^{\rL}_\mu(y^\inv z)\label{LeftCK}\\
	&\equiv d_\rL z \, \left(D^{\rL}_\nu*D^{\rL}_\mu\right)(z)\label{LeftCKlast}
\end{align}
while on the other
\begin{align}
	d_\rR z \, D^{\rR}_{\nu*\mu}(z)
	&\equiv d(\nu*\mu)(z)\\
	&= \int_{x\in\text{IG}} d\nu(z x^\inv) d\mu(x)\\
	&= \int_{x\in\text{IG}} d_\rR(z x^\inv) D^{\rR}_\nu(z x^\inv) d_\rR x \, D^{\rR}_\mu(x)\\
	&= \int_{x\in\text{IG}} d_\rR z \, D^{\rR}_\nu(z x^\inv) d_\rR x \, D^{\rR}_\mu(x)\\
	&= d_\rR z \, \int_{x\in\text{IG}} D^{\rR}_\nu(z x^\inv) d_\rR x \, D^{\rR}_\mu(x)\label{RightCK}\\
	&= d_\rR z \, \left(D^{\rR}_\nu*D^{\rR}_\mu\right)(z).
\end{align}

As seen in sections \ref{ContMeas} and \ref{SimultMeas} and to be explained further in section \ref{ODiff}, the samples of a weak-instrument Kraus-operator distribution are the increments of a continuous invariant Markov process, increments that are locally native to another space beside the IG, or rather tangent to it.
In particular, this means that it is usually easier for small times to work directly with the Kraus-operator distribution of the weak instrument and not the KOD.
Therefore, it is useful to notice that the KOD of a convolution can be computed by directly applying the Kraus-operator distribution of one instrument to the KOD of the other.
There are two such expressions in this regard, one for the KODs with respect to the left-invariant measure,
\begin{align}\label{left}
	D^{\rL}_{\nu*\mu}(z)
	&= \int_{y\in\text{IG}} d\nu(y)D^{\rL}_{\mu}(y^\inv z)\\
	&= \int_{x\in\text{IG}} d\nu(z x^\inv)D^{\rL}_{\mu}(x)
\end{align}
as can be seen from equation \ref{LeftCK},
and the other for the KODs with respect to the right-invariant measure,
\begin{align}\label{right}
	D^{\rR}_{\nu*\mu}(z)
	&= \int_{x\in\text{IG}} D^{\rR}_{\nu}(z x^\inv) d\mu(x)\\
	&= \int_{y\in\text{IG}} D^{\rR}_{\nu}(y) d\mu(y^\inv z)
\end{align}
as can be seen from equation \ref{RightCK}.\\

As time is considered to march from right to left, it is equation \ref{left} that will be relevant when the instrument sequence is continued forward in time.
Therefore, let it be understood that when the denominator of a KOD is not specified, it is implied to be \emph{the KOD with respect to the left-invariant Haar measure}:
\begin{equation}\label{KODLimply}
	\boxed{
		\vphantom{\Bigg(}
		\hspace{10pt}
		D_\mu(x) \equiv D_\mu^\rL(x).
		\hspace{10pt}
		\vphantom{\Bigg)}
	}
\end{equation}

\subsection{Translation Ultraoperators and their Invariant Derivatives}\label{TransDiff}

Every element of the IG defines a \emph{left-translation ultraoperator} that acts on the functions of the IG as
\begin{equation}\label{Ltrans}
\boxed{
	\vphantom{\Bigg(}
	\hspace{10pt}
	\fL_g[f](x) \equiv f(g^\inv x).
	\hspace{10pt}
	\vphantom{\Bigg)}
}
\end{equation}
It is easy to see that composition of these ultraoperators is a group homomorphism
\begin{align}
	\fL_h[\fL_g[f]](x)
	&=\fL_{g}[f](h^\inv x)\\
	&=f(g^\inv h^\inv x)\\
	&=f\left((hg)^\inv x\right)\\
	&=\fL_{hg}[f](x).
\end{align}
Acting on a given space of functions, the left-translation ultraoperators define the so-called \emph{left-regular representation} of the IG.
With the left-translation ultraoperators, one can extract from the convolution of two functions,
\begin{align}
	g*f(z)
	&\equiv \int_\text{IG} d_\rL y \, g(y)f(y^\inv z)\\
	&= \int_\text{IG} d_\rL y \, g(y)\fL_y[f](z),
\end{align}
and therefore define for every function $g$ the \emph{left-convolution ultraoperator}
\begin{align}
	\fZ^\rL_g
	&\equiv \int_\text{IG} d_\rL y \, g(y)\fL_y
\end{align}
for which it is easy to see that they form an algebra homomorphism | in particular
\begin{align}
	\fZ^\rL_g \circ \fZ^\rL_f = \fZ^\rL_{g*f}.
\end{align}
We can just as well define a left-convolution ultraoperator for any distribution $\mu$, which in turn means that for KODs (with respect to the left-invariant Haar measure)
\begin{equation}\label{LeftConvoOp}
\boxed{
	\vphantom{\Bigg(}
	\hspace{10pt}
	\fZ^{\rL}_\mu \equiv \fZ^\rL_{D^{\rL}_\mu}
	= \int_\text{IG} d\mu(x)\fL_x.
	\hspace{10pt}
	\vphantom{\Bigg)}
}
\end{equation}

Of course, there are also \emph{right-translation ultraoperators}
\begin{equation}
	\fR_g[f](x) \equiv f(x g^\inv),
\end{equation}
which instead compose to form a group anti-homomorphism,
\begin{align}
	\fR_h[\fR_g[f]](x)
	&=\fR_{g}[f](x h^\inv )\\
	&=f( x h^\inv g^\inv)\\
	&=f(x(gh)^\inv)\\
	&=\fR_{gh}[f](x).
\end{align}
Although it may be tempting to choose an alternative definition so to make the anti-homomorphism a plain homomorphism, we actually want to define the right-translation ultraoperators in this way because, like the left-translation ultraoperators, they transport a bump localized at the identity to a bump localized at the specified group element.
Here too one can extract from the definition of the convolution
\begin{align}
g*f(z)
&\equiv \int_\text{IG} d_\rR x \, g(zx^\inv)f(x)\\
&= \int_\text{IG} d_\rR x \, \fR_x[g](z)f(x)
\end{align}
and define for every function $f$ the \emph{right-convolution ultraoperator}
\begin{equation}
	\fZ^\rR_f = \int_\text{IG} d_\rR x \, f(x) \fR_x
\end{equation}
for which it is also easy to see they define an algebra anti-homomorphism | specifically
\begin{equation}
	\fZ^\rR_g \circ \fZ^\rR_f = \fZ^\rR_{f*g}.
\end{equation}
Again, we can just as well define a right-convolution ultraoperator for any distribution $\mu$ and for KODs with respect to the right-invariant Haar measure
\begin{align}
	\fZ^\rR_\mu \equiv \fZ^\rR_{D^{\rR}_\mu}
	&= \int_\text{IG} d\mu(x)\fR_x.
\end{align}
We will not have any need for the right-convolution ultraoperators.

Fantasticaly, the left- and right-translation ultraoperators commute,
\begin{equation}
	\fR_h \circ \fL_g = \fL_g \circ \fR_h.
\end{equation}
In other words, these translations are invariants of each other.
Because of this, left-translation is also called right-invariant translation and similarly right-translation is also called left-invariant translation.\\

If an IG is differentiable then it is called an \emph{instrumental Lie group} and has an associated \emph{instrumental Lie algebra} (ILA).  In the left-regular representation, the elements of the ILA correspond to the set of invariant differential ultraoperators on the functions of the IG, with the ultraoperator commutator representing the Lie bracket:
For every element $X\in$ ILA, one can differentiate a function $f$ at the point $x\in$ IG by left-translation,
\begin{equation}
\boxed{
	\vphantom{\Bigg(}
	\hspace{10pt}
	\left.\Rinv{X}\right|_x f \equiv \lim_{h\to0}\frac{f(e^{hX}x) - f(x)}{h}.
	\hspace{10pt}
	\vphantom{\Bigg)}
}
\end{equation}
In turn, one can define differential ultraoperators
\begin{equation}
	\Rinv{X}[f](x) \equiv \left.\Rinv{X}\right|_x f,
\end{equation}
the so-called \emph{right-invariant derivatives} because they have the property
\begin{equation}
	\Rinv{X} \circ \fR_g = \fR_g \circ \Rinv{X}
	\hspace{50pt}
	\text{or equivalently}
	\hspace{50pt}
	\left.\Rinv{X}\right|_x \circ \fR_g = \left.\Rinv{X}\right|_{xg}.
\end{equation}

\textcolor{red}{If a co\"ordinate system is established, these invariant derivatives can be expanded in terms of the partial derivatives with respect to those co\"ordinates.  Such expansions can be calculated with the use of the other tool of the IMP mentioned in the introduction, the so-called MMCSD.  However, this paper is intentionally focused on just one of the tools of the IMP, the KOD.  For further explanation, the most explicit instance of this can be found in \cite[eq.~164]{jackson2023weylheisenberg}}.

Given an IG element $e^X$, one can consider a path (a.k.a. one-parameter subgroup or flow line) $e^{tX}$ which at $t=0$ is the identity and at $t=1$ is the given IG element.
Similarly, any function $f$ of the IG can be considered as a function over the one-parameter subgroup $\phi^X(t) = f(e^{tX}x_o)$.
On any such one-parameter function one may consider Taylor expansions, the convergences of which correspond to properties of the translation ultraoperator $\phi^X(t+\delta) = e^{\delta \frac{d}{dt}}[\phi^X](t)$.
By an analytic continuation of Taylor's theorem on each one-parameter subgroup, one can therefore observe that the right-invariant derivatives generate the left-translation ultraoperators,
\begin{equation}
\boxed{
	\vphantom{\Bigg(}
	\hspace{10pt}
	\fL_{e^X}=e^{-\Rinv{X}}.
	\hspace{10pt}
	\vphantom{\Bigg)}
}
\end{equation}
Indeed, one can check that the map $X \mapsto \Rinv{X}$ is in fact a Lie algebra anti-homomorphism,
\begin{equation}
	[\Rinv{X},\Rinv{Y}]_\text{Comm} = -\Rinv{[X,Y]_\text{Abst}}
\end{equation}
where $[\mathfrak{U},\mathfrak{V}]_\text{Comm} \equiv \mathfrak{U}\circ\mathfrak{V} - \mathfrak{V}\circ\mathfrak{U}$ denotes the commutator of two ultraoperators on the space of functions on the IG and $[X,Y]_\text{Abst}$ denotes the abstract Lie bracket.

Similarly we can define for every element $X\in$ ILA the \emph{left-invariant derivative} at the point $x \in$ IG of a function $f$ as
\begin{align}
	\left.\Linv{X}\right|_x f \equiv  \lim_{h\to0}\frac{f(xe^{hX})-f(x)}{h}
\end{align}
and differential ultraoperators
\begin{equation}
	\Linv{X}[f](x) \equiv \left.\Linv{X}\right|_x f
\end{equation}
so named because they have the property
\begin{equation}
	\Linv{X} \circ \fL_g = \fL_g \circ \Linv{X}
	\hspace{50pt}
	\text{or equivalently}
	\hspace{50pt}
	\left.\Linv{X}\right|_x \circ \fL_g = \left.\Linv{X}\right|_{gx}.
\end{equation}
Sure enough, they too generate (over the vector space of analytic functions) the right-translation IG-ultraoperators
\begin{equation}
	\fR_{e^{tX}} = e^{-t\Linv{X}}
\end{equation}
and the map $X \mapsto \Linv{X}$ is a Lie-algebra homomorphism,
\begin{equation}
	[\Linv{X},\Linv{Y}]_{\text{Comm}}=\Linv{[X,Y]_{\text{Abst}}}.
\end{equation}

\subsection{The IG Algebra and Two Involutions:  Gelfand-Kolmogorov and Cartan-Schmidt}\label{AdjAlg}

Remembering that we are ultimately interested in continuous measurements, where time is represented by a product of Kraus operators proceeding from right to left, we focus again on the left-invariant Haar measure.\\

With the left-invariant Haar measure, one can define various norms and function spaces on the IG.
For any real number $p\ge1$, the Banach space $L^p(\text{IG})$ is the space of functions with norm
\begin{equation}
	||f||_{p,\text{IG}} \equiv \int_\text{IG} d_\rL x \, |f(x)|^p.
\end{equation}
Of these Banach spaces, only $L^1(\text{IG})$ is closed under convolution.\footnote{
	The simplest example of this is to consider an IG isomorphic to the real numbers, $\R$, where both Haar measures are equal to the standard differential of the Riemann integral, $d(x+a) = dx$.
For any real number $\epsilon > 0$, a bounded function $f$ of $\R$ with tails that fall off as $f(|x|\gg 1) \sim \frac{1}{|x|^{\frac{1}{p}+\epsilon}}$ is in $L^p(\R)$.
The convolution of such an $f$ with itself would therefore have tails that fall off as $f\!*\!f\,(|x|\gg 1) \sim \frac{1}{|x|^{\frac{2}{p}+2\epsilon-1}}$.
Therefore if $p>1$, then $f*f$ is not in $L^p(\R)$ for $2\epsilon \le 1 - \frac1p$.}
The Banach space $L^1(\text{IG})$ is therefore a Banach algebra called the group algebra or as we will call it the \emph{IG algebra} (IGA).
The IG algebra is the place where KODs live and closure under IG-convolution is simply a reflection of the fact that two instruments applied in sequence define an instrument.

In addition to norms, one can use the left-invariant measure of the IG to define the \emph{IG-inner product} between two complex-valued functions,
\begin{equation}
	(g,f)_\text{IG} \equiv \int_\text{IG} d_\rL x \,\; \overline{g(x)} f(x)
\end{equation}
where $\overline{c}$ denotes the complex conjugate of the complex number $c$.
With the IG-inner product, one can therefore define dual spaces of functions.
Most important for us is the space dual to the IGA, the space of bounded functions on the IG, sometimes called $L^\infty(\text{IG})$.
Since KODs live in the IGA, the instrument-element representation must consist of bounded superoperators for the total operation to exist.

There are a couple of important subspaces of the IGA that come up.
Although here is not the place to dig into them too much, they are mentioned because the KODs which solve the Kolmogorov equation to be introduced in section \ref{ODiff} are indeed in these subspaces for finite times.
The first important subspace is the closure of the space of infinitely-differentiable functions with compact support, sometimes called the Frechet space~\cite{helgason2008GASS}.
The dual space of this Frechet space is called the space of distributions.
The term ``distribution'' here refers to ``functions'' like Dirac delta-functions and their derivatives which although related are not to be confused with the Kraus-operator distribution, where the KOD is actually in the Frechet space, on the other side of the function duality from the ``distributions''.
The second important subspace is the so-called Schwartz space for semisimple IGs, their dual space known as the tempered distributions.~\cite{harish1966discrete}
If the KOD is in the Schwartz space, then the corresponding instrument-element representations are examples of the so-called tempered representations.~\cite{knapp1986representation}
However, for quantum systems the instrument-element representation is generally not a unitary representation as will be explained in just a moment.\\

The IG-inner product defines an ultraoperator adjoint
\begin{equation}\label{Kolmogorov}
	\boxed{
		\vphantom{\Bigg(}
		\hspace{10pt}
		\big(g,\mathfrak{U}^\ankh[f]\big)_\text{IG} \equiv \left(\mathfrak{U}[g],f\right)_\text{IG}
		\hspace{10pt}
		\vphantom{\Bigg)}
	}
\end{equation}
which will be called the \emph{Kolmogorov (ultraoperator) adjoint} because of its application in section \ref{ODiff}.
The use of an $\ankh$, called the ``Ankh'' or ``key to life'', is mostly because every obvious choice to denote an adjoint has already been taken | $*$ for convolution, $\dag$ for Hermitian conjugation, $\ddag$ for the Hilbert-Schmidt adjoint (equation \ref{HSadj}) , and $\qddag$ for the Choi-Jamiolkowski quasi-adjoint (equation \ref{CJquasiadj}).
As far as the author is aware, the Ankh does not have any political or historical associations that might make it a distastful symbol to use.
If anything, the Ankh has a fun turn-of-the-twentieth century Western alchemical vibe.

Of particular importance is the Kolmogorov adjoint of the left-translation ultraoperators and their right-invariant derivatives
\begin{equation}
	(\fL_g)^\ankh = \fL_g^\inv
	\hspace{50pt}
	\text{and}
	\hspace{50pt}
	(\Rinv{X})^\ankh = - \Rinv{X}.
\end{equation}
One thing this means is that the left-translation ultraoperators are \emph{stochastic}, meaning that they preserve the $L^1(\text{IG})$-norm for KODs $D(x)$ which are real and non-negative,
\begin{align}
	||\fL_g[D]||_{1,\text{IG}}
	&= (\fL_g[D],1)_{1,\text{IG}}\\
	&=(D,\fL_g^\ankh[1])_{1,\text{IG}}\\
	&=(D,1)_{1,\text{IG}}\\
	&=||D||_{1,\text{IG}}
\end{align}
where $1$ in the right argument of the IG-inner product stands for the constant function equal to unity over all of the IG.
While this also technically means the left-translation ultraoperators are $\ankh$-unitary in $L^2(\text{IG})$, this description should be avoided in this context because for quantum measurement they act on KODs and not wavefunctions (or density operators.)

Meanwhile, the right-translation ultraoperators are \emph{not} stochastic unless the IG is unimodular.
We will have no further use for the right-translation ultraoperators, so this point will not be discussed further.
Remember that the integration has been chosen with respect to the left-invariant Haar measure because the time of the measuring process is always considered to march from right to left, as explained at the end of section \ref{ODF}.\\

If the IG is unimodular, then the IGA is also an \emph{involutive Banach algebra}, actually in two different ways.
The first involution is what we will call the \emph{Gelfand involution}
\begin{equation}
	f^\ankh (x) \equiv \overline{f(x^\inv)}.
\end{equation}
Note that the symbol $\ankh$ has now been given a second definition, for elements of the IGA instead of the first definition for ultraoperators.
With the Gelfand involution, it is easy to see that the left-regular representation with the Kolmogorov ultraoperator adjoint is an $\ankh$-representation of the IGA, meaning it respects the Gelfand involution,
\begin{align}
	\left(\fZ^\rL_f\right)^\ankh
	&\equiv \left(\int d_\rL x \, f(x) \fL_x \right)^\ankh\\
	&= \int d_\rL x \, \overline{f(x)} \fL_x^\ankh\\
	&= \int d_\rL x \, \overline{f(x)} \fL_{x^\inv}\\
	&= \int d_\rL(x^\inv) \overline{f(x^\inv)} \fL_{x}\\
	&= \int d_\rR x \, f^\ankh(x) \fL_{x}\\
	&= \int d_\rL x \, f^\ankh(x) \fL_{x}\\
	&= \fZ^\rL_{f^\ankh}.
\end{align}
The IGA with the Gelfand involution is not quite a C*-algebra, or rather a ``$\text{C}^\ankh$-algebra'', because $||f^\ankh * f||_{1,\text{IG}} \neq ||f^\ankh||_{1,\text{IG}} ||f||_{1,\text{IG}}$.
However, this kind of involutive Banach algebra does in general define an enveloping C*-algebra by a standard construction.~\cite{dixmier1977c, dieudonne1979special}
In this case, I believe the left-regular representation continues to be an $\ankh$-representation.
On the other hand, the quantum instrument-element representations equipped with either $\ddag$ or $\qddag$ are \emph{not} $\ankh$-representations.
No deep use is made in this paper of the enveloping C*-algebraic aspects of the IGA, but this connection may nonetheless be important in other contexts where one might be interested in unitary representations of the IGA, such as would be the case in AdS-CFT.~\cite{kitaev2017notes}

The second involution is called the \emph{Cartan involution},
\begin{equation}
	f^\ddag (x) \equiv \overline{f(x^\dag)}
\end{equation}
where $\dag$ is a geometric abstraction of the Hermitian conjugate, also called a (type-IV) Cartan involution \cite{helgason2001differential,jackson2021spin,barut1986theory}.
The Kraus-operator representation of the IG is always assumed to be analytic in the sense that the actual Hermitian conjugate for a given representation $K_x$ respects the type-IV Cartan involution,
\begin{equation}
	(K_x)^\dag = K_{x^\dag}
\end{equation}
and therefore
\begin{equation}
	(\O_x)^\ddag = \O_{x^\dag},
\end{equation}
where recall the $\ddag$ of a superoperator denotes the Hilbert-Schmidt adjoint (equation \ref{HSadj}.)
Meanwhile, quantum operations such as equation \ref{quantumOperation} can be considered as superoperator analogs of the convolution ultraoperators,
\begin{equation}
	\Z_f \equiv \int d_\rL x \, \overline{f(x)} \O_x.
\end{equation}
With the Cartan involution, it is then easy to see that quantum instruments with the Hilbert-Schmidt adjoint are $\ddag$-representations of the IGA, meaning they respect the the Cartan involution,
\begin{align}
	(\Z_f)^\ddag
	&\equiv \left(\int d_\rL x \, \overline{f(x)} \O_x \right)^\ddag\\
	&= \int d_\rL x \, f(x)  \O_x^\ddag\\
	&= \int d_\rL x \, f(x)  \O_{x^\dag}\\
	&= \int d_\rL(x^\dag) f(x^\dag)  \O_{x}\\
	&= \int d_\rR x \, \overline{f^\ddag(x)} \O_{x}\\
	&= \int d_\rL x \, \overline{f^\ddag(x)} \O_{x}\\
	&= \Z_{f^\ddag}
\end{align}
where in particular $d_\rL(x^\dag) = d_\rR x \,$ because the right-invariant measure is unique  \cite[p.~26]{dieudonne1979special}.
In the same way as was explained for the $\ankh$, the IGA is not quite a ``$\text{C}^\ddag$-algebra'' either, but I believe it too defines another enveloping C*-algebra by the same standard construction \cite[p.~47]{dixmier1977c}.
The enveloping $\text{C}^\ddag$-algebra is clearly relevant to continuous quantum measurement.
Nevertheless, it is a technical perspective that does not seem to appear yet in quantum measurement circles.\\

The left-convolution ultraoperators are all that we will need to consider, so we will drop the $\rL$ but also replace it with a factor of the $\ankh$,
\begin{equation}
	\fZ_f^\ankh \equiv \fZ^\rL_f.
\end{equation}
This factor of the Kolmogorov adjoint ($\ankh$) appears quite consistently throughout the literature on Markov processes~\cite{feller1949theory} and we will see in the next section that it also intertwines nicely with the quantum channel.

\subsection{The Kolmogorov Equation for the KOD of a Continuous Measurement}\label{ODiff}

Let $\I_{dt}$ be a weak measurement such as those found in sections \ref{ContMeas} and \ref{SimultMeas} and $d\mu_T(x)$ be the Kraus-operator distribution of a continuous instrument
\begin{align}
	\I_T
	&\equiv \I_{dt} ^{*T/dt}\\
	&=\left\{d\mu_T(x) \O_x\right\}_{x\in\text{IG}}\label{It}
\end{align}
with corresponding total operation
\begin{align}
	\Z_T
	&\equiv \Z_{dt}^{\circ T/dt}\\
	&=\int_{\text{IG}} d\mu_T (x) \O_x.
\end{align}
Meanwhile, let the KOD | introduced in section \ref{ODF} (see equations \ref{KODL} and \ref{KODLimply}) | of the instrument at time $T$ be denoted as
\begin{equation}
	D_T(x) \equiv \frac{d\mu_T(x)}{d_\rL x \,}.
\end{equation}
Recall that not only do the instrument and total operation satisfy the one-parameter group property,
\begin{equation}
	\I_{t+dt} = \I_{dt} * \I_t
	\hspace{50pt}
	\text{and}
	\hspace{50pt}
	\Z_{t+dt} = \Z_{dt} \circ \Z_t
\end{equation}
as explained around equations \ref{OneParamSemiInst} and \ref{OneParamSemi},
but so do the Kraus-operator distribution and its KOD:
\begin{equation}\label{ChapKolm}
	\mu_{t+dt} = \mu_{dt} * \mu_t
	\hspace{50pt}
	\text{and}
	\hspace{50pt}
	D_{t+dt} = D_{dt} * D_t
\end{equation}
which can be thought of as special cases of equations \ref{SeqGrpPropInstr}, \ref{ConvoProp}, and \ref{LeftCKlast}.
Equations such as \ref{ChapKolm} are examples of so-called \emph{Chapman-Kolmogorov equations}.\cite{jackson2023positive, gardiner1986handbook, gardiner2019stochastic, gardiner2021elements}\\

Let $\O_x$ be an instrument-element representation of the IG (as explained in section \ref{OD}.)
Observe in particular that they intertwine the left-translation ultraoperators (introduced in section \ref{TransDiff}) with left composition by the corresponding superoperator from the instrument-element representation,
\begin{equation}
	\boxed{
	\hspace{10pt}
	\phantom{\Bigg(}
	\fL_g^\ankh[\O_x] = \O_{g} \circ \O_x.
	\phantom{\Bigg)}
	\hspace{10pt}
	}
\end{equation}
Let the weak instrument be now denoted
\begin{equation}
	\I_{dt} = \{d\mu_{dt}(v) \O_{\gamma(v)}\}_{v\in\Omega}
\end{equation}
with local sample space $\Omega$ (equal to either $\{0,1\}^{\times n}$ for locally Poisson processes as in equation \ref{Jump} or $\R^n$ for locally Wiener processes as in equation \ref{Diff}) and total operation
\begin{equation}
	\Z_{dt} \equiv \int_\Omega d\mu_{dt}(v) \O_{\gamma(v)}.
\end{equation}
By another simple observation,
\begin{align}
	\Z_{dt} \circ \O_x
	&= \int_\Omega d\mu_{dt}(v) \; \O_{\gamma(v)} \circ \O_x\\
	&= \int_\Omega d\mu_{dt}(v) \; \fL_{\gamma(v)^\inv}[\O_x]\\
	&= \fZ_{dt}[\O_x],
\end{align}
it is clear that the total operation superoperator $\Z_{dt}$ acting on the left of the instrument-element representation $\O_x$ by composition is equivalent to the left-convolution ultraoperator (also introduced in section \ref{TransDiff})
\begin{equation}
	\fZ_{dt} \equiv  \int_\Omega d\mu_{dt}(v) \; \fL_{\gamma(v)^\inv}
\end{equation}
acting on the instrument-element representation as a function of the IG.

Recall that the KOD of the instrument at time $t$ is $D_t(x)$, so equation \ref{It} may also be written as
\begin{equation}
	\I_t = \{d_\rL x \, D_t(x) \O_{x}\}_{x\in\text{IG}},
\end{equation}
with total operation
\begin{equation}
	\Z_t = \int_{\text{IG}} d_\rL x \, D_t(x) \O_x.
\end{equation}
Following this with an application of the weak instrument, the composite instrument at time $t+dt$ would be
\begin{align}
	\Z_{t+dt}
	&\equiv \Z_{dt}\circ\Z_t\\
	&= \int_{\text{IG}} d_\rL x \, D_t(x) \; \Z_{dt} \circ \O_x\\
	&= \int_{\text{IG}} d_\rL x \, D_t(x) \; \fZ_{dt}[\O_x]\\
	&= \int_{\text{IG}} d_\rL x \, \fZ_{dt}^\ankh[D_t](x) \; \O_x.
\end{align}
Therefore, the KOD of the composite instrument at time $t+dt$ would be
\begin{align}\label{prePDE}
	D_{t+dt} = \fZ_{dt}^\ankh[D_t]
\end{align}
where $\ankh$ is the Kolmogorov adjoint over the IG (introduced in section \ref{AdjAlg}.)
Meanwhile, the adjoint of the total operation ultraoperator of the weak instrument will have an infinitesimal generator,
\begin{align}
	\fZ_{dt}
	&=  \int_\Omega d\mu_{dt}(v) \fL_{\gamma(v)^\inv}\\
	&= e^{\fD \kappa dt}\\
	&= \mathfrak{I} + \fD \kappa dt
\end{align}
where $\mathfrak{I}$ denotes the identity ultraoperator and $\fD$ is the infinitesimal generator.
Therefore equation \ref{prePDE} can be expressed as a partial differential equation for the KOD, 
\begin{equation}\label{KolmogorovForward}
	\boxed{
		\vphantom{\Bigg(}
		\hspace{10pt}
		\frac1\kappa \frac{\partial}{\partial t} D_t = \fD^\ankh[D_t]
		\hspace{10pt}
		\vphantom{\Bigg)}
	}
\end{equation}
which is known as a \emph{Kolmogorov forward equation} for which we will therefore call $\fD^\ankh$ the \emph{(Kolmogorov) forward-generator}.
Naturally, this means $\fD$ is a Kolmogorov backward-generator, but we will have no use for backward equations in this paper.
In the diffusive case, the Kolmogorov forward equation is also sometimes called a \emph{diffusion equation} \cite{jackson2021spin} or a \emph{Fokker-Planck-Kolmogorov equation} \cite{jackson2023weylheisenberg,jackson2023positive}.\\

Recall that jump processes are measurements generated by the infinitesimal instrument elements of
\begin{equation}\label{JumpAgain}
	\I^{(L,\text{Jump})}_{dt} = \left\{ (\kappa dt)^{dN}\Odot\left[e^{-\frac12 L^\dag L \kappa dt}L^{dN}\right] \right\}_{dN\in\{0,1\}}
\end{equation}
and that diffusive processes are measurements generated by the infinitesimal instrument elements of
\begin{equation}\label{DiffAgain}
	\I^{(L,\text{Diff})}_{dt} = \left\{\frac{d(dW)}{\sqrt{2\pi dt}}e^{-\frac{dW^2}{2dt}}\Odot\left[e^{-\frac12(L^\dag L + L^2)\kappa dt + L\sqrt{\kappa}dW}\right]\right\}_{dW\in\R}
\end{equation}
as introduced in section \ref{ContMeas} (equations \ref{Jump} and \ref{Diff}.)
Simultaneous measurements can also be considered, but for simplicity only one Lindblad operator will be considered at first.
These weak instruments define IGs (also explained in section \ref{ContMeas}.)
Let the instrument elements of these weak instruments be ``minimally'' separated into a Kraus-operator distribution and an instrument-element representation in the following way:
\begin{equation}
	\mu_{dt}(dN) \equiv (\kappa dt)^{dN}
	\hspace{50pt}
	\text{and}
	\hspace{50pt}
	\O_{\gamma_{L,\text{Jump}}(dN)} \equiv \Odot\left[e^{-\frac12 \hat{L}^\dag \hat{L} \kappa dt}\hat{L}^{dN}\right]
\end{equation}
for jump processes and
\begin{equation}
	d\mu_{dt}(dW) \equiv \frac{d(dW)}{\sqrt{2\pi dt}}e^{-\frac{dW^2}{2dt}}
	\hspace{50pt}
	\text{and}
	\hspace{50pt}
	\O_{\gamma_{L,\text{Diff}}(dW)} \equiv \Odot\left[e^{-\frac12(\hat{L}^\dag \hat{L} + \hat{L}^2)\kappa dt + \hat{L}\sqrt{\kappa}dW}\right]
\end{equation}
for diffusive processes.

With these in mind, the total operation and Kolmogorov forward-generating ultraoperators may be calculated explicitly.
For the jump process we have
\begin{align}
	\mathfrak{Z}_{dt, L, \text{Jump}}^\ankh
	&= \sum_{dN} \mu_{dN}\fL_{\gamma_{L, \text{\text{Jump}(dN)}}}\\
	&= \fL_{\gamma_{L, \text{\text{Jump}(0)}}} + \kappa dt \fL_{\gamma_{L, \text{\text{Jump}(1)}}}\\
	&= \fL_{e^{-L^\dag L \frac12 \kappa dt}} + \kappa dt \fL_{L}\\
	&= e^{\Rinv{L^\dag L} \frac12 \kappa dt} + \kappa dt \fL_{L}\\
	&= 1+\kappa dt\left(\frac12 \Rinv{L^\dag L} + \fL_{L}\right)
\end{align}
and so the infinitesimal generator is
\begin{equation}
	\boxed{
		\vphantom{\Bigg(}
		\hspace{15pt}
		\fD_{L,\text{Jump}}^\ankh = \frac12 \Rinv{L^\dag L} + \fL_{L}.
		\hspace{10pt}
		\vphantom{\Bigg)}
	}
\end{equation}
For the diffusive process we have
\begin{align}
	\mathfrak{Z}_{dt,L, \text{Diff}}^\ankh
	&= \int d\mu(dW)\fL_{\gamma_\text{L,Diff}(dW)}\\
	&= \int d\mu(dW)e^{\Rinv{L^\dag L+L^2}\frac12\kappa dt - \Rinv{L} \sqrt\kappa dW}\\
	&= e^{\Rinv{L^\dag L+L^2}\frac12\kappa dt}\int d\mu(dW) e^{- \Rinv{L} \sqrt\kappa dW}\\
	&= e^{\Rinv{L^\dag L+L^2}\frac12\kappa dt}e^{\frac12 \Rinv{L} \Rinv{L} \kappa dt}\\
	&= e^{(\frac12\Rinv{L^\dag L+L^2}+\frac12 \Rinv{L} \Rinv{L}) \kappa dt}
\end{align}
and so the infinitesimal generator is
\begin{equation}
	\boxed{
		\vphantom{\Bigg(}
		\hspace{15pt}
		\fD_{L,\text{Diff}}^\ankh = \frac12\Rinv{L^\dag L+L^2}+\frac12 \Rinv{L} \Rinv{L}.
		\hspace{10pt}
		\vphantom{\Bigg)}
	}
\end{equation}
If we considered a set of Lindblad operators measured simultaneously, the Kolmogorov forward generator ultraoperator would simply be the sum of the individual ultraoperators,
\begin{equation}
	\fD_{\vec{L}}^\ankh = \sum_i \fD_{L_i}^\ankh.
\end{equation}\\

The Kolmogorov equation for the KOD (equation \ref{KolmogorovForward}) can be seen as a direct analog of the Lindblad master equation (equation \ref{LindbladMaster}).
Specifically, the superoperator dissipator $\D$ acting on the density operator $\rho_t$ in the dual of the von Neumann algebra corresponds exactly to the ultraoperator Kolmogorov forward generator $\fD^\ankh$ acting on the KOD $D_t(x)$ in the IGA.
As a final exercise, let us show that indeed
\begin{align}
	\fD[\O_x] = \D\circ \O_x,
\end{align}
for both jump and diffusive processes.
First, observe the following translation and differentiation identities
\begin{equation}
	\fL_g^\ankh [K_x] = K_g K_x
	\hspace{50pt}
	\text{and}
	\hspace{50pt}
	\Rinv{L}[K_x] = \hat{L} K_x
\end{equation}
and thus
\begin{equation}
	\fL_g^\ankh [\O_x] = (K_g \odot K_g^\dag) \circ \O_x
	\hspace{50pt}
	\text{and}
	\hspace{50pt}
	\Rinv{L}[\O_x] = (\hat{L} \odot \hat{1} + \hat{1} \odot \hat{L}^\dag)\circ \O_x.
\end{equation}
For jump processes we therefore have
\begin{align}
	\fD_{L,\text{Jump}}[\O_x]
	&= \left(-\frac12 \Rinv{L^\dag L} + \fL_{L}^\ankh \right)[\O_x]\\
	&= \left(-\frac12 (\hat{L}^\dag \hat{L} \odot \hat{1} + \hat{1} \odot \hat{L}^\dag \hat{L}) + \hat{L}\odot \hat{L}^\dag \right)\circ\O_x\\
	&=\D[\hat{L}]\circ \O_x
\end{align}
and for diffusive processes we have
\begin{align}
	\fD_{L,\text{Diff}}[\O_x]
	&= \left(-\frac12 \Rinv{L^\dag L+L^2} + \frac12 \Rinv{L} \Rinv{L} \right)[\O_x]\\
	&= \left(-\frac12 \Big((\hat{L}^\dag \hat{L} + \hat{L}^2) \odot \hat{1} + \hat{1} \odot (\hat{L}^\dag \hat{L}  +  \hat{L}^{2\dag})\Big) + \frac12(\hat{L} \odot \hat{1} + \hat{1} \odot \hat{L}^\dag)^2 \right)\circ\O_x\\
	&= \frac12 \left(-\hat{L}^\dag \hat{L} \odot \hat{1} - \hat{L}^2 \odot \hat{1} - \hat{1} \odot \hat{L}^\dag \hat{L} - \hat{1} \odot \hat{L}^{2\dag} + \hat{L}^2 \odot \hat{1} + 2 \hat{L} \odot \hat{L}^\dag + \hat{1}\odot \hat{L}^{2\dag} \right)\circ\O_x\\
	&= \frac12 \left(-\hat{L}^\dag \hat{L} \odot \hat{1} - \hat{1} \odot \hat{L}^\dag \hat{L} + 2 \hat{L} \odot \hat{L}^\dag \right)\circ\O_x\\
	&=\D[\hat{L}]\circ \O_x
\end{align}
as promised.

\section{Concluding Remarks}\label{Conclude}

In this paper, the instrument manifold\footnote{
	``Manifold'' is apparently a term from Kantian philosophy, where it is defined as ``The sum of the particulars furnished by sense before they have been unified by the synthesis of the understanding'', according to Oxford Languages.}
program was further developed by showing that the sequential application of measuring instruments corresponds to the instrumental-group (IG) \emph{convolution} of their Kraus-operator densities (KODs).
The IG and KOD were originally designed to analyze certain continuous measurements, the so-called principal instruments which generate finite-dimensional universal Lie groups~\cite{jackson2021spin,jackson2023weylheisenberg,jackson2023positive}.
In this paper, the perspective was shifted away from continuous measurement and towards the sequential, where by so doing, the presence of the convolution structure was revealed.
With the convolution at hand, it became clear that the space of absolutely integrable functions of the IG, among which KODs are a special case, was in fact an involutive Banach algebra, involutive in two ways, and we called this algebra the instrumental group algebra (IGA).\\

The convolution was developed in three phases.
We began by identifying how a sequence of two measuring instruments defines a new instrument,
the set of all pairs of elements, one from the first and the other from the second, composed.
This kind of set composition was the first, most basis type of convolution (equation \ref{convo}).
Then, the concepts of an instrumental group (IG) and the subsequent Kraus-operation distribution were introduced by recognizing that there is often enough structure to identify whether two distinct pairs of instrument elements produce the same composite instrument element or not,
\begin{equation}
	\Omega_x = \sqrt{d\mu(x)}K_x
	\hspace{25pt}
	\text{where}
	\hspace{25pt}
	K_y K_x = K_{yx}.
\end{equation}
This converted the structure of sequential measurement from combining sets of instrument elements to the convolution of their Kraus-operator distributions (equation \ref{convoMeas}),
\begin{equation}
	\I_1 * \I_0
	\hspace{25pt}
	\Longrightarrow
	\hspace{25pt}
	d(\mu_1*\mu_0)(z) \equiv \int_{y\in\text{IG}} d\mu_1(y)d\mu_0(y^\inv z).
\end{equation}
Finally, the IG was assumed to be ``topological'', meaning that it was assumed to have invariant measures,
\begin{equation}
	d\mu(x) = d_\rL x D(x)
	\hspace{25pt}
	\text{where}
	\hspace{25pt}
	d_\rL(gx) = d_\rL x.
\end{equation}
This converted the Kraus-operator distributions into the more functional Kraus-operator densities (KOD) which could be convolved in a style more typically seen in analysis (equation \ref{convoFunc}),
\begin{equation}
	d(\mu_1*\mu_0)(z)
	\hspace{25pt}
	\Longrightarrow
	\hspace{25pt}
	D_1*D_0(z) \equiv \int_{\text{IG}} d_{\rL}y \, D_1(y)D_0(y^\inv z).
\end{equation}

Having identified the convolution, the multiplication of the IGA, it was demonstrated that the Lindblad master equation which governs the evolution of the quantum channel superoperator and the previously discovered Kolmogorov equation which governs the evolution of the KOD,
\begin{equation}
	\frac1\kappa\frac{\partial}{\partial t}\Z_t^{(\lambda)} = \D^{(\lambda)} \circ \Z_t^{(\lambda)} 
	\hspace{25pt}
	\text{and}
	\hspace{25pt}
	\frac1\kappa\frac{\partial}{\partial t}D_t (x) = \fD^\ankh[D_t](x),
\end{equation}
are in fact intertwined by the instrument element representation of the IG,
\begin{equation}
	\fD[\O_x^{(\lambda)}] = \D^{(\lambda)} \circ \O_x^{(\lambda)}.
\end{equation}
The one-parameter group property characteristic of the superoperators that solve the Lindblad master equation were seen to lift to a one-parameter group property between KODs, the invariant Chapman-Kolmogorov equation,
\begin{equation}
	\Z^{(\lambda)}_{t+\Delta t} = \Z^{(\lambda)}_{\Delta t}\circ \Z^{(\lambda)}_t
	\hspace{25pt}
	\Longrightarrow
	\hspace{25pt}
	D_{t+ \Delta t} = D_{\Delta t} * D_t.
\end{equation}
Simultaneously, this showed that all quantum channel superoperators are in fact representations of IGA elements,
\begin{equation}
	\Z^{(\lambda)}_{D_1} \circ \Z^{(\lambda)}_{D_0} = \Z^{(\lambda)}_{D_1*D_0}
	\hspace{25pt}
	\text{and}
	\hspace{25pt}
	\Z^{(\lambda)}_D \equiv \int_{\text{IG}} d_\rL y \, \overline{D(y)} \O^{(\lambda)}_y,
\end{equation}
which generalized the (one-parameter) group property considerably.
In this way, similar to how the IG has the universal property of lifting the selective time-ordered exponentials off of the system Hilbert space, 
\begin{equation}
	K[d\vec{W}_{[0,T)}] = K_{\gamma[d\vec{W}_{[0,T)}]},
\end{equation}
the IGA has the universal property of lifting the time-dependence of the quantum channel evolution off of the system Hilbert space,
\begin{equation}
	\Z^{(\lambda)}_t = \Z^{(\lambda)}_{D_t},
\end{equation}
where the $\Z^{(\lambda)}_t$ on the left denotes the solution to the Lindblad equation and the $\Z^{(\lambda)}_D$ on the right denotes the representation of the KOD, $D \in \text{IGA}$.
In this way, every quantum channel superoperator becomes the noncommutative analog of a Laplace transform, evaluated at a particular irrep,
\begin{equation}
	\Z : D(x) \longmapsto \Z^{(\lambda)}_D.
\end{equation}

In addition to the Kolmogrov generator, further operators on the IGA, generally called ultraoperators, were also considered, allowing the analysis to be completely autonomous from the superoperators that originally represented the measuring process. 
Perhaps the most fundamental of the ultraoperators were the right-invariant derivatives and left-translations, denoted $\Rinv{X}$ and $\fL_x$.
With these, the left-convolution ultraoperators could be constructed in direct analogy with the quantum channel superoperators,
\begin{equation}
	\fZ^\ankh_{D_1} \circ \fZ^\ankh_{D_0} = \fZ^\ankh_{D_1*D_0}
	\hspace{25pt}
	\text{and}
	\hspace{25pt}
	\fZ^\ankh_D \equiv \int_{\text{IG}} d_\rL y \, D(y) \fL_y.
\end{equation}
Unlike the instrument elements $\O^{(\lambda)}_x$ which are not generally trace-preserving, the selective left-translation ultraoperators $\fL_x$ are always norm-preserving and therefore the left-convolution ultraoperators $\fZ^\ankh_D$ in the left-translation ultraoperator actually define bona fide stochastic/Markov matrices.
These operators all have superoperator counterparts with further intertwining relations which are summarized in Table~\ref{Intertwiners}.

\begin{table}	
	\begin{tabular}{r||cc|cc}
		\bf{Intertwining Relations}& integrated &&& differential\textreferencemark\\
		\hline\hline
		&&&&\\
		selective &
		$\fL_a^\ankh[\O^{(\lambda)}] = \O_a^{(\lambda)}\circ\O_x^{(\lambda)}$
		&\hspace{25pt}&\hspace{25pt}&
		$\Rinv{L}[\O_x^{(\lambda)}] = (\hat{L}^{(\lambda)}\odot 1 + 1 \odot \hat{L}^{(\lambda)\dag})\circ\O_x^{(\lambda)}$\\
		&&\\
		total &
		$\fZ_f[\O_x^{(\lambda)}] = \Z_f\circ\O_x^{(\lambda)}$
		&&&
		$\fD_L[\O_x^{(\lambda)}] = \D[\hat{L}^{(\lambda)}]\circ\O_x^{(\lambda)}$\\
		&&&&
	\end{tabular}
	\caption{{\bf Ultraoperator-Superoperator Intertwining Relations:} The instrument-element representations of the IG, $\O_x^{(\lambda)}$ define an intertwiner between the algebra of ultraoperators (acting on the IGA) and the algebra of superoperators (acting on the dual of the von Neumann algebra).  This table displays four pairs of operators, intertwined by the instrument-element representation, that are particularly relevant to continuous measurements and sequential measurements.  After the IGA and ultraoperator are introduced, these intertwining relations were derived and used in section \textsection\ref{ODiff}.  Concepts with the \textreferencemark \, mark were first introduced in previous work~\cite{jackson2023positive}.}
	\label{Intertwiners}
\end{table}

In addition to the translations, their derivatives, and their integrals (convolutions), also introduced were two involutive ultraoperators,
\begin{equation}
	D^\ddag(x) \equiv \overline{D(x^\dag)}
	\hspace{25pt}
	\text{and}
	\hspace{25pt}
	D^\ankh(x) \equiv \overline{D(x^\inv)},
\end{equation}
named after Cartan and Gelfand, respectively.
It was then identified that the Hilbert-Schmidt superoperator adjoint of the quantum channel representation represented the Cartan involution, while the Kolmogorov ultraoperator adjoint of the left-convolution representation represented the Gelfand involution,
\begin{equation}
	(\Z_D)^\ddag = \Z_{D^\ddag}
	\hspace{25pt}
	\text{and}
	\hspace{25pt}
	(\fZ_D)^\ankh = \fZ_{D^\ankh}.
\end{equation}

Finally, the KOD Kolmogorov forward-generating ultraoperators for diffusive processes and jump processes were explicitly calculated for general Lindblad operators.
For a single Lindblad operator the KOD Kolmogorov generators are
\begin{equation}
	\fD_{L,\text{Diff}}^\ankh = \frac12\Rinv{L^\dag L+L^2}+\frac12 \Rinv{L} \Rinv{L}
	\hspace{25pt}
	\text{and}
	\hspace{25pt}
	\fD_{L,\text{Jump}}^\ankh = \frac12 \Rinv{L^\dag L} + \fL_{L}.
\end{equation}
For multiple Lindblad operators (which may not commutte), one simply adds the individual forward generators,
\begin{equation}
	\fD_{(L_1,\ldots,L_n)} = \fD_{L_1} + \ldots + \fD_{L_n}.
\end{equation}\\

Perhaps the most important point about the IG is that because it is universal, it can be considered as a form of memory within the measuring instrument, making it a kind of automaton.
If the IG is principal or automatic, then the measuring process can be simulated with a classical automaton.
This will allow the IMP to take an enormous step:
From the perspective of the IG, it has become clear that the actual reason why a dissipative channel superoperator satisfies the group property is not exactly because the measuring process is Markov.
Rather, it is because in addition, the Markov kernel of the measuring process is right-invariant.

To break right-invariance would mean that the instrument would start to make changes based on its memory | that is, adapt.
If these adaptations are only based on the IG position at that moment in time, then although the time-dependence of the quantum channel superoperator would cease to be Markovian, the instrument evolution would still remain Markov!
From a technical point of view, what this is saying is that if the instrument were to adapt based on the current IG position, then although the Linblad master equation would cease to exist, there would still be a KOD Kolmogorov equation, but now with a Kolmogorov generator that depends on the IG position,
\begin{align}
	\frac1\kappa \frac{\partial}{\partial t}D_t(x) = \fD_x^\ankh [D_t](x)!
\end{align}

Having said all of this, it is crucial to return to the other primary tool of the IMP that was virtually abandoned at the beginning of this paper, the so-called \emph{modified Maurer-Cartan stochastic differential} or MMCSD.
The MMCSD is precisely what an automated measuring instrument would use to update its IG position as it continues to observe.
Further, such automation is only possible if the IG is automatable | that is, itself ``automatic''~(section \ref{ContMeas}).

It is therefore quite remarkable that the KOD, its Kolmogorov equation, the IGA, and the ultraoperator formalism, all still manage to exist, regardless of whether or not the IG was automatic to begin with.
Nevertheless, the fundamental interest is still with IGs that are automatic, especially IGs that are universally automatic such for ISM and SPQM~\cite{jackson2021spin,jackson2023weylheisenberg}.
Something about this seems to be exactly why classical and generalized phase-space correspondences define realistic, tomographically complete POVMs and also why the theory of quasiprobability manages to provide a completely autonomous, alternaive theory of quantum mechanics for systems such as those with spin or quadrature.

Currently under preparation is a paper that returns to the original, most fundamental of continuous measurements: the indirect single-mode photodetector \cite{srinivas1981photon} and heterodyne instrument \cite{Goetsch1994a, wiseman1994quantum}.
Indeed, these measurements have much simpler IGs than the ones that occur for ISM and SPQM.
By treating them systematically with these more general (but sophisticated) tools (the KOD and the MMCSD), perhaps it will become easier for more scientists to recognize the significance of the IMP.

\acknowledgements{
	The author would like to thank Carl Caves, Lucien Hardy, and Rob Spekkens for their support.
	Research at Perimeter Institute is supported in part by the Government of Canada through the Department of Innovation, Science and Economic Development Canada and by the Province of Ontario through the Ministry of Colleges and Universities.
}

\bibliography{ChrisJackson}

\begin{thebibliography}{87}%
\makeatletter
\providecommand \@ifxundefined [1]{%
 \@ifx{#1\undefined}
}%
\providecommand \@ifnum [1]{%
 \ifnum #1\expandafter \@firstoftwo
 \else \expandafter \@secondoftwo
 \fi
}%
\providecommand \@ifx [1]{%
 \ifx #1\expandafter \@firstoftwo
 \else \expandafter \@secondoftwo
 \fi
}%
\providecommand \natexlab [1]{#1}%
\providecommand \enquote  [1]{``#1''}%
\providecommand \bibnamefont  [1]{#1}%
\providecommand \bibfnamefont [1]{#1}%
\providecommand \citenamefont [1]{#1}%
\providecommand \href@noop [0]{\@secondoftwo}%
\providecommand \href [0]{\begingroup \@sanitize@url \@href}%
\providecommand \@href[1]{\@@startlink{#1}\@@href}%
\providecommand \@@href[1]{\endgroup#1\@@endlink}%
\providecommand \@sanitize@url [0]{\catcode `\\12\catcode `\$12\catcode
  `\&12\catcode `\#12\catcode `\^12\catcode `\_12\catcode `\%12\relax}%
\providecommand \@@startlink[1]{}%
\providecommand \@@endlink[0]{}%
\providecommand \url  [0]{\begingroup\@sanitize@url \@url }%
\providecommand \@url [1]{\endgroup\@href {#1}{\urlprefix }}%
\providecommand \urlprefix  [0]{URL }%
\providecommand \Eprint [0]{\href }%
\providecommand \doibase [0]{https://doi.org/}%
\providecommand \selectlanguage [0]{\@gobble}%
\providecommand \bibinfo  [0]{\@secondoftwo}%
\providecommand \bibfield  [0]{\@secondoftwo}%
\providecommand \translation [1]{[#1]}%
\providecommand \BibitemOpen [0]{}%
\providecommand \bibitemStop [0]{}%
\providecommand \bibitemNoStop [0]{.\EOS\space}%
\providecommand \EOS [0]{\spacefactor3000\relax}%
\providecommand \BibitemShut  [1]{\csname bibitem#1\endcsname}%
\let\auto@bib@innerbib\@empty
\bibitem [{\citenamefont {Jackson}\ and\ \citenamefont
  {Caves}(2022)}]{jackson2021spin}%
  \BibitemOpen
  \bibfield  {author} {\bibinfo {author} {\bibfnamefont {C.~S.}\ \bibnamefont
  {Jackson}}\ and\ \bibinfo {author} {\bibfnamefont {C.~M.}\ \bibnamefont
  {Caves}},\ }\bibfield  {title} {\bibinfo {title} {How to perform the coherent
  measurement of a curved phase space by continuous isotropic measurement. i.
  spin and the kraus-operator geometry of $\mathrm{SL}(2,\mathbb{C})$},\
  }\href@noop {} {\bibfield  {journal} {\bibinfo  {journal} {Quantum,
  arXiv:2107.12396}\ } (\bibinfo {year} {2022})}\BibitemShut {NoStop}%
\bibitem [{\citenamefont {Jackson}\ and\ \citenamefont
  {Caves}(2023{\natexlab{a}})}]{jackson2023weylheisenberg}%
  \BibitemOpen
  \bibfield  {author} {\bibinfo {author} {\bibfnamefont {C.~S.}\ \bibnamefont
  {Jackson}}\ and\ \bibinfo {author} {\bibfnamefont {C.~M.}\ \bibnamefont
  {Caves}},\ }\bibfield  {title} {\bibinfo {title} {Simultaneous momentum and
  position measurement and the instrumental weyl-heisenberg group},\
  }\href@noop {} {\bibfield  {journal} {\bibinfo  {journal} {Entropy,
  arXiv:2306.01045}\ } (\bibinfo {year} {2023}{\natexlab{a}})}\BibitemShut
  {NoStop}%
\bibitem [{\citenamefont {Jackson}\ and\ \citenamefont
  {Caves}(2023{\natexlab{b}})}]{jackson2023positive}%
  \BibitemOpen
  \bibfield  {author} {\bibinfo {author} {\bibfnamefont {C.~S.}\ \bibnamefont
  {Jackson}}\ and\ \bibinfo {author} {\bibfnamefont {C.~M.}\ \bibnamefont
  {Caves}},\ }\bibfield  {title} {\bibinfo {title} {Simultaneous measurements
  of noncommuting observables: Positive transformations and instrumental lie
  groups},\ }\href@noop {} {\bibfield  {journal} {\bibinfo  {journal} {Entropy,
  arXiv:2306.06167}\ } (\bibinfo {year} {2023}{\natexlab{b}})}\BibitemShut
  {NoStop}%
\bibitem [{\citenamefont {D'Ariano}\ \emph {et~al.}(2002)\citenamefont
  {D'Ariano}, \citenamefont {{Lo Presti}},\ and\ \citenamefont
  {Sacchi}}]{DAriano2002}%
  \BibitemOpen
  \bibfield  {author} {\bibinfo {author} {\bibfnamefont {G.~M.}\ \bibnamefont
  {D'Ariano}}, \bibinfo {author} {\bibfnamefont {P.}~\bibnamefont {{Lo
  Presti}}},\ and\ \bibinfo {author} {\bibfnamefont {M.~F.}\ \bibnamefont
  {Sacchi}},\ }\bibfield  {title} {\bibinfo {title} {A quantum measurement of
  the spin direction},\ }\href@noop {} {\bibfield  {journal} {\bibinfo
  {journal} {Physics Letters~A}\ }\textbf {\bibinfo {volume} {292}},\ \bibinfo
  {pages} {233} (\bibinfo {year} {2002})}\BibitemShut {NoStop}%
\bibitem [{\citenamefont {Massar}\ and\ \citenamefont
  {Popescu}(1995)}]{Massar1995}%
  \BibitemOpen
  \bibfield  {author} {\bibinfo {author} {\bibfnamefont {S.}~\bibnamefont
  {Massar}}\ and\ \bibinfo {author} {\bibfnamefont {S.}~\bibnamefont
  {Popescu}},\ }\bibfield  {title} {\bibinfo {title} {{Optimal extraction of
  information from finite quantum ensembles}},\ }\href@noop {} {\bibfield
  {journal} {\bibinfo  {journal} {Physical Review Letters}\ }\textbf {\bibinfo
  {volume} {74}},\ \bibinfo {pages} {1259} (\bibinfo {year}
  {1995})}\BibitemShut {NoStop}%
\bibitem [{\citenamefont {Shojaee}\ \emph {et~al.}(2018)\citenamefont
  {Shojaee}, \citenamefont {Jackson}, \citenamefont {Riofr\'{\i}o},
  \citenamefont {Kalev},\ and\ \citenamefont {Deutsch}}]{shojaee2018optimal}%
  \BibitemOpen
  \bibfield  {author} {\bibinfo {author} {\bibfnamefont {E.}~\bibnamefont
  {Shojaee}}, \bibinfo {author} {\bibfnamefont {C.~S.}\ \bibnamefont
  {Jackson}}, \bibinfo {author} {\bibfnamefont {C.~A.}\ \bibnamefont
  {Riofr\'{\i}o}}, \bibinfo {author} {\bibfnamefont {A.}~\bibnamefont
  {Kalev}},\ and\ \bibinfo {author} {\bibfnamefont {I.~H.}\ \bibnamefont
  {Deutsch}},\ }\bibfield  {title} {\bibinfo {title} {Optimal pure-state qubit
  tomography via sequential weak measurements},\ }\href@noop {} {\bibfield
  {journal} {\bibinfo  {journal} {Physical Review Letters}\ }\textbf {\bibinfo
  {volume} {121}},\ \bibinfo {pages} {130404} (\bibinfo {year}
  {2018})}\BibitemShut {NoStop}%
\bibitem [{\citenamefont {Deutsch}\ and\ \citenamefont
  {Jessen}(2010)}]{deutsch2010quantum}%
  \BibitemOpen
  \bibfield  {author} {\bibinfo {author} {\bibfnamefont {I.~H.}\ \bibnamefont
  {Deutsch}}\ and\ \bibinfo {author} {\bibfnamefont {P.~S.}\ \bibnamefont
  {Jessen}},\ }\bibfield  {title} {\bibinfo {title} {Quantum control and
  measurement of atomic spins in polarization spectroscopy},\ }\href@noop {}
  {\bibfield  {journal} {\bibinfo  {journal} {Optics Communications}\ }\textbf
  {\bibinfo {volume} {283}},\ \bibinfo {pages} {681} (\bibinfo {year}
  {2010})}\BibitemShut {NoStop}%
\bibitem [{\citenamefont {Leonhardt}(1997)}]{leonhardt1997measuring}%
  \BibitemOpen
  \bibfield  {author} {\bibinfo {author} {\bibfnamefont {U.}~\bibnamefont
  {Leonhardt}},\ }\href@noop {} {\emph {\bibinfo {title} {Measuring the Quantum
  State of Light}}},\ Cambridge Studies in Modern Optics\ (\bibinfo
  {publisher} {Cambridge University Press},\ \bibinfo {year}
  {1997})\BibitemShut {NoStop}%
\bibitem [{\citenamefont {Klauder}(1960)}]{klauder1960action}%
  \BibitemOpen
  \bibfield  {author} {\bibinfo {author} {\bibfnamefont {J.~R.}\ \bibnamefont
  {Klauder}},\ }\bibfield  {title} {\bibinfo {title} {The action option and a
  feynman quantization of spinor fields in terms of ordinary c-numbers},\
  }\href@noop {} {\bibfield  {journal} {\bibinfo  {journal} {Annals of
  Physics}\ }\textbf {\bibinfo {volume} {11}},\ \bibinfo {pages} {123}
  (\bibinfo {year} {1960})}\BibitemShut {NoStop}%
\bibitem [{\citenamefont {Husimi}(1940)}]{husimi1940formal}%
  \BibitemOpen
  \bibfield  {author} {\bibinfo {author} {\bibfnamefont {K.}~\bibnamefont
  {Husimi}},\ }\bibfield  {title} {\bibinfo {title} {Some formal properties of
  the density matrix},\ }\href@noop {} {\bibfield  {journal} {\bibinfo
  {journal} {Proceedings of the Physico-Mathematical Society of Japan. 3rd
  Series}\ }\textbf {\bibinfo {volume} {22}},\ \bibinfo {pages} {264} (\bibinfo
  {year} {1940})}\BibitemShut {NoStop}%
\bibitem [{\citenamefont {Arthurs}\ and\ \citenamefont
  {Kelly}(1965)}]{arthurs1965simultaneous}%
  \BibitemOpen
  \bibfield  {author} {\bibinfo {author} {\bibfnamefont {E.}~\bibnamefont
  {Arthurs}}\ and\ \bibinfo {author} {\bibfnamefont {J.~L.~J.}\ \bibnamefont
  {Kelly}},\ }\bibfield  {title} {\bibinfo {title} {On the simultaneous
  measurement of a pair of conjugate observables},\ }\href@noop {} {\bibfield
  {journal} {\bibinfo  {journal} {Bell Syst. Tech. J. Brief}\ }\textbf
  {\bibinfo {volume} {44}},\ \bibinfo {pages} {725} (\bibinfo {year}
  {1965})}\BibitemShut {NoStop}%
\bibitem [{\citenamefont {She}\ and\ \citenamefont
  {Heffner}(1966)}]{she1966simultaneous}%
  \BibitemOpen
  \bibfield  {author} {\bibinfo {author} {\bibfnamefont {C.~Y.}\ \bibnamefont
  {She}}\ and\ \bibinfo {author} {\bibfnamefont {H.}~\bibnamefont {Heffner}},\
  }\bibfield  {title} {\bibinfo {title} {Simultaneous measurement of
  noncommuting observables},\ }\href@noop {} {\bibfield  {journal} {\bibinfo
  {journal} {Physical Review}\ }\textbf {\bibinfo {volume} {152}},\ \bibinfo
  {pages} {1103} (\bibinfo {year} {1966})}\BibitemShut {NoStop}%
\bibitem [{\citenamefont {Personick}(1971)}]{personick1971image}%
  \BibitemOpen
  \bibfield  {author} {\bibinfo {author} {\bibfnamefont {S.~D.}\ \bibnamefont
  {Personick}},\ }\bibfield  {title} {\bibinfo {title} {An image band
  interpretation of optical heterodyne noise},\ }\href@noop {} {\bibfield
  {journal} {\bibinfo  {journal} {Bell Syst. Tech. J. Brief}\ }\textbf
  {\bibinfo {volume} {50}},\ \bibinfo {pages} {213} (\bibinfo {year}
  {1971})}\BibitemShut {NoStop}%
\bibitem [{\citenamefont {Lvovsky}\ and\ \citenamefont
  {Raymer}(2009)}]{lvovsky2009continuous}%
  \BibitemOpen
  \bibfield  {author} {\bibinfo {author} {\bibfnamefont {A.~I.}\ \bibnamefont
  {Lvovsky}}\ and\ \bibinfo {author} {\bibfnamefont {M.~G.}\ \bibnamefont
  {Raymer}},\ }\bibfield  {title} {\bibinfo {title} {Continuous-variable
  optical quantum-state tomography},\ }\href@noop {} {\bibfield  {journal}
  {\bibinfo  {journal} {Reviews of Modern Physics}\ }\textbf {\bibinfo {volume}
  {81}} (\bibinfo {year} {2009})}\BibitemShut {NoStop}%
\bibitem [{\citenamefont {Wiseman}(1996)}]{wiseman1996measurement}%
  \BibitemOpen
  \bibfield  {author} {\bibinfo {author} {\bibfnamefont {H.~M.}\ \bibnamefont
  {Wiseman}},\ }\bibfield  {title} {\bibinfo {title} {Quantum trajectories and
  quantum measurement theory},\ }\href@noop {} {\bibfield  {journal} {\bibinfo
  {journal} {Quantum and Semiclassical Optics: Journal of the European Optical
  Society Part B}\ }\textbf {\bibinfo {volume} {8}},\ \bibinfo {pages} {205}
  (\bibinfo {year} {1996})}\BibitemShut {NoStop}%
\bibitem [{\citenamefont {Goetsch}\ and\ \citenamefont
  {Graham}(1994)}]{Goetsch1994a}%
  \BibitemOpen
  \bibfield  {author} {\bibinfo {author} {\bibfnamefont {P.}~\bibnamefont
  {Goetsch}}\ and\ \bibinfo {author} {\bibfnamefont {R.}~\bibnamefont
  {Graham}},\ }\bibfield  {title} {\bibinfo {title} {Linear stochastic wave
  equations for continuously measured quantum systems},\ }\href
  {https://doi.org/10.1103/PhysRevA.50.5242} {\bibfield  {journal} {\bibinfo
  {journal} {Physical Review~A}\ }\textbf {\bibinfo {volume} {50}},\ \bibinfo
  {pages} {5242} (\bibinfo {year} {1994})}\BibitemShut {NoStop}%
\bibitem [{\citenamefont {Barchielli}\ \emph {et~al.}(1982)\citenamefont
  {Barchielli}, \citenamefont {Lanz},\ and\ \citenamefont
  {Prosperi}}]{barchielli1982model}%
  \BibitemOpen
  \bibfield  {author} {\bibinfo {author} {\bibfnamefont {A.}~\bibnamefont
  {Barchielli}}, \bibinfo {author} {\bibfnamefont {L.}~\bibnamefont {Lanz}},\
  and\ \bibinfo {author} {\bibfnamefont {G.~M.}\ \bibnamefont {Prosperi}},\
  }\bibfield  {title} {\bibinfo {title} {A model for the macroscopic
  description and continual observations in quantum mechanics},\ }\href@noop {}
  {\bibfield  {journal} {\bibinfo  {journal} {Il Nuovo Cimento}\ }\textbf
  {\bibinfo {volume} {72 B}} (\bibinfo {year} {1982})}\BibitemShut {NoStop}%
\bibitem [{\citenamefont {Barchielli}(2006)}]{barchielli2006continual}%
  \BibitemOpen
  \bibfield  {author} {\bibinfo {author} {\bibfnamefont {A.}~\bibnamefont
  {Barchielli}},\ }\bibfield  {title} {\bibinfo {title} {Continual measurements
  in quantum mechanics and quantum stochastic calculus},\ }in\ \href@noop {}
  {\emph {\bibinfo {booktitle} {Open quantum systems III: recent
  developments}}}\ (\bibinfo  {publisher} {Springer},\ \bibinfo {year} {2006})\
  pp.\ \bibinfo {pages} {207--292}\BibitemShut {NoStop}%
\bibitem [{\citenamefont {Barchielli}\ and\ \citenamefont
  {Gregoratti}(2009)}]{barchielli2009quantum}%
  \BibitemOpen
  \bibfield  {author} {\bibinfo {author} {\bibfnamefont {A.}~\bibnamefont
  {Barchielli}}\ and\ \bibinfo {author} {\bibfnamefont {M.}~\bibnamefont
  {Gregoratti}},\ }\href@noop {} {\emph {\bibinfo {title} {Quantum Trajectories
  and Measurements in Continuous Time: The Diffusive Case}}},\ Lecture Notes in
  Physics\ (\bibinfo  {publisher} {Springer-Verlag},\ \bibinfo {year}
  {2009})\BibitemShut {NoStop}%
\bibitem [{\citenamefont {Warszawski}\ \emph {et~al.}(2020)\citenamefont
  {Warszawski}, \citenamefont {Wiseman},\ and\ \citenamefont
  {Doherty}}]{Warszawski2020a}%
  \BibitemOpen
  \bibfield  {author} {\bibinfo {author} {\bibfnamefont {P.}~\bibnamefont
  {Warszawski}}, \bibinfo {author} {\bibfnamefont {H.~M.}\ \bibnamefont
  {Wiseman}},\ and\ \bibinfo {author} {\bibfnamefont {A.~C.}\ \bibnamefont
  {Doherty}},\ }\bibfield  {title} {\bibinfo {title} {Solving quantum
  trajectories for systems with linear {Heisenberg}-picture dynamics and
  {Gaussian} measurement noise},\ }\href@noop {} {\bibfield  {journal}
  {\bibinfo  {journal} {Physical Review~A}\ }\textbf {\bibinfo {volume}
  {102}},\ \bibinfo {pages} {042210} (\bibinfo {year} {2020})}\BibitemShut
  {NoStop}%
\bibitem [{\citenamefont {Kraus}(1983)}]{kraus1983states}%
  \BibitemOpen
  \bibfield  {author} {\bibinfo {author} {\bibfnamefont {K.}~\bibnamefont
  {Kraus}},\ }\href@noop {} {\emph {\bibinfo {title} {States, Effects, and
  Operations: Fundamental Notions of Quantum Theory}}},\ \bibinfo {series}
  {Lecture Notes in Physics}, Vol.\ \bibinfo {volume} {190, edited by
  A.~B{\"o}hm, J.~D. Dollard, and W.~H. Wootters}\ (\bibinfo  {publisher}
  {Springer},\ \bibinfo {year} {1983})\BibitemShut {NoStop}%
\bibitem [{\citenamefont {Von~Neumann}(1932)}]{vonneumann1932mathematical}%
  \BibitemOpen
  \bibfield  {author} {\bibinfo {author} {\bibfnamefont {J.}~\bibnamefont
  {Von~Neumann}},\ }\href@noop {} {\emph {\bibinfo {title} {Mathematical
  foundations of quantum mechanics}}}\ (\bibinfo  {publisher} {Princeton
  university press},\ \bibinfo {year} {1932})\BibitemShut {NoStop}%
\bibitem [{\citenamefont {Busch}\ and\ \citenamefont
  {Lahti}(2009)}]{busch2009luders}%
  \BibitemOpen
  \bibfield  {author} {\bibinfo {author} {\bibfnamefont {P.}~\bibnamefont
  {Busch}}\ and\ \bibinfo {author} {\bibfnamefont {P.}~\bibnamefont {Lahti}},\
  }\href@noop {} {\emph {\bibinfo {title} {Compendium of Quantum Physics}}},\
  edited by\ \bibinfo {editor} {\bibfnamefont {D.}~\bibnamefont {Greenberger}},
  \bibinfo {editor} {\bibfnamefont {K.}~\bibnamefont {Hentshel}},\ and\
  \bibinfo {editor} {\bibfnamefont {F.}~\bibnamefont {Weinert}}\ (\bibinfo
  {publisher} {Springer},\ \bibinfo {year} {2009})\ pp.\ \bibinfo {pages} {356
  -- 358}\BibitemShut {NoStop}%
\bibitem [{\citenamefont {DeBrota}\ and\ \citenamefont
  {Stacey}(2019)}]{debrota2019luders}%
  \BibitemOpen
  \bibfield  {author} {\bibinfo {author} {\bibfnamefont {J.~B.}\ \bibnamefont
  {DeBrota}}\ and\ \bibinfo {author} {\bibfnamefont {B.~C.}\ \bibnamefont
  {Stacey}},\ }\bibfield  {title} {\bibinfo {title} {Luders channels and the
  existence of symmetric-informationally-complete measurements},\ }\href@noop
  {} {\bibfield  {journal} {\bibinfo  {journal} {Physical Review~A}\ }\textbf
  {\bibinfo {volume} {100}},\ \bibinfo {pages} {062327} (\bibinfo {year}
  {2019})}\BibitemShut {NoStop}%
\bibitem [{\citenamefont {Shapiro}\ and\ \citenamefont
  {Wagner}(1984)}]{shapiro1984phase}%
  \BibitemOpen
  \bibfield  {author} {\bibinfo {author} {\bibfnamefont {J.~H.}\ \bibnamefont
  {Shapiro}}\ and\ \bibinfo {author} {\bibfnamefont {S.~S.}\ \bibnamefont
  {Wagner}},\ }\bibfield  {title} {\bibinfo {title} {Phase and amplitude
  uncertainties in heterodyne detection},\ }\href@noop {} {\bibfield  {journal}
  {\bibinfo  {journal} {IEEE Journal of Quantum Electronics}\ }\textbf
  {\bibinfo {volume} {QE-20}},\ \bibinfo {pages} {803 } (\bibinfo {year}
  {1984})}\BibitemShut {NoStop}%
\bibitem [{\citenamefont {Yuen}\ and\ \citenamefont
  {Lax}(1973)}]{yuen1973multiple}%
  \BibitemOpen
  \bibfield  {author} {\bibinfo {author} {\bibfnamefont {H.~P.}\ \bibnamefont
  {Yuen}}\ and\ \bibinfo {author} {\bibfnamefont {M.}~\bibnamefont {Lax}},\
  }\bibfield  {title} {\bibinfo {title} {Multiple-parameter quantum estimation
  and measurement of nonselfadjoint observables},\ }\href@noop {} {\bibfield
  {journal} {\bibinfo  {journal} {IEEE Transactions on Information Theory}\
  }\textbf {\bibinfo {volume} {IT-19}},\ \bibinfo {pages} {740} (\bibinfo
  {year} {1973})}\BibitemShut {NoStop}%
\bibitem [{\citenamefont {Gordon}\ and\ \citenamefont
  {Louisell}(1966)}]{gordon1966simultaneous}%
  \BibitemOpen
  \bibfield  {author} {\bibinfo {author} {\bibfnamefont {J.~P.}\ \bibnamefont
  {Gordon}}\ and\ \bibinfo {author} {\bibfnamefont {W.~H.}\ \bibnamefont
  {Louisell}},\ }\bibfield  {title} {\bibinfo {title} {Simultaneous
  measurements of noncommuting observables},\ }\href@noop {} {\ \bibinfo
  {series} {Physics of Quantum Electronics Conference Proceedings edited by
  {Kelley}, P.~L., {Lax}, B. and {Tannenwald}, P.~E.},\ \bibinfo {pages} {833 }
  (\bibinfo {year} {1966})}\BibitemShut {NoStop}%
\bibitem [{\citenamefont {Srinivas}\ and\ \citenamefont
  {Davies}(1981)}]{srinivas1981photon}%
  \BibitemOpen
  \bibfield  {author} {\bibinfo {author} {\bibfnamefont {M.}~\bibnamefont
  {Srinivas}}\ and\ \bibinfo {author} {\bibfnamefont {E.}~\bibnamefont
  {Davies}},\ }\bibfield  {title} {\bibinfo {title} {Photon counting
  probabilities in quantum optics},\ }\href@noop {} {\bibfield  {journal}
  {\bibinfo  {journal} {Optica Acta: International Journal of Optics}\ }\textbf
  {\bibinfo {volume} {28}},\ \bibinfo {pages} {981} (\bibinfo {year}
  {1981})}\BibitemShut {NoStop}%
\bibitem [{\citenamefont {Wiseman}(1994)}]{wiseman1994quantum}%
  \BibitemOpen
  \bibfield  {author} {\bibinfo {author} {\bibfnamefont {H.~M.}\ \bibnamefont
  {Wiseman}},\ }\emph {\bibinfo {title} {Quantum trajectories and feedback}},\
  \href@noop {} {Ph.D. thesis},\ \bibinfo  {school} {University of Queensland}
  (\bibinfo {year} {1994})\BibitemShut {NoStop}%
\bibitem [{\citenamefont {Silberfarb}\ \emph {et~al.}(2005)\citenamefont
  {Silberfarb}, \citenamefont {Jessen},\ and\ \citenamefont
  {Deutsch}}]{Silberfarb2005a}%
  \BibitemOpen
  \bibfield  {author} {\bibinfo {author} {\bibfnamefont {A.}~\bibnamefont
  {Silberfarb}}, \bibinfo {author} {\bibfnamefont {P.~S.}\ \bibnamefont
  {Jessen}},\ and\ \bibinfo {author} {\bibfnamefont {I.~H.}\ \bibnamefont
  {Deutsch}},\ }\bibfield  {title} {\bibinfo {title} {Quantum state
  reconstruction via continuous measurement},\ }\href@noop {} {\bibfield
  {journal} {\bibinfo  {journal} {Physical Review Letters}\ }\textbf {\bibinfo
  {volume} {95}},\ \bibinfo {pages} {030402} (\bibinfo {year}
  {2005})}\BibitemShut {NoStop}%
\bibitem [{\citenamefont {Jacobs}\ and\ \citenamefont
  {Steck}(2006)}]{jacobs2006straightforward}%
  \BibitemOpen
  \bibfield  {author} {\bibinfo {author} {\bibfnamefont {K.}~\bibnamefont
  {Jacobs}}\ and\ \bibinfo {author} {\bibfnamefont {D.~A.}\ \bibnamefont
  {Steck}},\ }\bibfield  {title} {\bibinfo {title} {A straightforward
  introduction to continuous quantum measurement},\ }\href@noop {} {\bibfield
  {journal} {\bibinfo  {journal} {Contemporary Physics}\ }\textbf {\bibinfo
  {volume} {47}},\ \bibinfo {pages} {279} (\bibinfo {year} {2006})}\BibitemShut
  {NoStop}%
\bibitem [{\citenamefont {Wiseman}\ and\ \citenamefont
  {Milburn}(2009)}]{wiseman2009quantum}%
  \BibitemOpen
  \bibfield  {author} {\bibinfo {author} {\bibfnamefont {H.~M.}\ \bibnamefont
  {Wiseman}}\ and\ \bibinfo {author} {\bibfnamefont {G.~J.}\ \bibnamefont
  {Milburn}},\ }\href@noop {} {\emph {\bibinfo {title} {Quantum Measurement and
  Control}}}\ (\bibinfo  {publisher} {Cambridge University Press},\ \bibinfo
  {year} {2009})\BibitemShut {NoStop}%
\bibitem [{\citenamefont {Carmichael}(2009)}]{carmichael2009open}%
  \BibitemOpen
  \bibfield  {author} {\bibinfo {author} {\bibfnamefont {H.}~\bibnamefont
  {Carmichael}},\ }\href@noop {} {\emph {\bibinfo {title} {An open systems
  approach to quantum optics: lectures presented at the Universit{\'e} Libre de
  Bruxelles, October 28 to November 4, 1991}}},\ Vol.~\bibinfo {volume} {18}\
  (\bibinfo  {publisher} {Springer Science \& Business Media},\ \bibinfo {year}
  {2009})\BibitemShut {NoStop}%
\bibitem [{\citenamefont {Cook}\ \emph {et~al.}(2014)\citenamefont {Cook},
  \citenamefont {Riofrio},\ and\ \citenamefont {Deutsch}}]{cook2014single}%
  \BibitemOpen
  \bibfield  {author} {\bibinfo {author} {\bibfnamefont {R.~L.}\ \bibnamefont
  {Cook}}, \bibinfo {author} {\bibfnamefont {C.~A.}\ \bibnamefont {Riofrio}},\
  and\ \bibinfo {author} {\bibfnamefont {I.~H.}\ \bibnamefont {Deutsch}},\
  }\bibfield  {title} {\bibinfo {title} {Single-shot quantum state estimation
  via a continuous measurement in the strong backaction regime},\ }\href@noop
  {} {\bibfield  {journal} {\bibinfo  {journal} {Physical Review~A}\ }\textbf
  {\bibinfo {volume} {90}},\ \bibinfo {pages} {032113} (\bibinfo {year}
  {2014})}\BibitemShut {NoStop}%
\bibitem [{\citenamefont {Gross}\ \emph {et~al.}(2018)\citenamefont {Gross},
  \citenamefont {Caves}, \citenamefont {Milburn},\ and\ \citenamefont
  {Combes}}]{gross2018qubit}%
  \BibitemOpen
  \bibfield  {author} {\bibinfo {author} {\bibfnamefont {J.~A.}\ \bibnamefont
  {Gross}}, \bibinfo {author} {\bibfnamefont {C.~M.}\ \bibnamefont {Caves}},
  \bibinfo {author} {\bibfnamefont {G.~J.}\ \bibnamefont {Milburn}},\ and\
  \bibinfo {author} {\bibfnamefont {J.}~\bibnamefont {Combes}},\ }\bibfield
  {title} {\bibinfo {title} {Qubit models of weak continuous measurements:
  Markovian conditional and open-system dynamics},\ }\href@noop {} {\bibfield
  {journal} {\bibinfo  {journal} {Quantum Science and Technology}\ }\textbf
  {\bibinfo {volume} {3}},\ \bibinfo {pages} {024005} (\bibinfo {year}
  {2018})}\BibitemShut {NoStop}%
\bibitem [{\citenamefont {Dixmier}(1977)}]{dixmier1977c}%
  \BibitemOpen
  \bibfield  {author} {\bibinfo {author} {\bibfnamefont {J.}~\bibnamefont
  {Dixmier}},\ }\href@noop {} {\emph {\bibinfo {title} {C*-Algebras, translated
  by J. Francis}}}\ (\bibinfo  {publisher} {North-Holland Publishing Co.},\
  \bibinfo {year} {1977})\BibitemShut {NoStop}%
\bibitem [{\citenamefont {Rotman}(1995)}]{rotman1995introduction}%
  \BibitemOpen
  \bibfield  {author} {\bibinfo {author} {\bibfnamefont {J.~J.}\ \bibnamefont
  {Rotman}},\ }\href@noop {} {\emph {\bibinfo {title} {An introduction to the
  theory of groups}}},\ \bibinfo {series} {Graduate texts in mathematics}\ No.\
  \bibinfo {number} {148}\ (\bibinfo  {publisher} {Springer-Verlag},\ \bibinfo
  {year} {1995})\BibitemShut {NoStop}%
\bibitem [{\citenamefont {Epstein}(1992)}]{epstein1992word}%
  \BibitemOpen
  \bibfield  {author} {\bibinfo {author} {\bibfnamefont {D.~B.~A.}\
  \bibnamefont {Epstein}},\ }\href@noop {} {\emph {\bibinfo {title} {Word
  Processing in Groups}}}\ (\bibinfo  {publisher} {CRC Press},\ \bibinfo {year}
  {1992})\BibitemShut {NoStop}%
\bibitem [{\citenamefont {Gromov}(1987)}]{gromov1987hyperbolic}%
  \BibitemOpen
  \bibfield  {author} {\bibinfo {author} {\bibfnamefont {M.}~\bibnamefont
  {Gromov}},\ }\bibfield  {title} {\bibinfo {title} {Hyperbolic groups},\
  }\href@noop {} {\ \bibinfo {series} {S.~M. Gersten, editor, Essays in group
  theory, MSRI Publ.},\ \textbf {\bibinfo {volume} {8}},\ \bibinfo {pages} {75}
  (\bibinfo {year} {1987})}\BibitemShut {NoStop}%
\bibitem [{\citenamefont {Cannon}(1984)}]{cannon1984combinatorial}%
  \BibitemOpen
  \bibfield  {author} {\bibinfo {author} {\bibfnamefont {J.}~\bibnamefont
  {Cannon}},\ }\bibfield  {title} {\bibinfo {title} {The combinatorial
  structure of cocompact discrete hyperbolic groups},\ }\href@noop {}
  {\bibfield  {journal} {\bibinfo  {journal} {Geom. Dedicata}\ }\textbf
  {\bibinfo {volume} {16}},\ \bibinfo {pages} {123} (\bibinfo {year}
  {1984})}\BibitemShut {NoStop}%
\bibitem [{\citenamefont {Wiener}()}]{wiener1948cybernetics}%
  \BibitemOpen
  \bibfield  {author} {\bibinfo {author} {\bibfnamefont {N.}~\bibnamefont
  {Wiener}},\ }\href@noop {} {\emph {\bibinfo {title} {Cybernetics or, Control
  and Communication in the Animal and the Machine}}}\BibitemShut {NoStop}%
\bibitem [{\citenamefont {Bellman}(1957)}]{bellman1957dynamic}%
  \BibitemOpen
  \bibfield  {author} {\bibinfo {author} {\bibfnamefont {R.}~\bibnamefont
  {Bellman}},\ }\href@noop {} {\emph {\bibinfo {title} {Dynamic Programming}}}\
  (\bibinfo  {publisher} {Princeton University Press},\ \bibinfo {year}
  {1957})\BibitemShut {NoStop}%
\bibitem [{\citenamefont {Sutton}\ and\ \citenamefont
  {Barto}(1998)}]{sutton1998reinforcement}%
  \BibitemOpen
  \bibfield  {author} {\bibinfo {author} {\bibfnamefont {R.~S.}\ \bibnamefont
  {Sutton}}\ and\ \bibinfo {author} {\bibfnamefont {A.~G.}\ \bibnamefont
  {Barto}},\ }\href@noop {} {\emph {\bibinfo {title} {Reinforcement
  Learning}}},\ A Bradford Book\ (\bibinfo  {publisher} {The MIT Press},\
  \bibinfo {year} {1998})\BibitemShut {NoStop}%
\bibitem [{\citenamefont {Bennett}\ \emph {et~al.}(1989)\citenamefont
  {Bennett}, \citenamefont {Hoffman},\ and\ \citenamefont
  {Prakash}}]{bennett1989observer}%
  \BibitemOpen
  \bibfield  {author} {\bibinfo {author} {\bibfnamefont {B.~M.}\ \bibnamefont
  {Bennett}}, \bibinfo {author} {\bibfnamefont {D.~D.}\ \bibnamefont
  {Hoffman}},\ and\ \bibinfo {author} {\bibfnamefont {C.}~\bibnamefont
  {Prakash}},\ }\href@noop {} {\emph {\bibinfo {title} {Observer Mechanics: A
  Formal Theory of Perception}}}\ (\bibinfo  {publisher} {Academic Press},\
  \bibinfo {year} {1989})\BibitemShut {NoStop}%
\bibitem [{\citenamefont {Dupont}\ \emph {et~al.}(2005)\citenamefont {Dupont},
  \citenamefont {Denis},\ and\ \citenamefont {Esposito}}]{dupont2005links}%
  \BibitemOpen
  \bibfield  {author} {\bibinfo {author} {\bibfnamefont {P.}~\bibnamefont
  {Dupont}}, \bibinfo {author} {\bibfnamefont {F.}~\bibnamefont {Denis}},\ and\
  \bibinfo {author} {\bibfnamefont {Y.}~\bibnamefont {Esposito}},\ }\bibfield
  {title} {\bibinfo {title} {Links between probabilistic automata and hidden
  markov models: probability distributions, learning models and induction
  algorithms},\ }\href@noop {} {\bibfield  {journal} {\bibinfo  {journal}
  {Pattern Recognition}\ }\textbf {\bibinfo {volume} {38}},\ \bibinfo {pages}
  {1349 } (\bibinfo {year} {2005})}\BibitemShut {NoStop}%
\bibitem [{\citenamefont {Bishop}(2006)}]{bishop2006pattern}%
  \BibitemOpen
  \bibfield  {author} {\bibinfo {author} {\bibfnamefont {C.~M.}\ \bibnamefont
  {Bishop}},\ }\href@noop {} {\emph {\bibinfo {title} {Pattern Recognition and
  Machine Learning}}}\ (\bibinfo  {publisher} {Springer},\ \bibinfo {year}
  {2006})\BibitemShut {NoStop}%
\bibitem [{\citenamefont {Karmakar}\ \emph {et~al.}(2022)\citenamefont
  {Karmakar}, \citenamefont {Lewalle},\ and\ \citenamefont
  {Jordan}}]{karmakar2022stochastic}%
  \BibitemOpen
  \bibfield  {author} {\bibinfo {author} {\bibfnamefont {T.}~\bibnamefont
  {Karmakar}}, \bibinfo {author} {\bibfnamefont {P.}~\bibnamefont {Lewalle}},\
  and\ \bibinfo {author} {\bibfnamefont {A.~N.}\ \bibnamefont {Jordan}},\
  }\bibfield  {title} {\bibinfo {title} {Stochastic path-integral analysis of
  the continuously monitored quantum harmonic oscillator},\ }\href@noop {}
  {\bibfield  {journal} {\bibinfo  {journal} {PRX quantum}\ }\textbf {\bibinfo
  {volume} {3}},\ \bibinfo {pages} {010327} (\bibinfo {year}
  {2022})}\BibitemShut {NoStop}%
\bibitem [{\citenamefont {Wiseman}(1995)}]{wiseman1995su2}%
  \BibitemOpen
  \bibfield  {author} {\bibinfo {author} {\bibfnamefont {H.~W.}\ \bibnamefont
  {Wiseman}},\ }\bibfield  {title} {\bibinfo {title} {Su(2) distribution
  functions and measurement of the fluorescence of a two-level atom},\
  }\href@noop {} {\bibfield  {journal} {\bibinfo  {journal} {Quantum and
  Semiclassical Optics}\ }\textbf {\bibinfo {volume} {7}},\ \bibinfo {pages}
  {569} (\bibinfo {year} {1995})}\BibitemShut {NoStop}%
\bibitem [{\citenamefont {von Neumann}(1999)}]{vonneumann1999invariant}%
  \BibitemOpen
  \bibfield  {author} {\bibinfo {author} {\bibfnamefont {J.}~\bibnamefont {von
  Neumann}},\ }\href@noop {} {\emph {\bibinfo {title} {Invariant Measures}}}\
  (\bibinfo  {publisher} {American Mathematical Society},\ \bibinfo {year}
  {1999})\ \bibinfo {note} {lectured at the IAS in 1940-1941}\BibitemShut
  {NoStop}%
\bibitem [{\citenamefont {Barut}\ and\ \citenamefont {Ra{\c
  c}szka}(1986)}]{barut1986theory}%
  \BibitemOpen
  \bibfield  {author} {\bibinfo {author} {\bibfnamefont {A.~O.}\ \bibnamefont
  {Barut}}\ and\ \bibinfo {author} {\bibfnamefont {R.}~\bibnamefont {Ra{\c
  c}szka}},\ }\href@noop {} {\emph {\bibinfo {title} {Theory of Group
  Representations and Applications}}}\ (\bibinfo  {publisher} {World
  Scientific},\ \bibinfo {year} {1986})\BibitemShut {NoStop}%
\bibitem [{\citenamefont {Dieudonne}(1979)}]{dieudonne1979special}%
  \BibitemOpen
  \bibfield  {author} {\bibinfo {author} {\bibfnamefont {J.}~\bibnamefont
  {Dieudonne}},\ }\href@noop {} {\emph {\bibinfo {title} {Special Functions and
  Linear Represenations of Lie Groups}}},\ \bibinfo {series} {regional
  conference series in mathematics}\ No.~\bibinfo {number} {42}\ (\bibinfo
  {publisher} {conference board of the mathematical sciences},\ \bibinfo {year}
  {1979})\BibitemShut {NoStop}%
\bibitem [{\citenamefont {Nachbin}(1965)}]{nachbin1965haar}%
  \BibitemOpen
  \bibfield  {author} {\bibinfo {author} {\bibfnamefont {L.}~\bibnamefont
  {Nachbin}},\ }\href@noop {} {\emph {\bibinfo {title} {The Haar Integral}}},\
  The University Series in Higher Mathematics\ (\bibinfo  {publisher} {D. Van
  Nostrand Co., Inc.},\ \bibinfo {year} {1965})\BibitemShut {NoStop}%
\bibitem [{\citenamefont {Hewitt}\ and\ \citenamefont
  {Ross}(1963)}]{hewitt1963abstract}%
  \BibitemOpen
  \bibfield  {author} {\bibinfo {author} {\bibfnamefont {E.}~\bibnamefont
  {Hewitt}}\ and\ \bibinfo {author} {\bibfnamefont {K.~A.}\ \bibnamefont
  {Ross}},\ }\href@noop {} {\emph {\bibinfo {title} {Abstract harmonic
  analysis, Volume I: Structure of Topological Groups, Integration Theory,
  Group Representations}}},\ \bibinfo {series} {Die Grundlehren Der
  Mathematischen Wissenschaften In Einzeldarstellungen}, Vol.\ \bibinfo
  {volume} {115}\ (\bibinfo  {publisher} {Springer-Verlag},\ \bibinfo {year}
  {1963})\BibitemShut {NoStop}%
\bibitem [{\citenamefont {Rudin}(1962)}]{rudin1962fourier}%
  \BibitemOpen
  \bibfield  {author} {\bibinfo {author} {\bibfnamefont {W.}~\bibnamefont
  {Rudin}},\ }\href@noop {} {\emph {\bibinfo {title} {Fourier Analysis on
  Groups}}},\ Interscience Tracts in Pure and Applied Mathematics\ (\bibinfo
  {publisher} {Interscience Publishers},\ \bibinfo {year} {1962})\BibitemShut
  {NoStop}%
\bibitem [{\citenamefont {Montgomery}\ and\ \citenamefont
  {Zippin}(1955)}]{montgomery1955topological}%
  \BibitemOpen
  \bibfield  {author} {\bibinfo {author} {\bibfnamefont {D.}~\bibnamefont
  {Montgomery}}\ and\ \bibinfo {author} {\bibfnamefont {L.}~\bibnamefont
  {Zippin}},\ }\href@noop {} {\emph {\bibinfo {title} {Topological
  Transformation Groups}}}\ (\bibinfo  {publisher} {Interscience Publishers,
  Inc.},\ \bibinfo {year} {1955})\BibitemShut {NoStop}%
\bibitem [{\citenamefont {Pontrjagin}()}]{pontryagin1946topological}%
  \BibitemOpen
  \bibfield  {author} {\bibinfo {author} {\bibfnamefont {L.}~\bibnamefont
  {Pontrjagin}},\ }\href@noop {} {\emph {\bibinfo {title} {Topological
  groups}}},\ Princeton Mathematical Series\ (\bibinfo  {publisher} {Princeton
  University Press})\ \bibinfo {note} {translated by Lehmer, E.}\BibitemShut
  {Stop}%
\bibitem [{\citenamefont {Berezin}(1967)}]{berezin1967remarks}%
  \BibitemOpen
  \bibfield  {author} {\bibinfo {author} {\bibfnamefont {F.~A.}\ \bibnamefont
  {Berezin}},\ }\bibfield  {title} {\bibinfo {title} {Some remarks about the
  associated envelope of a lie algebra},\ }\href@noop {} {\bibfield  {journal}
  {\bibinfo  {journal} {Funct Anal Its Appl}\ }\textbf {\bibinfo {volume}
  {1}},\ \bibinfo {pages} {91 } (\bibinfo {year} {1967})}\BibitemShut {NoStop}%
\bibitem [{\citenamefont {Kitaev}(2017)}]{kitaev2017notes}%
  \BibitemOpen
  \bibfield  {author} {\bibinfo {author} {\bibfnamefont {A.}~\bibnamefont
  {Kitaev}},\ }\bibfield  {title} {\bibinfo {title} {Notes on sl(2,r)
  representations},\ }\href@noop {} {\bibfield  {journal} {\bibinfo  {journal}
  {arXiv}\ } (\bibinfo {year} {2017})}\BibitemShut {NoStop}%
\bibitem [{\citenamefont {Helgason}(2008)}]{helgason2008GASS}%
  \BibitemOpen
  \bibfield  {author} {\bibinfo {author} {\bibfnamefont {S.}~\bibnamefont
  {Helgason}},\ }\href@noop {} {\emph {\bibinfo {title} {Geometric Analysis on
  Symmetric Spaces}}},\ \bibinfo {edition} {2nd}\ ed.,\ \bibinfo {series}
  {Mathematical Survey and Monographs}, Vol.~\bibinfo {volume} {39}\ (\bibinfo
  {publisher} {American Mathematical Society},\ \bibinfo {year}
  {2008})\BibitemShut {NoStop}%
\bibitem [{\citenamefont {Harish-Chandra}(1966)}]{harish1966discrete}%
  \BibitemOpen
  \bibfield  {author} {\bibinfo {author} {\bibnamefont {Harish-Chandra}},\
  }\bibfield  {title} {\bibinfo {title} {Discrete series for semisimple lie
  groups. ii. explicit determination of the characters},\ }\href@noop {}
  {\bibfield  {journal} {\bibinfo  {journal} {Acta Mathematica}\ }\textbf
  {\bibinfo {volume} {116}},\ \bibinfo {pages} {1 } (\bibinfo {year}
  {1966})}\BibitemShut {NoStop}%
\bibitem [{\citenamefont {Knapp}(1986)}]{knapp1986representation}%
  \BibitemOpen
  \bibfield  {author} {\bibinfo {author} {\bibfnamefont {A.~W.}\ \bibnamefont
  {Knapp}},\ }\href@noop {} {\emph {\bibinfo {title} {Representation Theory of
  Semisimple Groups: An Overview Based on Examples}}},\ \bibinfo {series}
  {Princeton Mathematical Series}\ No.~\bibinfo {number} {36}\ (\bibinfo {year}
  {1986})\BibitemShut {NoStop}%
\bibitem [{\citenamefont {Helgason}(2000)}]{helgason2001differential}%
  \BibitemOpen
  \bibfield  {author} {\bibinfo {author} {\bibfnamefont {S.}~\bibnamefont
  {Helgason}},\ }\href@noop {} {\emph {\bibinfo {title} {Differential Geometry
  and Symmetric Spaces}}}\ (\bibinfo  {publisher} {American Mathematical
  Society},\ \bibinfo {year} {2000})\BibitemShut {NoStop}%
\bibitem [{\citenamefont {Feller}(1949)}]{feller1949theory}%
  \BibitemOpen
  \bibfield  {author} {\bibinfo {author} {\bibfnamefont {W.}~\bibnamefont
  {Feller}},\ }\bibfield  {title} {\bibinfo {title} {On the theory of
  stochastic processes, with particular reference to applications},\ }in\
  \href@noop {} {\emph {\bibinfo {booktitle} {Proceedings of the [First]
  Berkeley Symposium on Mathematical Statistics and Probability}}},\ \bibinfo
  {series and number} {Berkeley Symposium on Mathematical Statistics and
  Probability}\ (\bibinfo  {publisher} {University of California Press},\
  \bibinfo {year} {1949})\ pp.\ \bibinfo {pages} {403--432}\BibitemShut
  {NoStop}%
\bibitem [{\citenamefont {Gardiner}(1986)}]{gardiner1986handbook}%
  \BibitemOpen
  \bibfield  {author} {\bibinfo {author} {\bibfnamefont {C.~W.}\ \bibnamefont
  {Gardiner}},\ }\bibfield  {title} {\bibinfo {title} {Handbook of stochastic
  methods for physics, chemistry and the natural sciences},\ }\href@noop {}
  {\bibfield  {journal} {\bibinfo  {journal} {Applied Optics}\ }\textbf
  {\bibinfo {volume} {25}},\ \bibinfo {pages} {3145} (\bibinfo {year}
  {1986})}\BibitemShut {NoStop}%
\bibitem [{\citenamefont {Gardiner}(2019)}]{gardiner2019stochastic}%
  \BibitemOpen
  \bibfield  {author} {\bibinfo {author} {\bibfnamefont {C.~W.}\ \bibnamefont
  {Gardiner}},\ }\href@noop {} {\emph {\bibinfo {title} {Stochastic Methods: A
  Handbook for the Natural and Social Sciences}}},\ \bibinfo {edition}
  {e-book}\ ed.\ (\bibinfo  {publisher} {Springer},\ \bibinfo {address}
  {Berlin},\ \bibinfo {year} {2019})\BibitemShut {NoStop}%
\bibitem [{\citenamefont {Gardiner}(2021)}]{gardiner2021elements}%
  \BibitemOpen
  \bibfield  {author} {\bibinfo {author} {\bibfnamefont {C.}~\bibnamefont
  {Gardiner}},\ }\href@noop {} {\emph {\bibinfo {title} {Elements of Stochastic
  Methods}}}\ (\bibinfo  {publisher} {AIP Publishing},\ \bibinfo {address}
  {Melville, New York},\ \bibinfo {year} {2021})\BibitemShut {NoStop}%
\bibitem [{\citenamefont {Peres}(2006)}]{peres2006quantum}%
  \BibitemOpen
  \bibfield  {author} {\bibinfo {author} {\bibfnamefont {A.}~\bibnamefont
  {Peres}},\ }\href@noop {} {\emph {\bibinfo {title} {Quantum Theory: Concepts
  and Methods}}}\ (\bibinfo  {publisher} {Kluwer Academic},\ \bibinfo {year}
  {2006})\BibitemShut {NoStop}%
\bibitem [{\citenamefont {Bengtsson}\ and\ \citenamefont
  {{\.Z}yczkowski}(2006)}]{bengtsson2006geometry}%
  \BibitemOpen
  \bibfield  {author} {\bibinfo {author} {\bibfnamefont {I.}~\bibnamefont
  {Bengtsson}}\ and\ \bibinfo {author} {\bibfnamefont {K.}~\bibnamefont
  {{\.Z}yczkowski}},\ }\href@noop {} {\emph {\bibinfo {title} {Geometry of
  quantum states: an introduction to quantum entanglement}}}\ (\bibinfo
  {publisher} {Cambridge university press},\ \bibinfo {year}
  {2006})\BibitemShut {NoStop}%
\bibitem [{\citenamefont {Holevo}(2011)}]{holevo2011probabilistic}%
  \BibitemOpen
  \bibfield  {author} {\bibinfo {author} {\bibfnamefont {A.}~\bibnamefont
  {Holevo}},\ }\href@noop {} {\emph {\bibinfo {title} {Probabilistic and
  Statsitical Aspects of Quantum Theory}}},\ \bibinfo {edition} {2nd}\ ed.\
  (\bibinfo  {publisher} {Scuola Normale Superiore Pisa},\ \bibinfo {year}
  {2011})\BibitemShut {NoStop}%
\bibitem [{\citenamefont {Davies}\ and\ \citenamefont
  {Lewis}(1970)}]{davies1970operational}%
  \BibitemOpen
  \bibfield  {author} {\bibinfo {author} {\bibfnamefont {E.~B.}\ \bibnamefont
  {Davies}}\ and\ \bibinfo {author} {\bibfnamefont {J.~T.}\ \bibnamefont
  {Lewis}},\ }\bibfield  {title} {\bibinfo {title} {An operational approach to
  quantum probability},\ }\href@noop {} {\bibfield  {journal} {\bibinfo
  {journal} {Commun. math. Phys}\ }\textbf {\bibinfo {volume} {17}},\ \bibinfo
  {pages} {239 } (\bibinfo {year} {1970})}\BibitemShut {NoStop}%
\bibitem [{\citenamefont {Jauch}\ and\ \citenamefont
  {Piron}(1967)}]{jauch1967generalized}%
  \BibitemOpen
  \bibfield  {author} {\bibinfo {author} {\bibfnamefont {J.~M.}\ \bibnamefont
  {Jauch}}\ and\ \bibinfo {author} {\bibfnamefont {C.}~\bibnamefont {Piron}},\
  }\bibfield  {title} {\bibinfo {title} {Generalized localizability},\
  }\href@noop {} {\bibfield  {journal} {\bibinfo  {journal} {Helv. Phys. Acta}\
  }\textbf {\bibinfo {volume} {40}},\ \bibinfo {pages} {559 } (\bibinfo {year}
  {1967})}\BibitemShut {NoStop}%
\bibitem [{\citenamefont {Berberian}(1966)}]{berberian1966spectral}%
  \BibitemOpen
  \bibfield  {author} {\bibinfo {author} {\bibfnamefont {S.~K.}\ \bibnamefont
  {Berberian}},\ }\href@noop {} {\emph {\bibinfo {title} {Notes on Spectral
  Theory}}}\ (\bibinfo  {publisher} {D. Van Nostrand Co., Inc., Princeton},\
  \bibinfo {year} {1966})\BibitemShut {NoStop}%
\bibitem [{\citenamefont {Sz.-Nagy}(1960)}]{nagy1960appendix}%
  \BibitemOpen
  \bibfield  {author} {\bibinfo {author} {\bibfnamefont {B.}~\bibnamefont
  {Sz.-Nagy}},\ }\href@noop {} {\emph {\bibinfo {title} {Appendix to Functional
  Analysis: Extensions of Linear Transformations in Hilbert Space Which Extend
  Beyond This Space}}}\ (\bibinfo  {publisher} {Frederick Ungar Publishing Co.
  New York},\ \bibinfo {year} {1960})\BibitemShut {NoStop}%
\bibitem [{\citenamefont {Stinespring}(1955)}]{stinespring1955positive}%
  \BibitemOpen
  \bibfield  {author} {\bibinfo {author} {\bibfnamefont {W.~F.}\ \bibnamefont
  {Stinespring}},\ }\bibfield  {title} {\bibinfo {title} {Positive functions on
  c*-algebras},\ }\href@noop {} {\bibfield  {journal} {\bibinfo  {journal}
  {Proceedings of the American Mathematical Society}\ }\textbf {\bibinfo
  {volume} {6}},\ \bibinfo {pages} {211} (\bibinfo {year} {1955})}\BibitemShut
  {NoStop}%
\bibitem [{\citenamefont {Neumark}(1943)}]{neumark1943representation}%
  \BibitemOpen
  \bibfield  {author} {\bibinfo {author} {\bibfnamefont {M.~A.}\ \bibnamefont
  {Neumark}},\ }\bibfield  {title} {\bibinfo {title} {On a representation of
  additive operator set functions},\ }\href@noop {} {\bibfield  {journal}
  {\bibinfo  {journal} {Comptes Rendus (Doklady) Acad. Sci. URSS}\ }\textbf
  {\bibinfo {volume} {41}},\ \bibinfo {pages} {359} (\bibinfo {year}
  {1943})}\BibitemShut {NoStop}%
\bibitem [{\citenamefont {Luders}(1952)}]{luders1951zustandsanderung}%
  \BibitemOpen
  \bibfield  {author} {\bibinfo {author} {\bibfnamefont {G.}~\bibnamefont
  {Luders}},\ }\bibfield  {title} {\bibinfo {title} {Uber die zustandsanderung
  durch den messprozess},\ }\href@noop {} {\bibfield  {journal} {\bibinfo
  {journal} {Ann. Phys.}\ }\textbf {\bibinfo {volume} {8}},\ \bibinfo {pages}
  {322 } (\bibinfo {year} {1952})}\BibitemShut {NoStop}%
\bibitem [{\citenamefont {Kirkpatrick}(2006)}]{kirkpatrick2006translation}%
  \BibitemOpen
  \bibfield  {author} {\bibinfo {author} {\bibfnamefont {K.~A.}\ \bibnamefont
  {Kirkpatrick}},\ }\bibfield  {title} {\bibinfo {title} {Translation of g.
  luders' uber die zustandsanderung durch den messprozess},\ }\href@noop {}
  {\bibfield  {journal} {\bibinfo  {journal} {Ann. Phys.}\ }\textbf {\bibinfo
  {volume} {15}},\ \bibinfo {pages} {663} (\bibinfo {year} {2006})}\BibitemShut
  {NoStop}%
\bibitem [{\citenamefont {Nielsen}\ and\ \citenamefont
  {Chuang}(2000)}]{Nielsen2000a}%
  \BibitemOpen
  \bibfield  {author} {\bibinfo {author} {\bibfnamefont {M.~A.}\ \bibnamefont
  {Nielsen}}\ and\ \bibinfo {author} {\bibfnamefont {I.~L.}\ \bibnamefont
  {Chuang}},\ }\href@noop {} {\emph {\bibinfo {title} {Quantum Computation and
  Quantum Information}}}\ (\bibinfo  {publisher} {Cambridge University Press},\
  \bibinfo {year} {2000})\BibitemShut {NoStop}%
\bibitem [{\citenamefont {De~la Madrid}(2005)}]{delaMadrid2005rigged}%
  \BibitemOpen
  \bibfield  {author} {\bibinfo {author} {\bibfnamefont {R.}~\bibnamefont
  {De~la Madrid}},\ }\bibfield  {title} {\bibinfo {title} {The role of the
  rigged hilbert space in quantum mechanics},\ }\href@noop {} {\bibfield
  {journal} {\bibinfo  {journal} {European journal of physics}\ }\textbf
  {\bibinfo {volume} {26}},\ \bibinfo {pages} {287} (\bibinfo {year}
  {2005})}\BibitemShut {NoStop}%
\bibitem [{\citenamefont {Roberts}(1966)}]{roberts1966rigged}%
  \BibitemOpen
  \bibfield  {author} {\bibinfo {author} {\bibfnamefont {J.~E.}\ \bibnamefont
  {Roberts}},\ }\bibfield  {title} {\bibinfo {title} {Rigged hilbert spaces in
  quantum mechanics},\ }\href@noop {} {\bibfield  {journal} {\bibinfo
  {journal} {Commun. math. Phys.}\ }\textbf {\bibinfo {volume} {3}} (\bibinfo
  {year} {1966})}\BibitemShut {NoStop}%
\bibitem [{\citenamefont {Gelfand}\ and\ \citenamefont
  {Vilenkin}(1966)}]{gelfand1964generalized}%
  \BibitemOpen
  \bibfield  {author} {\bibinfo {author} {\bibfnamefont {I.~M.}\ \bibnamefont
  {Gelfand}}\ and\ \bibinfo {author} {\bibfnamefont {N.~Y.}\ \bibnamefont
  {Vilenkin}},\ }\href@noop {} {\emph {\bibinfo {title} {Generalized Functions,
  Vol. IV: Applications of Harmonic Analysis, translated by Feinstein, A.}}}\
  (\bibinfo  {publisher} {American Mathematical Society},\ \bibinfo {year}
  {1966})\BibitemShut {NoStop}%
\bibitem [{\citenamefont {Kitaev}\ \emph {et~al.}(2002)\citenamefont {Kitaev},
  \citenamefont {Shen}, \citenamefont {Vyalyi},\ and\ \citenamefont
  {Vyalyi}}]{kitaev2002classical}%
  \BibitemOpen
  \bibfield  {author} {\bibinfo {author} {\bibfnamefont {A.~Y.}\ \bibnamefont
  {Kitaev}}, \bibinfo {author} {\bibfnamefont {A.}~\bibnamefont {Shen}},
  \bibinfo {author} {\bibfnamefont {M.~N.}\ \bibnamefont {Vyalyi}},\ and\
  \bibinfo {author} {\bibfnamefont {M.~N.}\ \bibnamefont {Vyalyi}},\
  }\href@noop {} {\emph {\bibinfo {title} {Classical and quantum
  computation}}},\ \bibinfo {number} {47}\ (\bibinfo  {publisher} {American
  Mathematical Soc.},\ \bibinfo {year} {2002})\BibitemShut {NoStop}%
\bibitem [{\citenamefont {Nielsen}\ and\ \citenamefont
  {Chuang}(2002)}]{nielsen2002quantum}%
  \BibitemOpen
  \bibfield  {author} {\bibinfo {author} {\bibfnamefont {M.~A.}\ \bibnamefont
  {Nielsen}}\ and\ \bibinfo {author} {\bibfnamefont {I.}~\bibnamefont
  {Chuang}},\ }\href@noop {} {\bibinfo {title} {Quantum computation and quantum
  information}} (\bibinfo {year} {2002})\BibitemShut {NoStop}%
\bibitem [{\citenamefont {Menicucci}(2005)}]{menicucci2005superoperator}%
  \BibitemOpen
  \bibfield  {author} {\bibinfo {author} {\bibfnamefont {N.}~\bibnamefont
  {Menicucci}},\ }\bibfield  {title} {\bibinfo {title} {Superoperator
  representation of higher dimensional bloch space transformations},\
  }\href@noop {} {\bibfield  {journal} {\bibinfo  {journal} {PhD Advanced
  Project, Princeton University}\ } (\bibinfo {year} {2005})}\BibitemShut
  {NoStop}%
\bibitem [{\citenamefont {Lindblad}(1976)}]{lindblad1976generators}%
  \BibitemOpen
  \bibfield  {author} {\bibinfo {author} {\bibfnamefont {G.}~\bibnamefont
  {Lindblad}},\ }\bibfield  {title} {\bibinfo {title} {On the generators of
  quantum dynamical semigroups},\ }\href@noop {} {\bibfield  {journal}
  {\bibinfo  {journal} {Communications in Mathematical Physics}\ }\textbf
  {\bibinfo {volume} {48}},\ \bibinfo {pages} {119} (\bibinfo {year}
  {1976})}\BibitemShut {NoStop}%
\bibitem [{\citenamefont {Albert}(2017)}]{albert2017thesis}%
  \BibitemOpen
  \bibfield  {author} {\bibinfo {author} {\bibfnamefont {V.~V.}\ \bibnamefont
  {Albert}},\ }\emph {\bibinfo {title} {Lindbladians with multiple steady
  states: theory and applications}},\ \href@noop {} {Ph.D. thesis},\ \bibinfo
  {school} {Yale University} (\bibinfo {year} {2017})\BibitemShut {NoStop}%
\bibitem [{\citenamefont {Strandberg}\ \emph {et~al.}()\citenamefont
  {Strandberg}, \citenamefont {Eriksson}, \citenamefont {Royer}, \citenamefont
  {Kervinen},\ and\ \citenamefont {Gasparinetti}}]{strandberg2024digital}%
  \BibitemOpen
  \bibfield  {author} {\bibinfo {author} {\bibfnamefont {I.}~\bibnamefont
  {Strandberg}}, \bibinfo {author} {\bibfnamefont {A.~M.}\ \bibnamefont
  {Eriksson}}, \bibinfo {author} {\bibfnamefont {B.}~\bibnamefont {Royer}},
  \bibinfo {author} {\bibfnamefont {M.}~\bibnamefont {Kervinen}},\ and\
  \bibinfo {author} {\bibfnamefont {S.}~\bibnamefont {Gasparinetti}},\
  }\bibfield  {title} {\bibinfo {title} {Digital homodyne and heterodyne
  detection for stationary bosonic modes},\ }\href@noop {} {\ }\BibitemShut
  {NoStop}%
\end{thebibliography}%

\appendix

\section{Quantum Measurement Theory: An Overview of Standard Knowledge}\label{StandardTheory}

This section is a summary of all the basic tools that will be needed in this paper which should already be familiar to most quantum scientists | instruments (\ref{POVMs}), channels (\ref{OperationChannels}), and Lindbladians (\ref{Decoherence}).

Section \ref{POVMs} defines the generalized measurement, its Kraus operators, and the generalized Born rule.
The positive-operator-valued measure (POVM) is introduced and the Luders rule is briefly mentioned.
Certain operations are then considered, definining the quantum \emph{instrument} and its subsequent total quantum operation.
Then the all-important concept of an indirect measurement is discussed.
The section finishes by addressing the position measurement.

Section \ref{OperationChannels} then moves into the topic of decoherence \emph{channels} and their usual superoperator properties, complete positivity and trace-preservation.
The Hilbert-Schmidt adjoint is defined along with the Choi map and the subsequent superoperator adjoint dual to the Hilbert-Schmidt.
Some care is then taken to ensure it is understood that the instrument behind the channel represents an observer and is not to be mistaken for an environment. 

Section \ref{Decoherence} then introduces the most basic tools of continuous measurement.
Lindblad operators and the \emph{Lindbladian} superoperator are discussed.
The one-parameter semigroup property is then explained as well as the subsequent Lindblad master equation.
Although the stochastic Lindblad master equation should also be considered if we were interested in the evolution of the state during measurement, the subject is neglected because this paper is about instrument evolution.\\

\subsection{Kraus operators, POVMs, and Instruments}\label{POVMs}

A \emph{generalized measurement} may be defined as a set of operators $\{\Omega_x\}_x$ which together satisfy a \emph{completeness relation},
\begin{equation}\label{completeness}
	\sum_x \Omega_x ^\dag \Omega_x = 1.
\end{equation}
The elements of a generalized measurement are called \emph{Kraus operators} or \emph{measurement operators}.
The index `$x$' is called the \emph{register} and corresponds to a classical piece of data.
The Kraus operators are understood to act on states represented by density operators in the dual of a von Neumann algebra, usually the set of bounded operators $\mathcal{B}(\Hb)$ of a given Hilbert space $\Hb$.
Upon measuring a quantum state with density operator $\rho \in \mathcal{B}(\Hb)^*$, the event of registering the value `$x$' is understood to be sampled from a probability distribution given by the \emph{generalized Born rule},
\begin{equation}
	P(x|\rho) = \Tr(\rho\, \Omega_x^\dag \Omega_x)/\Tr(\rho1),
\end{equation}
on which occasion the state is understood to simultaneously update to $\Omega_x \rho \Omega_x^\dag$.
Usually the density operator is normalized and renormalized, but this is not necessary and it is often far more convenient to let the normalization wander as it may.
The set of pure states $\{\proj\psi : \psi \in \Hb\}$ is closed under the action of a generalized measurement, in which case one may instead consider the update rule to be $\ket\psi \mapsto \Omega_x\ket\psi$ on the event of registering $x$.\\

Given a generalized measurement $\{\Omega_x\}_x$, the \emph{positive operator-valued measure} or ``\emph{POVM}'' is the set of operators
\begin{equation}\label{POVM}
	\E = \Big\{\Omega_x ^\dag \Omega_x\Big\}_x.
\end{equation}
The POVM is also called a \emph{generalized observable}.
The elements of the POVM are sometimes called \emph{effects} because they are imagined to leave an impression on the  observer, but they are more often just called \emph{POVM elements}.
The POVM is all that is needed to calculate the Born rule probability and can be studied in its own right, without reference to a full generalized measurement \cite{peres2006quantum, bengtsson2006geometry, holevo2011probabilistic, kraus1983states, davies1970operational, jauch1967generalized, berberian1966spectral, nagy1960appendix, stinespring1955positive, neumark1943representation}, so long as no further, post-measurement observations are made on the system.
Of course, the POVM does not by itself have the information needed to update the state.
However, given a POVM there can exist a generalized measurement with Kraus operators equal to the positive square-roots of the POVM elements, called the \emph{(generalized) Luders measurement} and the assumption to use such Kraus operators is called the \emph{generalized Luders rule} \cite{luders1951zustandsanderung, kirkpatrick2006translation, busch2009luders, debrota2019luders}.
The POVM of a position measurement is an essential example where the Luders measurement does not exist, as will be explained later in this subsection.

Operators on the von Neumann algebra or its dual are called \emph{superoperators}.
Associated with generalized measurements, we have already encountered a very fundamental type of superoperator, the so-called \emph{Kraus-rank-one operations}
\begin{equation}
	\Odot[\Omega] \equiv \Omega \!\odot\! \Omega^\dag
\end{equation}
where $\Omega$ is a Kraus operator and the $\odot$ is a tensor product with the extra implication that the superoperator acts on operators $\rho$ according to
\begin{equation}
	A \!\odot\! B^\dag \,(\rho) \equiv A\rho B^\dag.
\end{equation}
The \emph{instrument} of a generalized measurement is the set of Kraus-rank-one operations,
\begin{equation}\label{instrument}
	\mathcal{I} =\Big \{\Odot[\Omega_x]\Big\}_x.
\end{equation}
More generally, one could consider instrument elements that have Kraus-rank higher than one, but we will have no need for this.
In any case, the \emph{instrument elements} also go by the names of conditional operations or selective operations and the sum
\begin{equation}\label{CPTP}
	\Z \equiv \sum_x \Odot[\Omega_x]
\end{equation}
has been given corresponding names such as the \emph{total operation}, the unconditional operation, or the nonselective operation \cite{kraus1983states}.
It is also common that the unconditional operation is called \emph{the} quantum operation \cite{Nielsen2000a},
so to avoid confusion, we will mostly reserve the term ``operation'' for the \emph{total operation} of equation \ref{CPTP} and refer to the elements of equation \ref{instrument} as the \emph{instrument elements}.
For the instruments considered in this paper, where all instrument elements have Kraus-rank one,  the terms ``Kraus operator'' and ``instrument element'' will be used interchangably.

In most instances, we will be interested in a continuum of instrument elements.
In such a case each infinitesimal instrument element can be factored into a finite superoperator $\Odot[K_x]$ times an infinitesimal scalar measure $d\mu(x)$,
\begin{equation}\label{OVM}
	\Odot[\Omega_x] = d\mu(x) \Odot[K_x],
\end{equation}
and we will also call the $\Odot[K_x]$ and $K_x$ instrument elements and Kraus oepators.
Such an infinitesimal scalar measure is called an \emph{instrument-element distribution} or \emph{Kraus-operator distribution} \cite{jackson2023positive}.
Often it is also useful to think of the infinitesimal Kraus operators as factoring into the corresponding finite Kraus operator times the square-root of the Kraus-operator distribution,
\begin{equation}
	\Omega_x = \sqrt{d\mu(x)}K_x.
\end{equation}
In either case, the POVM of equation \ref{POVM} would therefore be denoted instead as
\begin{equation}
	\E = \{d\mu(x) K_x^\dag K_x\}_x
\end{equation}
and the sum in the total operation (equation \ref{CPTP}) would therefore be denoted instead as
\begin{equation}
	\Z = \int d\mu(x) \Odot[K_x].
\end{equation}

Without further context, the choice of how the Kraus-operator distribution is factored out of the instrument element is arbitrary or ``ostensible'' \cite{Warszawski2020a,wiseman1996measurement}.
In the context of continuous and sequential measurements, however, there is a distinct choice to make as will be explained in section \ref{OD}.\\

Any instrument can be ``dilated'' to a so-called indirect measurement \cite{kraus1983states, stinespring1955positive, vonneumann1932mathematical} where the system of interest interacts with a meter, after which the meter terminates in an eigenvalue measurement of some Hermitian meter observable.
This means the Kraus operators can be expanded into the form
\begin{equation}\label{Stinespring}
	\Omega_x = \bra{x}\UU_\text{int}\ket{0},
\end{equation}
where the meter pointer is prepared in the pure state $\ket{0}$ which then couples to the system via the interaction unitary $\UU_\text{int}$, terminating in the registration of an eigenvalue $x$ with linear functional $\bra{x}$.
Often the eigenvalue measurement of the meter is considered to be a ``von Neumann measurement.''
However, we will more often be interested in when the register $x$ corresponds to a continuous position observable, something that does not quite fit into the von Neumann paradigm but is a fundamental tool in the Dirac approach.

To be clear, the basic von Neumann-Dirac tension is only that the eigenfunctions of position are Dirac delta-functions and such functions (more rigorously known today as distributions~\cite{delaMadrid2005rigged,roberts1966rigged,gelfand1964generalized}) are not square-integrable and therefore it is literal nonsense to write $\ket{x}$ as a state because it cannot be plugged into the Born rule.
This is an important point and what it reflects is that the preparation of a pure position eigenstate cannot be physical.
Still, this does not exclude position as a legitimate measurement but rather only concludes that the Kraus operators of a position measurement cannot be position eigenstate projection operators.
Instead, the most ideal position measurement could only have Kraus operators of the form $\Omega^{\text{Dirac}}_x = \sqrt{dx}\ketbra{\psi_x}{x}$, where the $\ket{\psi_x}$ are post-measurement states (with square-integrable wavefunctions) that are not simply specified by the spectrum of the position operator alone.
For this reason a position measurement cannot be considered ``von Neumann'' while at the same time Dirac's completeness relation,
\begin{equation}
	\int_\R dx \proj{x} = \sum_{x\in \R} \Omega^{\text{Dirac}\dag}_x\Omega^{\text{Dirac}}_x = 1
\end{equation}
actually makes perfect sense.
Said succinctly, position as a POVM makes rigorous sense but the Luders-von Neumann update rule fails to give it a rigorous instrument.
Regardless, the post-measurement states of a Dirac position measurement are irrelevant to the dilation of the system Kraus operators in equation \ref{Stinespring} as it only uses the linear functional $\bra{x}$ after which no further consideration of the meter is required.\\

\subsection{Total Operations (a.k.a. Channels) and Decoherence}\label{OperationChannels}

Total operations can also be studied in their own right, without reference to an instrument.
In this case, they are defined as superoperators with two properties, usually referred to as ``CP'' and ``TP''.
A superoperator $\Z$ is \emph{completely positive} (CP) if not only does it map positive operators to positive operators, but so too does $\Z \otimes \mathcal{I}d$ for any appended Hilbert space with identity superoperator $\mathcal{I}d$ \cite{bengtsson2006geometry, kitaev2002classical, nielsen2002quantum}.
The superoperator is \emph{trace-preserving} (TP)  if $\tr(\Z(\rho)) = \tr\rho$ for all operators $\rho$.
Meanwhile, instrument elements are CP but not TP unless the total operation is the superoperator identity.

These properties can be elegantly captured in terms of a few superoperator involutions \cite{menicucci2005superoperator}, what we will call the \emph{Hilbert-Schmidt adjoint}
\begin{equation}\label{HSadj}
	\tr\left(Y^\dag\Z^\ddag(X)\right) \equiv \tr\left(\big(\Z(Y)\big)^\dag X\right),
\end{equation}
the \emph{Choi involution}
\begin{equation}\label{Choiinvo}
	(\ketbra{a}{c}\odot\ketbra{d}{b})^\sharp \equiv \ketbra{a}{b}\odot\ketbra{d}{c},
\end{equation}
and the \emph{Choi-Jamiolkowski quasi-adjoint}
\begin{equation}\label{CJquasiadj}
	\Z^\qddag \equiv \left(\left(\Z^\sharp\right)^\ddag\right)^\sharp.
\end{equation}
It is easy to then verify the following useful formulae
\begin{equation}
	(A \odot B^\dag )^\ddag \equiv A^\dag \odot B
	\hspace{25pt}
	\text{and therefore}
	\hspace{25pt}
	(A \odot B^\dag)^{\qddag} \equiv B \odot A^\dag
\end{equation}
for which it is in turn easy to observe the properties
\begin{equation}
	(\Y\circ\X)^\ddag = \X^\ddag \circ \Y^\ddag
	\hspace{25pt}
	\text{and}
	\hspace{25pt}
	(\Y\circ\X)^\qddag = \Y^\qddag \circ \X^\qddag
\end{equation}
where it should be noted in particular that the quasi-adjoint \emph{does not} reverse the order of the superoperator composition.
With these involutions, it is easy to see TP is simply the condition $\Z^\ddag(1)=1$.
Further, the \emph{Choi-Jamiolkowski theorem} may be phrased as the statement that $\Z$ is CP if and only if $\Z^\sharp$ is positive, for which it is a corollary that $\Z^\qddag=\Z$.

Of course, the total operation does not supply any information about which register may have been observed or even what kinds of registers are being observed.
In this case that the register becomes a hidden variable, and all the various features of the instrument take on different names because of their change in purpose:
The meter is called the \emph{environment}, the sytem is said to be \emph{open} to the environment, and the CPTP map is called a \emph{(decoherence) channel}.
Meanwhile, every channel can be ``dilated'' into an instrument like in equation~\ref{CPTP}, in which context equation~\ref{CPTP} is called a Kraus decomposition, an operator-sum decomposition, or an \emph{unraveling} \cite{wiseman2009quantum, carmichael2009open, kraus1983states}.
In the context of measurement theory and discussions of the observer, one must be careful when using these terms because of how they center the literal discussion around decoherence and dissipative processes, problems for which the instrument itself loses priority because the users of the channel usually don't know how much of the decoherence is due to the presence of an observer, called the evesdropper.
If the focus is on the measuring process and the observer (which it is for this paper) \textbf{one must remember that the value of the register is not hidden}.
Rather, it is quite the opposite.
The registered data is in fact the only thing the observer is able to see directly.\\

\subsection{Continuous-in-Time Decoherence/Dissipation}\label{Decoherence}

Infinitesimally generated channels have a particular form called a \emph{Lindbladian}.
Lindbladians include unitary evolution which is generated by superoperators of the form
\begin{equation}
	-\ad[iH] \equiv -iH \odot 1 + 1 \odot iH,
\end{equation}
for any Hermitian operator $H$, called the Hamiltonian. 
In addition, there are Lindbladians for dissipative processes, the generators of which are sometimes called \emph{dissipators} and are of the form
\begin{equation}\label{diss}
	\D[\hat{L}] = \hat{L} \odot \hat{L}^\dag - \frac12 (\hat{L}^\dag \hat{L} \odot 1 + 1 \odot \hat{L}^\dag \hat{L})
\end{equation}
where such an $\hat{L}$ is called a \emph{Lindblad operator} \cite{lindblad1976generators}.
\textcolor{red}{The $\hat{L}^\dag \hat{L}$ of the last two terms in equation \ref{diss} can be considered a non-Hermitian (see, for example \cite[section 2.1.3]{albert2017thesis}) perturbation to the Hamiltonian which corresponds to the shift and decay of the system's energy levels due to interacting with an environment or meter (as described in appendix~\ref{SMI_Models}.)}

General Lindbladians are the sum of a Hamiltonian superoperator and any number of dissipators.
Dissipative channels are therefore generated by infinitesimal channels of the form
\begin{equation}\label{DissChan}
	\Z^{(\hat{L})}_{dt} \equiv e^{\D[\hat{L}]\kappa dt}
\end{equation}
where $dt$ is an infinitesimal time and $\kappa$ is a rate with units of inverse time so that the Lindblad operators are unitless.
Such dissipative channels can obviously be extend to finite times $dt \rightarrow T$,
\begin{align}\label{DissChan2}
	\Z^{(\hat{L})}_{T}
	&= \left(\Z^{(\hat{L})}_{dt}\right)^{\circ T/dt}\\
	&= e^{\D[\hat{L}]\kappa T}
\end{align}
where $\Z^{\circ n}$ denotes the composition of $\Z$ with itself $n$ times.
Such superoperators are said to be Markovian and satify the \emph{(one-parameter) semigroup property},
\begin{equation}\label{OneParamSemi}
	\Z_{t+\Delta t} = \Z_{\Delta t}\circ\Z_t,
\end{equation}
where the semigroup can be considered to refer to the set of channels $\{\Z_t\}_{t\ge0}$,
but it often can also refer implicitly to the semigroup of composite instrument elements as will be explained later (in section \ref{ContMeas}, equation \ref{ConvoSemi}, and section \ref{OD}, equation \ref{ConvoProp}).
A state that evolves according to this channel $\rho_t = \Z_t(\rho_0)$ satisfies a partial differential equation
\begin{equation}\label{LindbladMaster}
	\frac{1}{\kappa}\frac{\partial}{\partial t} \rho_t = \D[\hat{L}](\rho_t)
\end{equation}
and equations of this type (which more generally include a Hamiltonian and mutiple Lindblad operators) are called \emph{Lindblad master equations}, the term ``master'' referring to the point that if one knew $\rho_t$ then they would inherently know the expectation value $\tr (X \rho_t)$ for all observables $X$.\\

\section{System-Meter-Interaction Models}\label{SMI_Models}

The Kraus operators of an instrument can be thought of as the back-actions on the system which are due to an indirect measurement where a meter is strongly measured after it has interacted with the system.
More precisely, a system-meter-interaction model consists of a meter with an pointer state that is measured after interacting with the system of interest.
Because we are interested in weak and continuous measurements, the meter will be modeled as a continuous variable, but one could also use qubits and it would make no difference because the interactions are weak.
(For a treatment with qubit meters, recommended sources are \cite{strandberg2024digital, gross2018qubit, wiseman1994quantum}.)
On the other hand, the system is treated generally.\\

The meter is considered to be a mode of the electromagnetic field, modeled by a standard continuous variable consisting of a pointer in an initial pure state $\ket{0}_\text{M}$ with quadrature observables $Q_\text{M}$ and $P_\text{M}$, satisfying the canonical commutation relation
\begin{equation}
	[Q_\text{M}, P_\text{M}]  = i \hbar 1_\text{M}.
\end{equation}
Let $\bra{q}_\text{M}$ be the basis of position quadrature functionals, upon which the actions of the quadrature observables are
\begin{align}
	\bra{q}_\text{M}Q_\text{M} = q\bra{q}_\text{M}
	\hspace{50pt}
	\text{and}
	\hspace{50pt}
	\bra{q}_\text{M}P_\text{M} = -i\hbar\frac{\partial}{\partial q}\bra{q}_\text{M}.
\end{align}
The initial pointer state is assumed to be vacuum, with Gaussian wavefunction
\begin{align}
	\sqrt{dq}\,\psi_\text{M}(q)
	&\equiv\sqrt{dq}\braket{q}{0}_{\text{M}}\\
	&= \sqrt{\frac{dq}{\sqrt{2\pi \sigma^2}}e^{-q^2/2\sigma^2}}
\end{align}
and annihilator
\begin{equation}
	a_\text{M} = \frac{1}{2\sigma}Q_\text{M} + i \frac{\sigma}{\hbar}P_\text{M}
\end{equation}
defined by the property
\begin{equation}
	a_\text{M}\ket{0}_{\!\text{M}} = 0.
\end{equation}

The initial pointer then interacts with the system of interest according to a unitary transformation
\begin{align}
	\UU_\text{int}
	&= e^{-i\sqrt{\kappa dt}\left(X_\text{S}\otimes 2\frac{\sigma}{\hbar} P_\text{M} -Y_\text{S}\otimes \frac{1}{\sigma}Q_\text{M}\right)}
\end{align}
which can be interpreted as the system observables $X_\text{S}$ and $Y_\text{S}$ driving a translation (change in position) and a boost (change in momentum) in the meter, considered to occur for a time $dt$ at a rate $\kappa$.
The two system observables can be put together to define a Lindblad operator
\begin{align}
	L_\text{S} = X_\text{S} + i Y_\text{S}
\end{align}
with which the interaction unitary affords the alternative expression
\begin{align}
	\UU_\text{int}
	= e^{\sqrt{\kappa dt}\left(L_\text{S}\otimes a_\text{M}^\dag-L_\text{S}^\dag\otimes a_\text{M} \right)}.
\end{align}
This expression may then be processed into a form which will clarify the action of the system-meter unitary on the initial pointer state:
\begin{align}
	\UU_\text{int}
	&= e^{\sqrt{\kappa dt} (L_\text{S}\otimes a_\text{M}^\dag-L_\text{S}^\dag\otimes a_\text{M})}\\
	&= e^{\sqrt{\kappa dt} L_\text{S}\otimes a_\text{M}^\dag}e^{-\sqrt{\kappa dt}L_\text{S}^\dag\otimes a_\text{M}}e^{-\frac12 \kappa dt [L_\text{S}\otimes a_\text{M}^\dag,-L_\text{S}^\dag\otimes a_\text{M}]}\\
	&= e^{\sqrt{\kappa dt} L_\text{S}\otimes a_\text{M}^\dag}e^{-\sqrt{\kappa dt}L_\text{S}^\dag\otimes a_\text{M}}
	e^{\frac12 \kappa dt \left([L_\text{S},L_\text{S}^\dag]\otimes a_\text{M}^\dag a_\text{M}+L_\text{S}^\dag L_\text{S}\otimes [a_\text{M}^\dag,a_\text{M}]\right)}\\
	&= e^{\sqrt{\kappa dt} L_\text{S}\otimes a_\text{M}^\dag}e^{-\sqrt{\kappa dt}L_\text{S}^\dag\otimes a_\text{M}}
	e^{\frac12 \kappa dt [L_\text{S},L_\text{S}^\dag]\otimes a_\text{M}^\dag a_\text{M}}e^{-\frac12 \kappa dt L_\text{S}^\dag L_\text{S}\otimes 1_M}.
\end{align}
Acting on the initial pointer state,
\begin{align}
	\UU_\text{int}\ket{0}_{\!\text{M}}
	&= e^{\sqrt{\kappa dt} L_\text{S}\otimes a_\text{M}^\dag}\ket{0}_{\!\text{M}} e^{-\frac12 \kappa dt L_\text{S}^\dag L_\text{S}}.
\end{align}

\subsection{Jump Processes}\label{JumpPro}
For jump processes, the meter finally terminates in a number measurement.
The Kraus operators are therefore
\begin{align}
	\bra{n}_\text{M}\UU_\text{int}\ket{0}_{\!\text{M}}
	&=\bra{n}_\text{M}e^{\sqrt{\kappa dt} L_\text{S}\otimes a_\text{M}^\dag}\ket{0}_{\!\text{M}} e^{-\frac12 \kappa dt L_\text{S}^\dag L_\text{S}}\\
	&=\bra{0}_\text{M}\frac{a_\text{M}^n}{\sqrt{n!}}e^{\sqrt{\kappa dt} L_\text{S}\otimes a_\text{M}^\dag}\ket{0}_{\!\text{M}} e^{-\frac12 \kappa dt L_\text{S}^\dag L_\text{S}}\\
	&=\bra{0}_\text{M}\frac{(1_\text{S}\otimes a_\text{M})^n}{\sqrt{n!}}e^{\sqrt{\kappa dt} L_\text{S}\otimes a_\text{M}^\dag}\ket{0}_{\!\text{M}} e^{-\frac12 \kappa dt L_\text{S}^\dag L_\text{S}}\\
	&=\bra{0}_\text{M}e^{\sqrt{\kappa dt} L_\text{S}\otimes a_\text{M}^\dag}\frac{(1_\text{S}\otimes a_\text{M} + \sqrt{\kappa dt} L_\text{S}\otimes1_\text{M})^n}{\sqrt{n!}}\ket{0}_{\!\text{M}} e^{-\frac12 \kappa dt L_\text{S}^\dag L_\text{S}}\\
	&=\bra{0}_\text{M}\frac{(1_\text{S}\otimes a_\text{M} + \sqrt{\kappa dt} L_\text{S}\otimes1_\text{M})^n}{\sqrt{n!}}\ket{0}_{\!\text{M}} e^{-\frac12 \kappa dt L_\text{S}^\dag L_\text{S}}\\
	&=\bra{0}_\text{M}\frac{(\sqrt{\kappa dt} L_\text{S}\otimes1_\text{M})^n}{\sqrt{n!}}\ket{0}_{\!\text{M}} e^{-\frac12 \kappa dt L_\text{S}^\dag L_\text{S}}\\
	&=\frac{(\sqrt{\kappa dt} L_\text{S})^n}{\sqrt{n!}} e^{-\frac12 \kappa dt L_\text{S}^\dag L_\text{S}}\\
	&=\begin{cases}
		e^{-\frac12 \kappa dt L_\text{S}^\dag L_\text{S}}& : n=0\\
		\sqrt{\kappa dt} L_\text{S}& : n=1\\
		\sqrt{\OO\big((\kappa dt)^2\big)}& : n\ge2
	\end{cases}\\
	&=\sqrt{(\kappa dt)^{dN}} e^{-\frac12 \kappa dt L_\text{S}^\dag L_\text{S}}L_\text{S}^{dN}
\end{align}
as was to be shown, where $dN = n$ denotes a standard Poisson increment.
The Kraus operators for $n\ge2$ make no contribution because the corresponding instrument elements are $\OO\big((\kappa dt)^2\big)$.

\subsection{Diffusive Processes}\label{DiffPro}

For diffusive processes, the meter terminates in a position quadrature measurement.
The Kraus operators are therefore
\begin{align}
	\sqrt{dq}\bra{q}_\text{M}\UU_\text{int}\ket{0}_{\!\text{M}}
	&=\sqrt{dq}\bra{q}_\text{M}e^{\sqrt{\kappa dt} L_\text{S}\otimes a_\text{M}^\dag}\ket{0}_{\!\text{M}}e^{-\frac12 \kappa dt L_\text{S}^\dag L_\text{S}}\\
	&= \sqrt{dq}e^{\sqrt{\kappa dt} L_\text{S}\left(\frac{q}{2\sigma} - \sigma\partial_q\right)}\braket{q}{0}_{\!\text{M}}e^{-\frac12 \kappa dt L_\text{S}^\dag L_\text{S}}\\
	&= \sqrt{dq}e^{-\frac12\kappa dtL_\text{S}^2[\frac{q}{2},-\partial_q]}e^{\sqrt{\kappa dt} L_\text{S}\frac{q}{2\sigma}}e^{-\sqrt{\kappa dt} L_\text{S} \sigma\partial_q}\psi_\text{M}(q)e^{-\frac12 \kappa dt L_\text{S}^\dag L_\text{S}}\\
	&= \sqrt{dq}e^{-\frac14\kappa dtL_\text{S}^2}e^{\sqrt{\kappa dt} L_\text{S}\frac{q}{2\sigma}}\psi_\text{M}\left(q-\sigma\sqrt{\kappa dt} L_\text{S}\right)e^{-\frac12 \kappa dt L_\text{S}^\dag L_\text{S}}\\
	&= \sqrt{\frac{dq}{\sqrt{2\pi \sigma^2}}}e^{-\frac14\kappa dtL_\text{S}^2}e^{\sqrt{\kappa dt} L_\text{S}\frac{q}{2\sigma}}e^{-\frac{1}{4\sigma^2}(q-\sigma\sqrt{\kappa dt} L_\text{S})^2}e^{-\frac12 \kappa dt L_\text{S}^\dag L_\text{S}}\\
	&= \sqrt{\frac{dq}{\sqrt{2\pi \sigma^2}}}e^{-\frac{q^2}{4\sigma^2}}e^{-\frac12\kappa dtL_\text{S}^2}e^{\sqrt{\kappa dt} L_\text{S}\frac{q}{\sigma}}e^{-\frac12 \kappa dt L_\text{S}^\dag L_\text{S}}\\
	&= \sqrt{\frac{dq}{\sqrt{2\pi \sigma^2}}e^{-\frac{q^2}{2\sigma^2}}}e^{-\frac12\kappa dt(L_\text{S}^\dag L_\text{S} + L_\text{S}^2)}e^{\sqrt{\kappa dt} L_\text{S}\frac{q}{\sigma}}\\
	&= \sqrt{\frac{d(dW)}{\sqrt{2\pi dt}}e^{-\frac{dW^2}{2dt}}}e^{-\frac12\kappa dt(L_\text{S}^\dag L_\text{S} + L_\text{S}^2)}e^{L_\text{S}\sqrt\kappa dW}
\end{align}
where the standard Wiener increment has been defined.
\begin{equation}
	dW \equiv \sqrt{dt}\frac{q}{\sigma}.
\end{equation}
In this case, it is also instructive to express the Kraus operators in terms of the system observables $X_\text{S}$ and $Y_\text{S}$,
\begin{align}
	\sqrt{\frac{d(dW)}{\sqrt{2\pi dt}}e^{-\frac{dW^2}{2dt}}}e^{-\frac12\kappa dt(L_\text{S}^\dag L_\text{S} + L_\text{S}^2)}e^{L_\text{S}\sqrt\kappa dW}
	&= \sqrt{\frac{d(dW)}{\sqrt{2\pi dt}}e^{-\frac{dW^2}{2dt}}}e^{-\frac12\kappa dt\left(X_\text{S}^2+Y_\text{S}^2+[X_\text{S},iY_\text{S}] + X_\text{S}^2-Y_\text{S}^2+\{X_\text{S},iY_\text{S}\} \right)}e^{(X_\text{S} + iY_\text{S})\sqrt\kappa dW}\\
	&= \sqrt{\frac{d(dW)}{\sqrt{2\pi dt}}e^{-\frac{dW^2}{2dt}}}e^{-X_\text{S}^2\kappa dt - i\{X_\text{S},Y_\text{S}\}\frac12\kappa dt}e^{iY_\text{S}\sqrt\kappa dW+X_\text{S}\sqrt\kappa dW+[iY_\text{S},X_\text{S}]\frac12\kappa dt}\\
	&= \sqrt{\frac{d(dW)}{\sqrt{2\pi dt}}e^{-\frac{dW^2}{2dt}}}e^{-X_\text{S}^2\kappa dt - i\{X_\text{S},Y_\text{S}\}\frac12\kappa dt}e^{iY_\text{S}dW}e^{X_\text{S}\sqrt\kappa dW}\\
	&= e^{- i\{X_\text{S},Y_\text{S}\}\frac12\kappa dt+iY_\text{S}dW}\sqrt{\frac{d(dW)}{\sqrt{2\pi dt}}e^{-\frac{dW^2}{2dt}}}e^{-X_\text{S}^2\kappa dt + X_\text{S}\sqrt\kappa dW}.\label{quadratures}
\end{align}
In this form, there are at least three basic features that can be understood.
The first is that it is only the observable $X_\text{S}$ that generates POVM | that is, the POVM has no dependence on the observable $Y_\text{S}$.
This in turn leads to the second feature, that the back-action generated $Y_\text{S}$ is instead an inverse post-measurement unitary transformation conditioned on the latest Wiener increment, $dW$.
This conditional unitary is automatic, simply due to the interaction, but it could also be considered equivalent to a basic feedforward mechanism.
Third is that if $X_\text{S}=0$, then the instrument amounts to just a random unitary.
Otherwise, the conditional unitary transformation is accompanied by an unconditional drift generated by the quadratic anticommutator $\frac12\{X_\text{S},Y_\text{S}\}$.

In practice, the meter-quadrature measurement involves mixing the meter with a local oscillator and the phase of the meter-quadrature measured is determined by the relative phase between the meter and the local oscillator.
As the above calculation assumed a position quadrature measurement, the relative phase has been chosen to be zero.
If the phase of the local oscillator is changed to $\phi$, then this would be equivalent to multiplying the Lindblad operator of the system to $e^{-i\phi}L_\text{S}$, as can be easily shown:
\begin{align}
	\bra{q}_\text{M} e^{-ia^\dag_\text{M}a_\text{M}\phi}\UU_\text{int}\ket{0}_\text{M}
	&=\bra{q}_\text{M} e^{-ia^\dag_\text{M}a_\text{M}\phi}e^{\sqrt{\kappa dt} (L_\text{S}\otimes a_\text{M}^\dag-L_\text{S}^\dag\otimes a_\text{M})}\ket{0}_\text{M}\\
	&=\bra{q}_\text{M} e^{\sqrt{\kappa dt} (L_\text{S}\otimes a_\text{M}^\dag e^{-i\phi} - L_\text{S}^\dag\otimes a_\text{M}e^{i\phi})}e^{-ia^\dag_\text{M}a_\text{M}\phi}\ket{0}_\text{M}\\
	&=\bra{q}_\text{M} e^{\sqrt{\kappa dt} (L_\text{S}\otimes a_\text{M}^\dag e^{-i\phi} - L_\text{S}^\dag\otimes a_\text{M}e^{i\phi})}\ket{0}_\text{M}\\
	&=\bra{q}_\text{M} e^{\sqrt{\kappa dt} (e^{-i\phi}L_\text{S}\otimes a_\text{M}^\dag - (e^{-i\phi}L_\text{S})^\dag\otimes a_\text{M})}\ket{0}_\text{M}.
\end{align}

\section{Invariant One-Forms, Haar Measures, and the Modular Function}\label{OneForms}

Every finite-dimensional Lie group has two vector-valued one-forms called the so-called Maurer-Cartan (MC) forms.
The left-invariant MC form is
\begin{align}
	\omega_{\rL}(x)
	&\equiv x^\inv dx\\
	&= x^\inv g^\inv g dx\\
	&= (gx)^\inv d(gx)\\
	&= \omega_{\rL}(gx)
\end{align}
and the right-invariant MC form is
\begin{align}
	\omega_{\rR}(x)
	&\equiv dx \, x^\inv\\
	&= dx \,gg^\inv x^\inv\\
	&= d(xg) \, (xg)^\inv\\
	&= \omega_{\rR}(xg).
\end{align}
Fixing a basis $\{X_\alpha\}$ of the Lie algebra, the differential may be expanded in terms of either their left- or right-invariant derivatives
\begin{align}
	dx
	&= \theta_{\rL}^\mu(x) \Linv{X_\alpha}[x]\\
	&= \theta_{\rL}^\mu(x) x X_\alpha\label{LeftInvOneForms}\\
	&= \theta_{\rR}^\mu(x) \Rinv{X_\alpha}[x]\\
	&= \theta_{\rR}^\mu(x) X_\alpha x\label{RightInvOneForms}
\end{align}
which means that the left- and right-invariant MC forms can be expanded as
\begin{align}
	\omega_{\rL}(x) = X_\alpha \theta^\alpha_\rL(x)
	\hspace{50pt}
	\text{and}
	\hspace{50pt}
	\omega_{\rR}(x)
	= X_\alpha \theta^\alpha_\rR(x).
\end{align}
The one-forms $\theta^\alpha_\rL(x)$ and $\theta^\alpha_\rR(x)$, dual to the left- and right-invariant derivatives are also left- and right-invariant respectively,
\begin{align}
	\theta^\alpha_\rL(gx) = \theta^\alpha_\rL(x)
	\hspace{50pt}
	\text{and}
	\hspace{50pt}
	\theta^\alpha_\rR(xg) = \theta^\alpha_\rR(x).
\end{align}
These one-forms are examples of what are more generally called solder forms. 
The right-invariant MC form is related to the left-invariant MC form by conjugation
\begin{equation}
	\omega_{\rR} = x \omega_{\rL} x^\inv
\end{equation}
which in turn implies the right-invariant solder forms can be expanded simply in terms of the left-invariant solder forms
\begin{equation}\label{Adj}
	\theta^\beta_\rR(x) = {(\Ad_x)^\beta}_\alpha\theta^\alpha_\rL(x)
\end{equation}
where $\Ad_x$ denotes the adjoint representation
\begin{equation}
	x X_\alpha x^\inv = X_\beta {(\Ad_x)^\beta}_\alpha.
\end{equation}\\

The \emph{left-} and \emph{right-invariant Haar measures} are therefore
\begin{equation}
	d_\rL x \equiv C|\theta^1_\rL(x)\wedge \cdots \wedge \theta^n_\rL(x)|
\end{equation}
and
\begin{equation}
	d_\rR x \equiv C|\theta^1_\rR(x)\wedge \cdots \wedge \theta^n_\rR(x)|
\end{equation}
where $\wedge$ denotes the wedge product, $|\;|$ denotes taking an absolute value of the Jacobian determinant, and $C$ is an arbitary normalization constant (chosen to be the same for both Haar measures.)
From equation \ref{Adj} it is clear
\begin{equation}\label{modular}
	d_\rR x = \Delta(x) d_\rL x
\end{equation}
where defined is the \emph{modular function}
\begin{equation}
	\Delta(x) \equiv |\det \Ad_x|.
\end{equation}
Clearly, the modular function is also a representation of the group
\begin{equation}
	\Delta(xy) = \Delta(x)\Delta(y).
\end{equation}
The group is said to be \emph{unimodular} if the modular function is identically equal to one.
Compact groups, Abelian groups, and nilpotent groups are unimodular.\\

With equation \ref{modular}, it is easy to see the Haar measures are quasi-invariant with respect to their opposite translation,
\begin{equation}
	d_{\rR}(gx) = \Delta(g)d_{\rR}x
	\hspace{50pt}
	\text{and}
	\hspace{50pt}
	d_{\rL}(xg) = \frac{d_{\rL}x}{\Delta(g)}.
\end{equation}
Some further properties of the Haar measures and the group convolution are discussed in Appendix \ref{delta}.

\section{The $\delta$-Function of a Group}\label{delta}

Let $G$ be a finite-dimensional Lie group with left- and right-invariant Haar measures $d_\rL x \,$ and $d_\rR x \,$ as defined in section \ref{ODF}.\\

The \emph{Dirac delta-distribution} $\delta(x)$ of a group $G$ will first be defined with respect to the left-invariant measure to be the \emph{unique} distribution with the property
\begin{equation}
	\int_G d_\rL x \, \delta(x)f(x) = f(e)
\end{equation}
for all functions $f$ over $G$, where $e \in G$ denotes the identity group element.
It follows that the $\delta$ has the same property with respect to the right-invariant measure as well,
\begin{align}
	\int_G d_\rR x \, \delta(x)f(x)
	&= \int_G d_\rL x \, \delta(x)\Delta(x)f(x)\\
	&= \Delta(e)f(e)\\
	&= f(e).
\end{align}
With these properties, we can imagine the $\delta$ as a certain ``function'' sharply peaked at the identity.
Inverting at the identity reveals
\begin{align}
	\int_G d_\rL x \, \delta(x^\inv)f(x)
	&= \int_G d_\rL(\xi^\inv) \delta(\xi) f(\xi^\inv)\\
	&= \int_G d_\rR \xi\, \delta(\xi) f(\xi^\inv)\\
	&= f(e^\inv)\\
	&= f(e)
\end{align}
and so it follows that the delta is symmetric under inversion,
\begin{equation}
	\delta(x^\inv) = \delta(x).
\end{equation}
Conjugating at the identity reveals
\begin{align}
	\int_G d_\rL x \, \delta(g x g^\inv)f(x)
	&= \int_G d_\rL(g^\inv \xi g) \delta(\xi) f(g^\inv \xi g)\\
	&= \int_G d_\rL\xi \delta(\xi) \frac{f(g^\inv \xi g)}{\Delta(g)}\\
	&= \frac{f(g^\inv e g)}{\Delta(g)}\\
	&=\frac{f(e)}{\Delta(g)}
\end{align}
and so it follows that the delta is only quasi-invariant under conjugation if the group is not unimodular
\begin{equation}
	\delta(g x g^\inv) = \frac{\delta(x)}{\Delta(g)}.
\end{equation}
In light of this, it is also interesting to observe the Haar measures are equally quasi-invariant under conjugation
\begin{equation}
	d_\rR(gx g^\inv) = \Delta(g)d_\rR x \,
	\hspace{50pt}
	\text{and}
	\hspace{50pt}
	d_\rL(gxg^\inv) =\Delta(g)d_\rL x \,.
\end{equation}

The $\delta$ can also be translated appropriately to evaluate the function at any other point,
\begin{align}
	\int_G d_\rL x \, \delta(x_0^\inv x)f(x)
	&= \int_G d_\rL(x_0 \xi) \delta(\xi)f(x_0\xi)\\
	&= \int_G d_\rL \xi\, \delta(\xi)f(x_0\xi)\\
	&= f(x_0 e)\\
	&= f(x_0)
\end{align}
and similarly
\begin{align}
	\int_G d_\rR x \, \delta(x x_0^\inv)f(x)
	&= f(x_0).
\end{align}

The $\delta$ can also be considered as a trikernal of the group convolution,
\begin{align}
	g*f (z)
	&\equiv \int d_\rL y \,d_\rR x \, \delta\!\left(y^\inv z x^\inv\right) g(y)f(x)\\
	&= \int d_\rL y \,d_\rL x \, \delta\!\left((yx)^\inv z\right) g(y)f(x)\\
	&= \int d_\rR y \,d_\rR x \, \delta\!\left(z (yx)^\inv\right) g(y)f(x).
\end{align}

\end{document}